# Zero-Field Partition Function and Free Energy Density of the Two-Dimensional Heisenberg Classical Square Lattice

Jacques Curély[1]

**Abstract** We rigorously examine $2d$ square lattices composed of $N_S$ classical spins isotropically coupled. If $H_{i,j}^{ex}$ is the local exchange Hamiltonian each operator $\exp(-\beta H_{i,j}^{ex})$ is expanded on the basis of spherical harmonics $Y_{l_{i,j}, m_{i,j}}$. We derive selection rules for the $l_{i,j}$'s and $m_{i,j}$'s. For infinite $N_S$ the value $m = 0$ is selected. We obtain an exact $l$-polynomial for the zero-field partition function, *valid for any temperature*. Its thermal study allows to point out *crossovers* between the $l$-eigenvalues. Near $T_c = 0$ K we derive a diagram showing three magnetic phases, each one being characterized by the low-temperature behavior of the correlation length. At $T_c = 0$ K, we retrieve the critical exponent $\nu = 1$. We identify three regimes: the renormalized classical, the quantum disordered and the quantum critical regimes. We exactly express the free energy density $\mathcal{F}$. For each result we retrieve the corresponding one derived from the renormalization approach.

**Keywords** Lattice theory and statistics · Classical spin models · Ferrimagnetics

## 1 Introduction

Since the middle of the eighties with the discovery of high-temperature superconductors [1], the nonlinear σ-model in $D = 2 + 1$ dimensions has known a new interest for it notably allows to describe the properties of two-dimensional ($d = 2$) quantum antiferromagnets such as $La_2CuO_4$ [2-5]. Indeed these antiferromagnets, when properly doped, become superconductors up to a critical temperature $T_c$ notably high compared to other types of superconducting materials. In this respect, the nonlinear σ-model has been conjectured to be equivalent at low temperatures to the $2d$-Heisenberg model [6], which in turn can be derived from the Hubbard model in the large $U$-limit [7]. In a seminal paper, Chakravarty *et al.* [4] have studied this model using the method of one-loop renormalization group (RG) improved perturbation theory initially developed by Nelson and Pelkovits [8]. These authors have related the σ-model to the spin-1/2 Heisenberg model by simply considering: (i) a nearest-neighbor $s = 1/2$ antiferromagnetic Heisenberg Hamiltonian on a square lattice characterized by a large exchange energy; (ii) very small interplanar couplings and spin anisotropies. In addition, they have pointed out that the long-wavelength, low-energy properties are well described by a mapping to a *2d-classical Heisenberg* magnet because all the effects of quantum fluctuations can be resorbed by means of adapted renormalizations of the coupling constants. Thus, the particular and important conclusion of Chakravarty's work motivated us to focus again on the 2d-classical O(3) model developed on a square lattice [9,10].

✉ Jacques Curély
  jacques.curely@u-bordeaux.fr

[1] Laboratoire Ondes et Matière d'Aquitaine (LOMA), UMR 5798, Université de Bordeaux, 351 Cours de la Libération, 33405 Talence Cedex, France

In this article we reconsider this study but, now, we develop a *complete* treatment for a 2d square lattice showing edges (case *a*) or cyclic boundary conditions (case *b*) i.e., a 2d lattice wrapped on a torus. $d$ represents the crystallographic dimension and $D = d + 1$ is the space time one. Both types of lattices are composed of classical spins isotropically coupled between first-nearest neighbors. If $H_{i,j}^{ex}$ is the corresponding local exchange Hamiltonian associated with each lattice site $(i, j)$ the evaluation of the zero-field partition function $Z_N(0)$ necessitates to expand each local operator $\exp(-\beta H_{i,j}^{ex})$ on the infinite basis of spherical harmonics.

Each harmonics is characterized by a couple of integers $(l, m)$, with $l \geq 0$ and $m \in [-l, +l]$. The corresponding eigenvalue of each operator $\exp(-\beta H_{i,j}^{ex})$ is nothing but the Bessel function of the first kind $(\pi/2\beta J)^{1/2} I_{l+1/2}(-\beta J)$ where $\beta = 1/k_B T$ is the Boltzmann factor and $J$ the exchange energy between consecutive spin neighbors whereas each spherical harmonics is the eigenfunction. A polynomial expansion can be obtained for the zero-field partition function $Z_N(0)$. Each term of the expansion appears as a product of two parts: i) a radial part involving itself the product of all the $l_{i,j}$-eigenvalues characterizing all the operators $\exp(-\beta H_{i,j}^{ex})$ for the whole lattice; ii) an angular part composed of a product of all the integrals $F_{i,j}$ (with one integral per lattice site $(i, j)$). Each of these integrals $F_{i,j}$ is composed of a product of spherical harmonics with one spherical harmonics per bond connecting site $(i, j)$ to its first nearest neighbors. As a result the polynomial expansion describing the zero-field partition function $Z_N(0)$ is nothing but the characteristic polynomial.

The non-vanishing condition of each integral $F_{i,j}$ leads to a first set of selection rules with a first subset for the *l*'s and a second one for the *m*'s involved. For the whole lattice and for each type of integer *l* or *m*, we obtain a system of imbricate equations (with one equation per lattice site) i.e., equations involving *l*'s on the one hand and *m*'s on the other one due to the fact that each classical spin interacts with its first-nearest neighbors $(i + 1, j)$, $(i, j - 1)$, $(i - 1, j)$ and $(i, j + 1)$. At first sight, when the lattice is finite, it seems that the solution of this system is not unique. In other words no unique closed-form expression exists for the characteristic polynomial associated with the zero-field partition function $Z_N(0)$ so that the statistical problem seems unsolvable from a mathematical point of view.

Thus, in order to obtain a full set of selection rules which allow to find an analytical expression for $Z_N(0)$, we are led to introduce the concept of *decorated lattice* whose existence is duly justified from physical reasons: it is similar to the crystallographic lattice i.e., it is characterized by a square unit cell and composed of two sublattices. Each bond of the first square sublattice is characterized by an integer *l* or *l'* and the whole square sublattice by the set $(…, l_{i,j},…, l'_{i,j}, …)$. Similarly each bond of the second square sublattice is described by a relative integer *m* or *m'* (with $m \in [-l, +l]$, respectively $m' \in [-l', +l']$) and the whole square sublattice by the set $(\ldots, m_{i,j}, \ldots, m'_{i,j}, \ldots)$. Under these conditions and as for the crystallographic space, we examine the effects of symmetries characterizing each decorated sublattice over coefficients $m_{i,j}$ and $m'_{i,j}$ (respectively, coefficients $l_{i,j}$ and $l'_{i,j}$) for deriving more precise selection rules. For doing this work we use a *duality principle* and we define the conditions of validity which must be fulfilled, thus leading to list the nature of the allowed spin couplings and the structure of corresponding spin arrangements. We conclude that *only the 2d classical Heisenberg model is integrable in the case of finite lattices*. Finally we show that, in the thermodynamic limit, the duality principle becomes irrelevant, thus allowing to reconsider all the discarded cases (notably the 2d classical *xy* model) when applying this principle to finite lattices. As a result *all the 2d classical spin systems are fully integrable in the case of infinite lattices*. More particularly, it means that the 2d classical *xy* model, which is nonintegrable on a finite lattice, becomes integrable in the thermodynamic limit.

As a result, if using the symmetries of the decorated lattice i.e., the duality principle, we obtain a second set of selection rules. i) For a finite lattice showing edges (case *a*), all the sites of the vertical or horizontal lines are characterized by the same integer $l_0$ but two different lines show different values $l_0$, except for those ones which are similar under lattice symmetries. As for the *m*'s we rigorously have $m = 0$ for the whole lattice; the same reasoning can be extended to infinite lattices. ii) For a finite lattice wrapped on a torus (case *b*), all the *l*'s (respectively all the *m*'s) show a common value $l_0$ (respectively, $m_0$). iii) Separately, in the thermodynamic limit (i.e., for an infinite lattice), the value $m = 0$ is numerically selected for the torus characterized by two infinite radii of curvature, all the *l*'s remaining equal to the common value $l_0$. This allows to prove that *the application of the duality principle permits to obtain the good value for the m's*, thus bringing a strong validation for this new concept. For the lattice showing edges, only the term of the characteristic polynomial involving *l*'s and *l*"s equal to a common value $l_0$ remains (the *m*'s being already null) so that, as expected, we deal with the same type of lattice as the torus presenting infinite radii of curvature which represents the most interesting physical case. Consequently a very simple closed-form expression can jointly be obtained in cases *a* and *b* for $Z_N(0)$.

In the case of an infinite lattice for which $m = 0$ we report a further thermal study of the *l*-current term of $Z_N(0)$. For sake of simplicity it is restricted to the case of similar exchange energies $J = J_1 = J_2$ but the result can be applied to the general case $J_1 \neq J_2$. As a result, if $J_1 = J_2$, the current *l*-term is reduced to $F_{i,j} \left[ (\pi/2\beta J)^{1/2} I_{l+1/2}(-\beta J) \right]^2$ where the integral $F_{i,j}$ is composed of four spherical harmonics $Y_{l,0}$. Thus we show that it appears *crossovers* between two consecutive terms of the polynomial expansion due to the presence of integrals $F_{i,j} \neq 1$ (whereas for spin chains we always have $F_{i,j} = 1$). Coming from high temperatures, the value $l = 0$ characterizes the dominant term for reduced temperatures such as $k_B T/|J| \geq 0.255$. For $0.255 \geq k_B T/|J| \geq 0.043$ we have $l = 1$ and so on. As a result, *l*-eigenvalues, with increasing $l > 0$, are successively dominant. In the vicinity of absolute zero the dominant term is characterized by $l \to +\infty$ whereas, for consecutive terms characterized by integers $l$ and $l + 1$, the ratio of integrals $F_{i,j}$ tends towards unity. As all the Bessel functions of the first kind $(\pi/2\beta J)^{1/2} I_{l+1/2}(-\beta J)$ have a close behavior when $\beta \to +\infty$ (i.e., $T \to 0$) we derive that, near absolute zero, all the *l*-eigenvalues become equivalent, confirming the fact that the critical temperature is $T_c = 0$ K, in agreement with Mermin-Wagner's theorem [11].

From this low-temperature study we obtain a diagram exhibiting three magnetic phases: it is exactly similar to the one derived from the renormalization group technique [4,5]. In the vicinity of $T_c$ we show that the square root of the convergence ratio of the characteristic poly-nomial associated with $Z_N(0)$ is equal to the spin-spin correlation $|<S_{0,0}.S_{0,1}>|$ (or equivalently $|<S_{0,0}.S_{1,0}>|$ as $J_1 = J_2$) between first-nearest neighbors belonging to the same lattice line (or row). Then, owing to the Ornstein-Zernike theorem which gives the decay law of the spin-spin correlation *vs* distance, we derive three low-temperature behaviors of the correlation length i.e., one per magnetic phase of the diagram, in agreement with previous results obtained from a renormalization technique [5]. At $T_c = 0$ K the critical exponent is $\nu = 1$, as expected. As a result we identify three regimes: the Renormalized Classical Regime (RCR), the Quantum Disordered Regime (QDR) and the Quantum Critical Regime (QCR). Finally, from the exact closed-form expression of $Z_N(0)$, we express the free energy density $\mathcal{F}$. Near $T_c = 0$ K and for each magnetic phase, we respectively obtain the same expression than the corresponding one derived from the renormalization group approach [5].

## 2 Evaluation of the Zero-Field Partition Function

### 2.1 Definitions

We start from the general Hamiltonian describing a lattice characterized by edges, a square unit cell and $(2N + 1)^2$ sites, each one being the carrier of a classical spin $S_{i,j}$:

$$H = \sum_{i=-N}^{N} \sum_{j=-N}^{N} (H_{i,j}^{ex} + H_{i,j}^{mag}), \quad (2.1)$$

with:

$$H_{i,j}^{ex} = (J_1 S_{i,j+1} + J_2 S_{i+1,j}) S_{i,j}, \quad (2.2)$$

$$H_{i,j}^{mag} = -G_{i,j} S_{i,j}^z B, \quad (2.3)$$

where:

$$G_{i,j} = G \ (i + j \text{ even or null}), \ G_{i,j} = G' \ (i + j \text{ odd}). \quad (2.4)$$

In (2.2) we recall that $J_1$ and $J_2$ refer to the exchange interaction between nearest neighbors belonging to the same horizontal lines and vertical rows of the lattice, respectively. In addition $J_i > 0$ (respectively, $J_i < 0$, with $i = 1, 2$) denotes an antiferromagnetic (respectively, ferromagnetic) coupling. $G_{i,j}$ is the Landé factor characterizing each spin $S_{i,j}$ and is expressed in $\mu_B/\hbar$ unit. For the absent edge bonds the exchange Hamiltonian $H_{i,j}^{ex}$ must be corrected by assuming vanishing exchange energies. Finally we consider that the classical spins $S_{i,j}$ are unit vectors so that the exchange energy $JS^2$ is written $J$. It means that we do not take into account the number of spin components in the normalization of $S_{i,j}$'s so that $S^2 = 1$.

The square lattice showing edges (case *a*) can also be wrapped on a torus (case *b*) so that we deal with cyclic boundary conditions. There are three possibilities: (i) the horizontal lines $i = N$ and $i = -N$ (respectively, the vertical rows $j = N$ and $j = -N$) are linked owing to $2N + 1$ vertical (respectively, horizontal) extra bonds (case *b*1); we always have $(2N + 1)^2$ sites as well as $(2N + 1)^2$ horizontal (respectively, vertical) bonds; (ii) the horizontal lines $i = N$ and $i = -N$ (respectively, the vertical rows $j = N$ and $j = -N$) coincide (case $b2$) so that we deal with $(2N)^2$ sites as well as $(2N)^2$ horizontal (respectively, vertical) bonds; iii) the last solution consists in a mix of cases $b1$ and $b2$: for instance the horizontal lines $i = N$ and $i = -N$ coincide whereas the vertical rows $j = N$ and $j = -N$ are linked by means of $N_e$ extra horizontal bonds on line $i = 0$ ($2NN_e$ for the $2N$ horizontal lines $i$); we then deal with $2N(2N + N_e)$ sites as well as $2N + N_e$ horizontal ($2N$ vertical) bonds. From now, in this article, we shall exclusively consider case $b3$ that we shall now label case $b$. This choice will be justified later. In addition, in the infinite-lattice limit (i.e., in the thermodynamic limit $N \to +\infty$), we expect that cases $a$ and $b$ show the same asymptotic partition function because edge effects are ineffective.

For the general lattice previously described the field-dependent partition function $Z_N(B)$ is defined as:

$$Z_N(B) = \int d\mathbf{S}_{-N,-N} \ldots \int d\mathbf{S}_{i,j} \ldots \int d\mathbf{S}_{N,N} \exp\left(-\beta \sum_{i=-N}^{N} \sum_{j=-N}^{N} \left(H_{i,j}^{\text{ex}} + H_{i,j}^{\text{mag}}\right)\right), \qquad (2.5)$$

where the Boltzmann factor $\beta = 1/k_B T$ must not be confused with the critical exponent $\beta_c$. At this step it must be noticed that the calculation of the field-dependent partition function $Z_N(B)$ is plainly more complicated because of the presence of the further term $H_{i,j}^{\text{mag}}$ in the exponential argument, for each site $(i, j)$. This aspect is not examined in the present article. Under these conditions, the zero-field partition $Z_N(0)$ is simply obtained by integrating the operator $\exp(-\beta H^{\text{ex}})$ over all the angular variables characterizing the states of all the classical spins belonging to the lattice.

## 2.2 Zero-Field Partition Function of a Finite Lattice

### 2.2.1 Generalities

Because of the presence of classical spin momenta, all the operators $H_{i,j}^{\text{ex}}$ commute and the corresponding exponential factor appearing in the integrand of (2.5) considered in the zero-field limit can be written:

$$\exp\left(-\beta \sum_{i=-N}^{N} \sum_{j=-N}^{N} H_{i,j}^{\text{ex}}\right) = \prod_{i=-N}^{N} \prod_{j=-N}^{N} \exp\left(-\beta H_{i,j}^{\text{ex}}\right). \qquad (2.6)$$

As a result, the particular nature of $H_{i,j}^{\text{ex}}$ given by (2.2) allows one to separate the contributions corresponding to the exchange coupling involving classical spins belonging to the same horizontal line $i$ of the layer (i.e., $\mathbf{S}_{i,j-1}$, $\mathbf{S}_{i,j+1}$ and $\mathbf{S}_{i,j}$) or to the same vertical row $j$ (i.e., $\mathbf{S}_{i-1,j}$, $\mathbf{S}_{i+1,j}$ and $\mathbf{S}_{i,j}$). In fact, for each of the four contributions (one per bond connected to the site $(i, j)$ carrying the spin $\mathbf{S}_{i,j}$), we have to expand a term such as $\exp(-A\mathbf{S}_1 \cdot \mathbf{S}_2)$ where $A$ is $\beta J_1$ or $\beta J_2$ (the classical spins $\mathbf{S}_1$ and $\mathbf{S}_2$ being considered as unit vectors). If we call $\Theta_{1,2}$ the angle between vectors $\mathbf{S}_1$ and $\mathbf{S}_2$, respectively characterized by the couples of angular variables $(\theta_1, \varphi_1)$ and $(\theta_2, \varphi_2)$, it is possible to expand the operator $\exp(-A\cos\Theta_{1,2})$ on the infinite basis of spherical harmonics which are also eigenfunctions of the angular part of the Laplacian operator on the sphere of unit radius $\mathcal{S}^2$:

$$\exp(-A\cos\Theta_{1,2}) = 4\pi \sum_{l=0}^{+\infty} \left(\frac{\pi}{2A}\right)^{1/2} I_{l+1/2}(-A) \sum_{m=-l}^{+l} Y_{l,m}^*(\mathbf{S}_1) Y_{l,m}(\mathbf{S}_2). \tag{2.7}$$

In the previous equation $\mathbf{S}_1$ and $\mathbf{S}_2$ symbolically represent the couples of angles $(\theta_1, \varphi_1)$ and $(\theta_2, \varphi_2)$ which can be defined in the classical spin space $\mathcal{S}$. $l$ and $m$ (with $m \in [-l, +l]$) are quantum numbers characterizing the classical spin states i.e., the orientation of the spin momentum in $\mathcal{S}$. If we set:

$$\lambda_l(-\beta j) = \left(\frac{\pi}{2\beta j}\right)^{1/2} I_{l+1/2}(-\beta j), \; j = J_1 \text{ or } J_2, \tag{2.8}$$

each local operator $\exp(-\beta H_{i,j}^{ex})$ is finally expanded on the basis of eigenfunctions (the spherical harmonics), whereas the $\lambda_l$'s are nothing but the associated eigenvalues. Under these conditions, the zero-field partition function $Z_N(0)$ directly appears as a characteristic polynomial and can be written as:

$$Z_N(0) = (4\pi)^{f(N)} \prod_{i=-N}^{N} \prod_{j=-N}^{N} \sum_{l_{i,j}=0}^{+\infty} \lambda_{l_{i,j}}(-\beta J_1) \sum_{l'_{i,j}=0}^{+\infty} \lambda_{l'_{i,j}}(-\beta J_2) \sum_{m_{i,j}=-l_{i,j}}^{+l_{i,j}} \sum_{m'_{i,j}=-l'_{i,j}}^{+l'_{i,j}} F_{i,j} \tag{2.9}$$

with:

$$f(N) = 4N(2N+1) \text{ case } a \text{ (lattice with edges)},$$

$$f(N) = 4N(2N + N_e) \text{ case } b \text{ (lattice wrapped on a torus)} \tag{2.10}$$

where $f(N)$ represents the total number of lattice bonds.

### 2.2.2 Finite Lattice Showing Edges (case *a*)

In the case of a lattice showing edges (case *a*) we recall that, in (2.9), some contributions must be dropped when the spin momenta are carried by one of the lattice edge sites. In other words, the spherical harmonics corresponding to the absent bond(s) must be replaced by unity i.e., they are characterized by $l = 0$ and $m = 0$. Consequently, if we define

$$F_{i,j} = \int d\mathbf{S}_{i,j} Y_{l'_{i+1,j}, m'_{i+1,j}}(\mathbf{S}_{i,j}) Y_{l_{i,j-1}, m_{i,j-1}}(\mathbf{S}_{i,j}) Y_{l_{i,j}, m_{i,j}}^*(\mathbf{S}_{i,j}) Y_{l'_{i,j}, m'_{i,j}}^*(\mathbf{S}_{i,j}) \tag{2.11}$$

as the current integral appearing in (2.9) it can involve the product of two, three or four spherical harmonics (one per lattice bond linked with the current site $(i, j)$) according to that the considered site belongs to a corner, an edge or to the inside part of the lattice. For two corners of the lattice i.e., for sites $(-N, -N)$ and $(N, N)$, $F_{i,j}$ reduces to:

$$\int d\mathbf{S} \; Y_{l',m'}^*(\mathbf{S}) Y_{l,m}(\mathbf{S}) = \delta_{l,l'} \delta_{m,m'} \tag{2.12}$$

where $\delta_{l,l'}$ (respectively, $\delta_{m,m'}$) is the Kronecker symbol. For site $(N, -N)$ $F_{i,j}$ is composed of the product of two conjugated spherical harmonics whereas at site $(-N, N)$ the product only involves two non-conjugated spherical harmonics. But, globally, owing to a well-known analytical property between a spherical harmonics and its conjugated form

$$Y_{l,m}^{*}(\mathbf{S}) = (-1)^{m} Y_{l,-m}(\mathbf{S}), \tag{2.13}$$

it is always possible to have an integral such as the one given by (2.12). Under that condition we directly have the selection rules for the lattice corners:

$$l_{-N,-N} = l'_{-(N-1),-N}, \; m_{-N,-N} = m'_{-(N-1),-N}, \; \text{site } (-N, -N),$$

$$l_{N,-N} = l'_{N,-N}, \; m_{N,-N} = -m'_{N,-N}, \; \text{site } (N, -N),$$

$$l_{-N,N-1} = l'_{-(N-1),N}, \; m_{-N,N-1} = -m'_{-(N-1),N}, \; \text{site } (-N, N), \tag{2.14}$$

$$l_{N,N-1} = l'_{N,N}, \; m_{N,N-1} = m'_{N,N}, \; \text{site } (N, N).$$

Using the following formula (i.e., a Clebsch-Gordan series) [12]

$$Y_{l_1,m_2}(\mathbf{S})Y_{l_2,m_2}(\mathbf{S}) = \sum_{L=|l_1-l_2|}^{l_1+l_2} \sum_{M=-L}^{+L} \left[\frac{(2l_1+1)(2l_2+1)}{4\pi(2L+1)}\right]^{1/2} C_{l_1\;0\;l_2\;0}^{L\;0} C_{l_1\;m_1\;l_2\;m_2}^{L\;M} Y_{L,M}(\mathbf{S}) \tag{2.15}$$

for any pair of spherical harmonics appearing in the integrand of $F_{i,j}$ we have for the lattice in-sites characterized by four spherical harmonics:

$$F_{i,j} = \frac{1}{4\pi}\left[(2l'_{i+1,j}+1)(2l_{i,j-1}+1)(2l_{i,j}+1)(2l'_{i,j}+1)\right]^{1/2} \sum_{L_{i,j}=L_<}^{L_>} \frac{1}{2L_{i,j}+1}$$

$$\times \sum_{M_{i,j}=-L_{i,j}}^{+L_{i,j}} C_{l'_{i+1,j}\;0\;l_{i,j-1}\;0}^{L_{i,j}\;0} C_{l'_{i+1,j}\;m'_{i+1,j}\;l_{i,j-1}\;m_{i,j-1}}^{L_{i,j}\;M_{i,j}} \times$$

$$\times C_{l_{i,j}\;0\;l'_{i,j}\;0}^{L_{i,j}\;0} C_{l_{i,j}\;m_{i,j}\;l'_{i,j}\;m'_{i,j}}^{L_{i,j}\;M_{i,j}}. \tag{2.16}$$

In the previous equations $C_{l\;m\;l'\;m'}^{L\;M}$ is a Clebsch-Gordan (CG) coefficient. The CG coefficients $C_{l_{i,j}\;m_{i,j}\;l'_{i,j}\;m'_{i,j}}^{L_{i,j}\;M_{i,j}}$ and $C_{l'_{i+1,j}\;m'_{i+1,j}\;l_{i,j-1}\;m_{i,j-1}}^{L_{i,j}\;M_{i,j}}$ (with $M_{i,j} \neq 0$ or $M_{i,j} = 0$) appearing in (2.16) do not vanish if the triangular inequalities $|l_{i,j} - l'_{i,j}| \leq L_{i,j} \leq l_{i,j} + l'_{i,j}$ and $|l'_{i+1,j} - l_{i,j-1}| \leq L_{i,j} \leq l'_{i+1,j} + l_{i,j-1}$ are fulfilled, respectively. As a result, we must have $L_< = \max(|l'_{i+1,j} - l_{i,j-1}|, |l_{i,j} - l'_{i,j}|)$ and $L_> = \min(l'_{i+1,j} + l_{i,j-1}, l_{i,j} + l'_{i,j})$. For an integral involving three spherical harmonics (and characterizing an edge site), the $L_{i,j}$- and $M_{i,j}$-summations disappear and $F_{ij}$ reduces to the product of two CG coefficients, the two remaining ones being equal to unity because of the non-existing bond of the lattice characterized by vanishing coefficients $l_{i,j}$ and $m_{i,j}$ (or $l'_{i,j}$ and $m'_{i,j}$) [12].

### 2.2.3 Finite Lattice Wrapped on a Torus (case *b*)

In the case of a lattice wrapped on a torus (case *b*3 labeled *b*) the zero-field partition function $Z_N(0)$ is always given by (2.9). The $2N(2N + N_e)$ lattice sites are all equivalent ($N_e$ being the number of extra bonds added to each horizontal line *i*) : there are no more edges and each site is always an in-site characterized by four neighbors. The lattice can finally be constructed by only considering the translation of two bonds per site characterized by the exchange energies $J_1$ and $J_2$, respectively. As a result the integral $F_{i,j}$ is always composed of four spherical harmonics and is exclusively given by (2.16). A similar reasoning can be achieved for cases *b*1 and *b*2.

However in both cases the non-vanishing condition of CG coefficients give more general selection rules than those given by (2.14), for in-sites coefficients $l_{i,j}$, $l'_{i,j}$, $m_{i,j}$ and $m'_{i,j}$. They are derived in a forthcoming subsection (*cf* Sect. 2.2.5).

### 2.2.4 Principles of Construction of the Characteristic Polynomial Associated with the Zero-Field Partition Function

The examination of (2.9) giving the polynomial expansion of the zero-field partition function $Z_N(0)$ allows one to say that its writing is nothing but that one derived from the formalism of the transfer-matrix technique. Each term of (2.9) appears as a product of two subterms: (i) a *temperature-dependent radial factor* containing a product of the various eigenvalues $\lambda_l(-\beta j)$, $j = J_1$ or $J_2$ (with one eigenvalue per bond); (ii) an *angular factor* containing a product of integrals $F_{i,j}$ describing all the spin states of all the lattice sites (with one integral per site). This last factor seems to be temperature-independent. This vision is wrong because the magnetic ordering is described by the set of all the couples of angles $(\theta_{i,j}, \varphi_{i,j})$ describing all the spin states and is highly temperature-dependent (through the set of coefficients $l_{i,j}$ and $l'_{i,j}$, *cf* (2.7)).

As the square lattice contains $4N(2N + 1)$ bonds in case *a* (lattice with edges) i.e., $2N(2N + 1)$ horizontal and vertical bonds or $4N(2N + N_e)$ bonds in case *b* (when the lattice is wrapped on a torus) i.e., $2N(2N + N_e)$ horizontal and vertical ones, each eigenvalue $\lambda_l(-\beta j)$ appearing in each radial factor is characterized by a superscript giving the number of similar bonds showing the *same* positive (or null) value $l$ or $l'$. Thus, the sum of all these superscripts is equal to $2N(2N + 1)$ in case *a* (respectively, $2N(2N + N_e)$ in case *b*) for the total contribution of the horizontal (respectively, vertical) lattice lines (respectively, rows). Under these conditions the evaluation of the set of integers $(\ldots, l_{i,j}, \ldots, l'_{i,j}, \ldots)$ for the whole lattice allows the complete determination of each radial factor of the characteristic polynomial.

The angular factor is determined by the simultaneous knowledge of the sets $(\ldots, l_{i,j}, \ldots, l'_{i,j}, \ldots)$ and $(\ldots, m_{i,j}, \ldots, m'_{i,j}, \ldots)$, with $m_{i,j} \in [-l_{i,j}, +l_{i,j}]$ and $m'_{i,j} \in [-l'_{i,j}, +l'_{i,j}]$, due to the fact that each current integral $F_{i,j}$ closely depends on these integers. The set of integers $(\ldots, l_{i,j}, \ldots, l'_{i,j}, \ldots)$ is shared with the radial factor, thus ensuring the link with temperature for the local magnetic ordering characterized by the couple of angles $(\theta_{i,j}, \varphi_{i,j})$, as noted above. In fact, $F_{i,j}$ is described by the datum of four couples of integers (respectively, three or two for an edge or a corner site). Two couples $(l_{i,j}, m_{i,j})$ and $(l'_{i,j}, m'_{i,j})$ characterize the current site $(i, j)$ whereas the two other ones $(l'_{i+1,j}, m'_{i+1,j})$ and $(l_{i,j-1}, m_{i,j-1})$ refer to the first-nearest neighbor sites, thus conferring an imbricate character to the product of integrals $F_{i,j}$ constituting the final angular factor.

In other words, it means that the determination of the set $(\ldots, l_{i,j}, \ldots, l'_{i,j}, \ldots)$ must be achieved in a first step, for the whole lattice, even if this latter is infinite, and the set of relative integers $(\ldots, m_{i,j}, \ldots, m'_{i,j}, \ldots)$ must then be obtained in a separate second step (or vice versa). Thus, the non-vanishing conditions of each current integral $F_{i,j}$ (one for integers $l_{i,j}$, $l'_{i,j}$, $l'_{i+1,j}$ and $l_{i,j-1}$ and one for relative integers $m_{i,j}$, $m'_{i,j}$, $m'_{i+1,j}$ and $m_{i,j-1}$) are going to permit the introduction of *selection rules* over the various sets of integers $(\ldots, l_{i,j}, \ldots, l'_{i,j}, \ldots)$ and $(\ldots, m_{i,j}, \ldots, m'_{i,j}, \ldots)$.

A numerical argument must then be used to classify the various terms of the characteristic polynomial in order to write them in the decreasing modulus order. Notably this is an important turning point in the infinite-lattice limit where the term of higher degree must be selected.

In the case of a chain ($J_1 = 0$ or $J_2 = 0$) this work is easy. The corresponding current integral $F_{i,j}$ reduces to $F_i$ and we always have $F_i = 1$. Indeed, for a finite chain ($\forall i \in [-N, +N]$) we have $m_{i-1} = m_i = m_0$ and $l_{i-1} = l_i = l_0$ (cyclic boundary conditions). Each term of the characteristic polynomial only reduces to the radial factor $\left[\lambda_{l_0}(-\beta j)\right]^{2N+1}$, with $j = J_1$ or $J_2$. If the chain shows free extremities (free boundary conditions), we now have for each in-site $m_{i-1} = m_i = m_0 = 0$ and $l_{i-1} = l_i = l_0 = 0$. In the thermodynamic limit ($N \to +\infty$), for a cyclic chain, we can use the following numerical property: for non-zero temperatures and for a similar argument $\beta j$, the functions $\lambda_l(-\beta j)$ (which are modified Bessel functions of the first kind $(\pi/2\beta j)^{1/2} I_{l+1/2}(-\beta j)$) rapidly decrease (in absolute value) when $l$ increases. As a result the highest eigenvalue $\lambda_0(-\beta j)$ is selected, like for the case of free boundary conditions, thus imposing $m = 0$ and $Z_N^{free}(0) \approx Z_\infty(0) \approx \left[\lambda_0(-\beta j)\right]^{2N+1}$.

In the case of a 2$d$ lattice the situation is more complicated. We have to consider the whole term $F_{i,j}\lambda_{l_{i,j}}(-\beta J_1)\lambda_{l'_{i,j}}(-\beta J_2)$, with now $F_{i,j} \neq 1$. Thus, taken into account all the previous comments and using the analogy with a chain, the zero-field partition $Z_N(0)$ of a 2$d$ lattice can be written in a first step under the general formal form:

$$Z_N(0) = (4\pi)^{f(N)} \left[ \sum_{l=0}^{+\infty} (\lambda_l(-\beta J_1))^{f_h(N)} (\lambda_l(-\beta J_2))^{f_v(N)} \sum_{\{m\},\{m'\}} \right.$$

$$\left. + \sum_{\substack{\{l_1,\ldots,l_K\},\\ \{l'_1,\ldots,l'_{K'}\}}} (\lambda_{l_1}(-\beta J_1))^{a_{1,l_1}} \ldots (\lambda_{l_K}(-\beta J_2))^{a_{2,l'_{K'}}} \sum_{\substack{\{m_1,\ldots,m_K\},\\ \{m'_1,\ldots,m'_{K'}\}}} \right] \prod_{i=-N}^{N} \prod_{j=-N}^{N} F_{i,j} \quad (2.17)$$

with:

$$\sum_{i=1}^{K} a_{1,l_i} = f_h(N), \quad \sum_{i=1}^{K'} a_{2,l'_i} = f_v(N), \quad f(N) = f_h(N) + f_v(N). \quad (2.18)$$

where $f_h(N)$ and $f_v(N)$ are the number of bonds for all the horizontal (respectively, vertical) lines; the total number of lattice bonds $f(N)$ is given by (2.10) and the current integral per site $F_{i,j}$ by (2.16). *This expression can indifferently be employed for a lattice showing edges* (case a) *or a lattice wrapped on a torus* (case b). In the first term, the symbolical notations $\{m\}$ and $\{m'\}$ respectively refer to the $f_h(N)$ summations over the relative integers $m_{i,j}$ and the $f_v(N)$ summations over $m'_{i,j}$'s. For the second term the notation $\{l_1,\ldots, l_K\},\{l'_1,\ldots, l'_{K'}\}$ (respectively, $\{m_1, \ldots, m_K\}, \{m'_1, \ldots, m'_{K'}\}$) symbolically represents the set of authorized integers $l_{i,j}$ and $l'_{i,j}$ (respectively, $m_{i,j}$ and $m'_{i,j}$) over which the $l_1$-, …, $l_K$-, $l'_1$-, …, $l'_{K'}$- (respectively, the $m_1$-, …, $m_K$-, $m'_1$-, …, $m'_{K'}$-) summations must be achieved. This time, a numerical study is then unavoidable. This is done in the following subsections. When temperature is cooling down we notably observe *crossovers* between two consecutive terms of the polynomial expansion due to the presence of integrals $F_{i,j} \neq 1$. In addition we show that there is a strong analogy between a 2$d$ lattice and a 1$d$ chain regarding the boundary conditions.

## 2.2.5 General Selection Rules for the Whole Lattice; Consequences

For a finite or an infinite lattice, the non-vanishing condition of each current integral $F_{i,j}$ given by (2.16) is mainly due to that of Clebsch-Gordan (CG) coefficients for the in- and edge sites. The first CG coefficient $C_{l'_{i+1,j}\ m'_{i+1,j}\ l_{i,j-1}\ m_{i,j-1}}^{L_{i,j}\ M_{i,j}}$ does not vanish if $M_{i,j} = m'_{i+1,j} + m_{i,j-1}$ whereas for the second one $C_{l_{i,j}\ m_{i,j}\ l'_{i,j}\ m'_{i,j}}^{L_{i,j}\ M_{i,j}}$ we have $M_{i,j} = m_{i,j} + m'_{i,j}$ so that we finally derive:

$$m_{i,j-1} + m'_{i+1,j} - m_{i,j} - m'_{i,j} = 0 \quad (SRm). \qquad (2.19)$$

This is the *first selection rule* labeled from now *SRm*: it exclusively concerns the various coefficients $m_{i,j}$ and $m'_{i,j}$ characterizing each site $(i, j)$. At this step we must note that, if each spherical harmonics $Y_{l,m}(\theta,\varphi)$ appearing in the integrand of $F_{i,j}$ is replaced by its own definition i.e., $C_l^m \exp(im\varphi) P_l^m(\cos\theta)$ where $C_l^m$ is a constant depending on coefficients $l$ and $m$ and $P_l^m(\cos\theta)$ is the associated Legendre polynomial, the non-vanishing condition of the $\varphi$-part directly leads to (2.19). Finally, we can make three remarks. *(i) The SRm relation is unique and temperature-independent. (ii) The selection rules obtained for the corner sites in case a (cf (2.14)) can simply be derived from the previous general SRm by assuming vanishing coefficients m and m' for the absent bonds. (iii) Due to the fact that the $\varphi$-part of the $F_{i,j}$-integrand reduces to unity, $F_{i,j}$ is a pure real number.*

Under these conditions, we have to solve: (i) in case *a* (lattice with edges), a linear system of $2N(2N+1)$ unknowns $m_{i,j}$ or $m'_{i,j}$ (one per lattice bond) for $(2N+1)^2$ equations such as (2.19) (one per site); (ii) in case *b* (lattice wrapped on a torus) a similar linear system of $2N(2N+N_e)$ equations but with $2N(2N+N_e)$ unknowns $m_{i,j}$ or $m'_{i,j}$'s. As it remains $4N(2N+1) - (2N+1)^2 = 4N^2 - 1$ independent solutions over the set $\mathbb{Z}$ of relative integers $m_{i,j}$ and $m'_{i,j}$ in case *a* or $4N(2N+N_e) - 2N(2N+N_e) = 2N(2N+N_e)$ ones in case *b*, *it means that there are $4N^2 - 1$ (case a) or $2N(2N+N_e)$ (case b) different expressions for each local angular factor appearing in each term of the same characteristic polynomial giving $Z_N(0)$ so that the statistical problem remains unsolved.* Thus, at first sight, this result means that *there is no unique expression for $Z_N(0)$, except if $|m_{i,j}| = |m'_{i,j}| \neq 0$ or $m_{i,j} = m'_{i,j} = 0$. In a forthcoming section we show that, in the thermodynamic limit ($N \to +\infty$), the values $m_{i,j} = m'_{i,j} = 0$ are selected so that $Z_N(0)$ has an analytical expression. As a result it would mean that there is no analytical expression of the zero-field partition function for a finite lattice whereas it becomes analytical when the lattice becomes infinite.*

However, if we take into account the spin lattice symmetries they must allow one to introduce new extra relations between coefficients $m_{i,j}$ and $m'_{i,j}$ so that the number of unknowns becomes equal or lower than the number of equations (2.19). Consequently it then becomes possible to derive a general selection rule for all the coefficients $m$ and $m'$, for the whole lattice, and a single expression for the characteristic polynomial.

The second non-vanishing condition of CG coefficients appearing in each current integral $F_{i,j}$ concerns the positive (or null) integers $l_{i,j}$ and $l'_{i,j}$ intervening in each radial factor of the characteristic polynomial. In addition to the triangular inequalities $|l_{i,j} - l'_{i,j}| \leq L_{i,j} \leq l_{i,j} + l'_{i,j}$ and $|l_{i,j-1} - l'_{i+1,j}| \leq L_{i,j} \leq l_{i,j-1} + l'_{i+1,j}$ respectively followed by the CG coefficients $C_{l_{i,j}\ m_{i,j}\ l'_{i,j}\ m'_{i,j}}^{L_{i,j}\ M_{i,j}}$

and $C_{l'_{i+1,j}\ m'_{i+1,j}\ l_{i,j-1}\ m_{i,j-1}}^{L_{i,j}\ M_{i,j}}$ (with $M_{i,j} \neq 0$ or $M_{i,j} = 0$), we have a more restrictive vanishing condition when the involved $m_{i,j}$'s, $m'_{i,j}$'s and $M_{i,j}$'s vanish [12]:

$$C_{l_1\ 0\ l_2\ 0}^{l_3\ 0} = 0, \text{ if } l_1 + l_2 + l_3 = 2g + 1,$$

$$C_{l_1\ 0\ l_2\ 0}^{l_3\ 0} = (-1)^{g-l_3}\sqrt{2l_3+1}\ K, \text{ if } l_1 + l_2 + l_3 = 2g, \tag{2.20}$$

with:

$$K = \frac{g!}{(g-l_1)!(g-l_2)!(g-l_3)!}\left[\frac{(2g-2l_1)!(2g-2l_2)!(2g-2l_3)!}{(2g+1)!}\right]^{1/2}. \tag{2.21}$$

We can note that $K$ is unchanged under the permutation of integers $l_1$, $l_2$ and $l_3$. $C_{l_{i,j}\ 0\ l'_{i,j}\ 0}^{L_{i,j}\ 0}$ does not vanish if $l_{i,j} + l'_{i,j} + L_{i,j} = 2A_{i,j} \geq 0$ whereas, for $C_{l'_{i+1,j}\ 0\ l_{i,j-1}\ 0}^{L_{i,j}\ 0}$, we must have $l_{i,j-1} + l'_{i+1,j} + L_{i,j} = 2A'_{i,j} \geq 0$. Thus, if we sum the two previous equations over $l$ and $l'$, we have $(2N+1)^2$ equations in case *a* and $2N(2N+N_e)$ in case *b* (one per lattice site) such as:

$$l_{i,j-1} + l'_{i+1,j} + l_{i,j} + l'_{i,j} = 2g_{i,j}\ , \ g_{i,j} = A_{i,j} + A'_{i,j} - L_{i,j} \geq 0\ (SRl1) \tag{2.22}$$

or, for instance, by assuming the difference:

$$l_{i,j-1} + l'_{i+1,j} - l_{i,j} - l'_{i,j} = 2g'_{i,j}\ , \ g'_{i,j} = A'_{i,j} - A_{i,j}\ (SRl2). \tag{2.23}$$

We obtain two types of equations which are similar to (2.19) but now, in (2.22) and (2.23), instead of having a null right member like in (2.19), we can have a positive, null or negative but always even second member $2g_{i,j}$. This is the *second selection rule*: it exclusively concerns the various coefficients $l_{i,j}$ and $l'_{i,j}$ characterizing each site $(i, j)$ and, from now, it is labeled *SRl1* (*cf* (2.22)) and *SRl2* (*cf* (2.23)). *(i) As previously remarked for the first selection rule SRm it is temperature-independent although each $l_{i,j}$ and $l'_{i,j}$ is linked to a temperature-dependent function $\lambda_l(-\beta j)$. (ii) The selection rules obtained for the corner sites in case a (cf (2.14)) can simply be derived from the previous general SRl1 or SRl2 by assuming vanishing coefficients l and l' for the absent bonds.* In addition, as for coefficients $m_{i,j}$ and $m'_{i,j}$, we expect that the spin lattice symmetries must allow one to introduce new extra relations between coefficients $l_{i,j}$ and $l'_{i,j}$ so that the number of these unknowns becomes equal or lower than the number of equations (2.22) and (2.23).

### 2.2.6 Influence of the Spin Lattice Symmetries on the Selection Rules

### 2.2.6.1 Introduction of the Concept of Decorated Spin Lattice

Usually, in Solid State Physics, the study of symmetries is achieved in the crystallographic space and often leads to the determination of a point group (if existing) which characterizes the system under consideration. In the present case we focus on the quantum numbers $m_{i,j}$ and $m'_{i,j}$ on the one hand, and $l_{i,j}$ and $l'_{i,j}$ on the other one, which characterize the spin orientations i.e., the spin states. Indeed, their respective knowledge is of the highest importance for deriving a closed-form expression for the zero-field partition function $Z_N(0)$.

As previously observed, all the selection rules have led to an incomplete knowledge of the set of coefficients $m_{i,j}$ and $m'_{i,j}$ (respectively, $l_{i,j}$ and $l'_{i,j}$). As a result and in order to obtain the full set of these quantum numbers we are compelled to introduce *a new concept*, that of *decorated lattice*. This lattice is similar to the crystallographic lattice, i.e., it is composed of two square sublattices, each one having a square unit cell. The properties of the associated space $\mathcal{D}$ must take into account both properties characterizing the crystallographic space $\mathcal{C}$ and the spin space $\mathcal{S}$. As a result, the $\mathcal{D}$ space is defined by the bijection

$$\mathcal{C} \otimes \mathcal{S} \rightarrow \mathcal{D}.$$

The $\mathcal{D}$ elements are spin quantum numbers $l$ and $m$ (respectively $l'$ and $m'$) which characterize the spin states $(l, m)$ (respectively, $(l', m')$) for a given position $(i, j)$ of the spin carrier in space $\mathcal{C}$. Consequently each quantum number is characterized by a general index $(i, j)$ which indicates the position of the spin carrier in $\mathcal{C}$. Thus, the set of all elements is constituted by the mapping $(…, (l_{i,j}, m_{i,j}),…, (l'_{i,j}, m'_{i,j}), …)$.

The deep reasons leading to the introduction of this decorated lattice are based on the following arguments which allow to define the characteristics of this lattice:

i) each lattice bond is due to the interaction between two spins: the first spin is located on a given lattice site and the second one on a first-nearest neighbor; each spin $S_i$ is carried by an electron involved in the chemical bond whose position in $\mathcal{C}$ is $r_i$; the corresponding wave function of these two interacting electrons is $\Psi(r_1, S_1; r_2, S_2)$; when there is no spin-orbit coupling (as set in this article) i.e., no physical relationship between spaces $\mathcal{C}$ and $\mathcal{S}$, we can write

$$\Psi(r_1, S_1; r_2, S_2) = \Phi(r_1, r_2)\chi(S_1, S_2);$$

$\Phi(r_1, r_2)$ is called an orbital and $\chi(S_1, S_2)$ is the spin contribution (chemists often speak of "magnetic orbital" for labeling $\Psi(r_1, S_1; r_2, S_2)$); these functions are defined in spaces $\mathcal{C}$ and $\mathcal{S}$, respectively, and separately obey the symmetry properties of each corresponding space;

ii) the mechanism of direct exchange (or superexchange) imposes the physical presence of first-nearest neighbor orbitals $\Phi(r_1, r_2)$ which directly overlap (or couples of orbitals characterizing spin carrier sites separated by a nonmagnetic site and overlapping with its orbital); their distribution is related to the location of spin carriers within the lattice; as a result the symmetry elements of $\mathcal{C}$ play a role in describing the distribution of exchange between nearest-neighbor spins;

iii) each bond of the decorated lattice is characterized by the datum of a couple $(l_{i,j}, m_{i,j})$ or $(l'_{i,j}, m'_{i,j})$ for any site $(i, j)$; these quantum numbers come from the corresponding spherical harmonics appearing in the characteristic polynomial associated with $Z_N(0)$ (*cf* (2.7)-(2.11)); these couples allow to describe the spin states of each lattice site characterized by four bonds, each one linked with the first-nearest spin neighbor: this is given by the selection rules per site *SRm* (*cf* (2.19)), *SRl*1 and *SRl*2 (*cf* (2.22), (2.23));

iv) the space $\mathcal{D}$ can then be mathematically decomposed into two subspaces, each subspace being associated with a sublattice: the $\mathcal{L}$ subspace corresponding to all the bonds character-ized by *l* (or *l'*) and the $\mathcal{M}$ subspace whose bonds are characterized by *m* (or *m'*); in other words, the $\mathcal{D}$ space is the direct product of subspaces $\mathcal{L}$ and $\mathcal{M}$, i.e.,

$$\mathcal{D} = \mathcal{L} \otimes \mathcal{M};$$

v) as a result the first decorated sublattice is described by the set $(\ldots, l_{i,j}, \ldots, l'_{i,j}, \ldots)$ and the second one by $(\ldots, m_{i,j}, \ldots, m'_{i,j}, \ldots)$; it means that each element of $(\ldots, l_{i,j}, \ldots, l'_{i,j}, \ldots)$ is an eigenvalue of $\mathcal{L}$ whereas each element of $(\ldots, m_{i,j}, \ldots, m'_{i,j}, \ldots)$ is an eigenvalue of $\mathcal{M}$;

vi) as we deal with spin arrangements, some conditions must be fulfilled independently of the ferromagnetic ‑ F ‑ or antiferromagnetic ‑ AF ‑ nature of couplings between nearest neighbors, in the whole range of temperature; thus, it is the nature of the spin arrangement which essentially determines all symmetry elements of $\mathcal{D}$ i.e, the *corresponding point group if existing, independently of the crystallographic space symmetries*;[1] in particular, for ensuring a symmetry of translation in $\mathcal{D}$ space as in $\mathcal{C}$,

- spin orientations must be parallel or antiparallel,
- spin carriers must be close to each others and no notable change in the orientation of two spins must occur in the first-nearest spin neighboring due to exchange i.e., there must exist a distance-independent correlation, for a given temperature *T*, within this small correlation domain;
- these properties must also be fulfilled inside the first-nearest neighbor correlation domains; these co-operative effects are described in $\mathcal{D}$ space by the imbricate character of selection rules characterizing each spin state of each lattice site;

these two important points are detailed in Sec. 2.2.6.3;

vii) as a consequence, due the nature of spin orientations described by quantum numbers *l* and *m* (respectively, *l'* and *m'*) and by construction, the decorated lattice shows a geometrical structure similar to that of $\mathcal{C}$ (which is here a pure coincidence, see footnote 1);
for instance, for a lattice showing edges (case *a*), the corresponding symmetry elements of the associated $\mathcal{D}$ space are mirror symmetries with respect to the medians (respective traces of mirrors $\mathcal{M}_1$ and $\mathcal{M}_2$ perpendicular to the decorated lattice plane) and the diagonals (respective traces of mirrors $\mathcal{M}_1'$ and $\mathcal{M}_2'$ perpendicular to the same plane); both systems are derived from each other through a rotation of $\pm\pi/4$ around the axis perpendicular to the lattice and located at the intersection of $\mathcal{M}_1$ and $\mathcal{M}_2$ (or $\mathcal{M}_1'$ and $\mathcal{M}_2'$); the corresponding point group is *p*4*m*;
thus, in case *a*, this allows to define the operators acting in $\mathcal{D}$; the situation is more complicated for a lattice wrapped on a torus (case *b*) and it is examined later (see footnote 3).

Furthermore, for any operator acting in $\mathcal{D}$ (i.e., for any symmetry operation in each subspace $\mathcal{L}$ or $\mathcal{M}$), the corresponding result must be valid:

viii) for both types of lattice configurations (cases *a* and *b*) and their associated decorated lattices (including their common infinite limit i.e., the thermodynamic case);

---

[1] Without spin-orbit couplings the only case for which the geometrical structure of the crystallographic space influences the nature of spin arrangements concerns frustrated magnetic systems; it is out of the scope of the present article. Thus, for any type of unit cell (rectangular or square) in $\mathcal{C}$, the corresponding unit cell in $\mathcal{D}$ has a square structure by construction.

ix) in the whole temperature range, from absolute zero (F or AF arrangements, strong couplings) up to high temperature ($k_B T \sim |J|$);[2] it means that the symmetries of the spin arrangement must not change as it can occur for instance when temperature varies between a value $T < T_c$ and a value $T > T_c$ where $T_c$ is the phase transition temperature, if one wishes to obtain a single closed-form expression of the zero-field partition function $Z_N(0)$; in other words *the point group if existing must remain unchanged in the whole temperature range once determined*;[3] if this situation cannot occur there are two possibilities: a) a single type of spin arrangement exists within a restricted temperature range; it is then always possible to obtain a single closed-form expression for $Z_N(0)$ but its validity is restricted to this range characterizing this magnetic phase; b) within the restricted temperature range the spin arrangement is composed of at least two different spin structures and it is impossible to derive a closed-form expression for $Z_N(0)$;[4]

x) for parallel or antiparallel spin arrangements, the reasoning must be independent of the spin orientation angles appearing in the various integrals $F_{i,j}$ (*cf* (2.16)); thus, *under any symmetry operation, the selection rules SRm (cf (2.19)), SRl1 and SRl2 (cf (2.22), (2.23)) must remain unchanged.*

Under these conditions, for each decorated sublattice, we are led to examine the effects of symmetries defined in vii) with respect to each spin state (characterizing a bond in $\mathcal{D}$ space). In other words, for each given site $(i, j)$ described by the quantum numbers $l$ or $l'$ in $\mathcal{L}$ (respectively, $m$ or $m'$ in $\mathcal{M}$), we have to examine the effects of symmetries on the selection rules concerning the quantum numbers $l_{i,j}$ and $l'_{i,j}$ (respectively, $m_{i,j}$ and $m'_{i,j}$).

What are the consequences for the function $Z_N(0)$ which describes all lattice states? We have seen that $Z_N(0)$ is expressed as a characteristic polynomial (*cf* (2.17)). Each eigenvalue $\lambda_{l_{i,j}}(-\beta J_1)$ or $\lambda_{l'_{i,j}}(-\beta J_2)$ is characterized by an index $l$ or $l'$; the combination of eigenfunctions given by spherical harmonics $Y_{l_{i,j-1}, m_{i,j-1}}(\theta_{i,j}, \varphi_{i,j})$, $Y_{l'_{i+1,j}, m'_{i+1,j}}(\theta_{i,j}, \varphi_{i,j})$, $Y^*_{l_{i,j}, m_{i,j}}(\theta_{i,j}, \varphi_{i,j})$ and $Y^*_{l'_{i,j}, m'_{i,j}}(\theta_{i,j}, \varphi_{i,j})$ appears in the integral $F_{i,j}$ (*cf* (2.16)) which is a numerical factor giving the spatial average of spin orientations, for a given lattice site $(i, j)$. In (2.17) the first term is characterized by the coefficients $l_{i,j} = l'_{i,j} = l$; the coefficients $m_{i,j}$ and $m'_{i,j}$ are not necessarily equal (but some of them can be equal). For the second term we have the same situation for the $m_{i,j}$'s and $m'_{i,j}$'s as well as for $l_{i,j}$'s and $l'_{i,j}$'s. As a result, studying the effects of symmetries on $l_{i,j}$ and $l'_{i,j}$ (respectively, on $m_{i,j}$ and $m'_{i,j}$) must permit to derive the full set of selection rules for the whole lattice and then allows to obtain a simpler closed-form expression for $Z_N(0)$.

---

[2] In the next subsection we show that we directly have $l = 0$, $m = 0$ as soon as $T$ reaches the paramagnetic regime temperature i.e., $k_B T \sim |J|$.

[3] It can occur that, even if symmetry elements of a system are fully determined, there is no sense to define a point group; this is the case when, at least, two different types of scales appear in the system; for instance, in the case of a lattice wrapped on a torus (case b), we have a first type of scale on the surface of the torus where the lattice is characterized by a square unit cell and the corresponding point group is *p4m*; there is a second type of scale if examining the toroidal surface characterized by a revolution symmetry; as a result there is no global point group and only the full collection of symmetry elements is useful in that case (see Sec. 2.2.8.2).

[4] This is the case if considering the 2*d* classical *xy* model; if $T_{KT}$ is the Kosterlitz-Thouless transition temperature we deal with two different types of spin arrangements according to whether $T < T_{KT}$ (aligned or antialigned spins with few pairs of spin vortices-antivortices when $T$ approaches $T_{KT}$) or $T > T_{KT}$ (free vortices, exclusively).

For fixing ideas, we reconsider the simple case of a lattice showing edges (case *a*). In a first step, in the decorated lattice space $\mathcal{D}$ and due to the nature of the spin arrangement, we determine the action of mirrors $\mathcal{M}_1$ and $\mathcal{M}_2$ whose respective traces are the horizontal and vertical medians. We must have

$$\mathcal{M}_1(F_{i,j}) = F_{-i,j}, \quad \mathcal{M}_2(F_{i,j}) = F_{i,-j}.$$

This relationship is exclusively valid for correlated spins inside the same correlation domain i.e., spins showing the same orientation in $\mathcal{S}$. As each of integrals $F_{i,j}$, $F_{-i,j}$ or $F_{i,-j}$ gives the spatial average of the spin orientation in $\mathcal{S}$ and is a real number expressed *vs* quantum numbers *l*, *l'*, *m* and *m'* appearing in the integrand (*cf* (2.16)), its corresponding image in $\mathcal{D}$ i.e., the $\mathcal{D}$-image, is by construction the same real number in order to ensure the conservation of the spin orientation while passing from $\mathcal{S}$ to $\mathcal{D}$ i.e., the spin state conservation. In other words, in $\mathcal{D}$, the mirror operators exclusively act on quantum numbers with the law imposing the spin state conservation. As a result these operations allow to derive relationships between the *l*'s, *l"*'s, *m*'s and *m"*'s involved in integrals $F_{i,j}$ and $F_{-i,j}$ (respectively, $F_{i,j}$ and $F_{i,-j}$). There are completely detailed in the next section.

Through a rotation of $\pi/4$ around the axis located at the intersection of $\mathcal{M}_1$ and $\mathcal{M}_2$, we have two other mirrors $\mathcal{M}_1'$ and $\mathcal{M}_2'$ whose respective traces are the first and second diagonals ($i = j$ and $i = -j$, respectively). In a second step, if wishing to express each operator $\mathcal{M}_1$, $\mathcal{M}_2$, $\mathcal{M}_1'$ or $\mathcal{M}_2'$, it becomes necessary to use the result given by each of these operators. For instance the symmetry with respect to the first diagonal i.e., $\mathcal{M}_1'(F_{i,j}) = F_{j,i}$ leads to the following relationships for sites $(i, j)$ and $(j, i)$

$$m'_{i+1,j} = m_{j,i},\ m_{i,j-1} = m'_{j,i},\ m'_{i,j} = m_{j,i-1},\ m_{i,j} = m'_{j+1,i},\ (\mathcal{M}\text{ space})$$

$$l'_{i+1,j} = l_{j,i},\ l_{i,j-1} = l'_{j,i},\ l'_{i,j} = l_{j,i-1},\ l_{i,j} = l'_{j+1,i},\ (\mathcal{L}\text{ space})$$

so that the operator $\mathcal{M}_1'$ can be written as

$$\mathcal{M}_1' = \mathcal{M}_1'(\mathcal{L}) \otimes \mathcal{M}_1'(\mathcal{M});$$

the corresponding elements are

$$\left(\mathcal{M}_1'(\mathcal{L})\right)_{i,j} = \delta_{l'_{i+1,j},l_{j,i}} \delta_{l_{i,j-1},l'_{j,i}} \delta_{l'_{i,j},l_{j,i-1}} \delta_{l_{i,j},l'_{j+1,i}},$$

$$\left(\mathcal{M}_1'(\mathcal{M})\right)_{p,q} = \delta_{m'_{p+1,q},m_{q,p}} \delta_{m_{p,q-1},m'_{q,p}} \delta_{m'_{p,q},m_{q,p-1}} \delta_{m_{p,q},m'_{q+1,p}}.$$

A similar operation can be achieved for the remaining operators $\mathcal{M}_1$, $\mathcal{M}_2$ or $\mathcal{M}_2'$. Finally, using the selection rules *SRm* (*cf* (2.19)), *SRl*1 and *SRl*2 (*cf* (2.22), (2.23)) and their imbricate character from site to site, can allow the determination of coefficients *l* and *m* (*l'* and *m'*) for all lattice lines and rows.

All the previous work has concerned integrals $F_{i,j}$. It leads to the full determination of the sets $(\ldots, l_{i,j}, \ldots, l'_{i,j}, \ldots)$ and $(\ldots, m_{i,j}, \ldots, m'_{i,j}, \ldots)$, thus allowing to write the angular factor of the characteristic polynomial given by (2.17). The radial factor, composed of a product of eigenvalues $\lambda_{l_{i,j}}(-\beta J_1)$ and $\lambda_{l'_{i,j}}(-\beta J_2)$, just necessitates the knowledge of the set $(\ldots, l_{i,j}, \ldots, l'_{i,j}, \ldots)$; thus it is fully determined.

How can we verify that the previous concept of decorated lattice leads to exact results allowing to write the correct closed-form expression of the zero-field partition function $Z_N(0)$?

We then must examine the action of mirrors $\mathcal{M}_1'$ and $\mathcal{M}_2'$ on the eigenvalue $\lambda_{l_{i,j}}(\beta|J|)$ (respectively, $\lambda_{l'_{j+1,i}}(\beta|J|)$). As integrals $F_{i,j}$, $F_{-i,j}$ or $F_{i,-j}$ which are functions of quantum numbers $l$, $l'$, $m$ and $m'$ whose value is a real number, $\lambda_{l_{i,j}}(\beta|J|)$ (respectively, $\lambda_{l'_{j+1,i}}(\beta|J|)$) is also expressed as a real number which is a $l_{i,j}$-series of argument $|J|$ (respectively, a $l'_{j+1,i}$-one). For a given temperature and the same exchange energy $|J|$, $\lambda_{l_{i,j}}(\beta|J|)$ (respectively, $\lambda_{l'_{j+1,i}}(\beta|J|)$) gives the nature of the spin arrangement (ferromagnetic if $J < 0$ or antiferromagnetic if $J > 0$). The law imposing the spin state conservation between $\mathcal{S}$ and $\mathcal{D}$ automatically imposes the conservation of the nature of the spin arrangement. As a result the corresponding $\mathcal{D}$-image of $\lambda_{l_{i,j}}(\beta|J|)$ (respectively, $\lambda_{l'_{j+1,i}}(\beta|J|)$) is the same real number while passing from $\mathcal{S}$ to $\mathcal{D}$. As a result, in the case of a lattice characterized by the same exchange energies $J_1 = J_2 = J$ (case $a$) and due to the fact that $\mathcal{M}_1'(F_{i,j}) = F_{j,i}$ notably leads to write $l_{i,j} = l'_{j+1,i}$, if considering $\lambda_{l_{i,j}}(\beta|J|)$ and $\lambda_{l'_{j+1,i}}(\beta|J|)$, $\mathcal{M}_1'$ mirror exclusively acts on quantum number $l_{i,j}$ (respectively, $l'_{j+1,i}$) for any argument $|J|$ so that we can write

$$\mathcal{M}_1'\left(\lambda_{l_{i,j}}(\beta|J|)\right) = \lambda_{l'_{j+1,i}}(\beta|J|).$$

This remains true in the general case $J_1 \neq J_2$ on condition that the sign of $J_1$ and $J_2$ is conserved.

$$\mathcal{M}_1'\left(\lambda_{l_{i,j}}(\beta|J_1|)\right) = \lambda_{l'_{j+1,i}}(\beta|J_1|), \quad \mathcal{M}_1'\left(\lambda_{l'_{j+1,i}}(\beta|J_2|)\right) = \lambda_{l_{i,j}}(\beta|J_2|)$$

and similar relationships concerning the other $l$-(or $l'$)-quantum numbers characterizing sites $(i, j)$ and $(j, i)$.

In other words the previous formalism of decorated lattice leads to the duality principle i.e., the necessity of exchanging the exchange energies $J_1$ and $J_2$ characterizing the horizontal and vertical lines of the lattice.

**Duality Principle** *The closed-form expression obtained for the zero-field partition function $Z_N(0)$ must be invariant under the permutation of the exchange energies $J_1$ and $J_2$.*

If this principle leads to a correct closed-form expression of the zero-field partition function $Z_N(0)$ we must be able to verify that the set of coefficients $(…, l_{i,j},…, l'_{i,j}, …)$ and/or $(…, m_{i,j},…, m'_{i,j},…)$ obtained through the formalism of decorated lattice can also be derived through a totally different method (notably, a numerical method). This must be true for a finite lattice but also for an infinite one (thermodynamic limit).

### 2.2.6.2 The Duality Principle: Framework and Conditions of Validity

We must define the general framework within which the duality principle is fulfilled before considering the case of $2d$ isotropic couplings. For sake of simplicity we restrict the study to the case $J = J_1 = J_2$ but the reasoning can be extended to the general case $J_1 \neq J_2$. Similarly we only consider the case of a finite lattice showing edges (case $a$) but the reasoning can be adapted to a finite lattice wrapped on a torus (case $b$).

The case of dipole-dipole interactions and spin-spin exchange interactions depending on the distance *r* between spins (RKKY couplings [13]) must be excluded as well as the cases of multiple exchange interactions occurring between first-nearest neighbor spins but also with spins which are not first-nearest neighbors, and double exchange mechanism [14]. Under these conditions we restrict the study to *direct exchange* or *superexchange* between nearest spin neighbors. For superexchange we have generalized the mechanism proposed by Anderson [15]. Our model leads to constant exchange energies regularly distributed over the lattice. As a result, we assume here that exchange energies are temperature-independent so that the nature of coupling between first-nearest neighbor spins does not change from 0 K to high temperatures. An extension to alternating exchange couplings characterized by an exchange energy *J* of constant absolute value (or not) but regularly distributed is possible. However the case of randomly distributed couplings (spin glasses) is discarded because the distribution of exchange energies along the crystallographic axes is naturally different.

There is no anisotropy, notably due to local crystal fields (this particular case will be detailed below). Moreover there is no antisymmetric contribution to exchange such as Dzialoshinskii-Moriya couplings which can induce canted spin arrangements (alternating distribution of the sense of the Dzialoshinskii vector ***D***) or helical ones (all the vectors ***D*** show the same sense) [16,17].

As a result, we exclusively consider the role of exchange between first-nearest neighbor spins and the co-operative effects in the spin neighborhood. Finally we have to examine the *local spin ordering* i.e., the size of domains within which spins are correlated for any lattice size.

For 1*d* spin chains (if exclusively considering constant exchange energies), the critical temperature is always $T_c = 0$ K whatever the isotropic or anisotropic character of exchange couplings, with or without an antisymmetric contribution [17]. In the present case it is well-known [4,5] that, when dealing with 2*d* isotropic spin couplings characterized by a regular distribution (showing or not a regular alternation of *J* sign), the critical temperature is $T_c = 0$ K as for 1*d* spin chains. When $d = 2$, $T_c \neq 0$ K in all the other cases characterized by anisotropic couplings (*zz* or *xy* model depending on whether anisotropy favors the *z*-axis or the *xy* plane), with or without an antisymmetric contribution. When $d > 2$ we always have $T_c \neq 0$ K for any type of couplings.

For isotropic couplings, at $T_c = 0$ K ($d = 2$) or below $T_c \neq 0$ K ($d > 2$), there is an ordered spin arrangement and above $T_c$ we deal with the *critical domain* for any *d*. Thus, in this domain, the spin arrangements can be described as independent blocks of sides $\xi$, the correlation length (if $J = J_1 = J_2$) which becomes infinite at $T_c$: those are the well-known *Kadanoff blocks* composed of parallel (F couplings) or antiparallel (AF couplings) spin alignments. As $\xi$ only depends on exchange energies and temperature it means that, for a given couple (*J*, *T*), the value of $\xi$ is fixed so that all the blocks show the same size and the same surface $\xi^2$.

At $T_c$, these blocks show an infinite size because $\xi$ diverges at this particular temperature. They are quasi infinite nearby $T \geq T_c$: we have a *long-range order*. When *T* increases from $T_c$ the size of Kadanoff blocks decreases and spins become less and less correlated: there is a *local order* i.e., a *short-range order* due to the competition between the magnetic order and the thermal disorder. It is precisely described by the thermal behavior of the correlation length $\xi$. When $k_B T \sim |J|$ we approach the *paramagnetic regime*: spins are more and more independent from each other over the whole lattice and exchange couplings are becoming more and more inefficient. As a result, as soon as temperature deviates from $T_c$ and reaches the range $k_B T > |J|$, the correlation length tends to decrease down to the value $\xi \sim a$ where *a* is the lattice spacing.

When $k_BT \gg |J|$ we reach the pure paramagnetic regime: without an external magnetic field, irrespectively of the existence of an easy natural axis of magnetization, the totality of spin orientations over the full lattice is randomly distributed and there is no more local order. In the local equation (2.7), when $k_BT \gg |J|$ ($\beta|J| \ll 1$) the modified Bessel functions of the first kind $(\pi/2\beta J)^{1/2}I_{l+1/2}(-\beta J)$ (which are the eigenvalues $\lambda_l(-\beta J)$ - cf (2.8) - of the local operator $\exp(-\beta H^{ex})$) have a small argument and rapidly decrease (in absolute value) when $l$ increases continuously, for the same argument (i.e., the dominant eigenvalue is characterized by $l = 0$). As the lattice can be described from site to site by the translation of a couple of bonds (a horizontal bond and a vertical one, with one eigenvalue per bond) the contribution per site to the characteristic polynomial giving $Z_N(0)$ is proportional to $[\lambda_l(-\beta J)]^2$. As a result, when $k_BT \gg |J|$ ($\beta|J| \ll 1$), the total contribution per site asymptotically tends towards $[\lambda_0(-\beta J)]^2 = [(\pi/2\beta J)^{1/2}I_{1/2}(-\beta J)]^2$, thus selecting the values $l = 0$, $m = 0$. Finally it means that the study can be limited to the upper temperature value $k_BT \sim |J|$ for determining the set of coefficients $l$ and $l'$ (respectively, $m$ and $m'$), for the whole lattice.

In summary, below $T_c$ ($d > 2$) or at $T_c = 0$ K ($d = 2$), for isotropic couplings, we always have sublattices of aligned spins and a long-range order. Above $T_c$ i.e., inside the critical domain, we have a short-range order which disappears little by little when $T$ increases. In a first step, if excluding the particular case of paramagnetic regime for which $l$ and $m$ are determined, it is always possible to find domains (Kadanoff blocks) characterized by locally correlated spins (i.e., aligned spins − F couplings − or anti-aligned spins − AF couplings). Due to the fact that all spins are strongly correlated, conditions are fulfilled for the application of the duality principle inside these domains and outside them, for the first-nearest neighbor correlation domains.

Why the case of anisotropic couplings cannot enter this framework? Let us restrict to the *xy* couplings [18] in two dimensions ($d = 2$). Kosterlitz and Thouless [18] have first shown that, for this model, there exists a special phase transition. i) Below the critical temperature i.e., the Kosterlitz-Thouless transition temperature $T_{KT}$, there is a low-temperature phase where most spins are aligned or antialigned (the correlation function decays with a power law); thus we have a long-range order within correlation domains of large size co-existing with few pairs of vortices/antivortices which exclusively intervene on very short distances when $T$ approaches $T_{KT}$. When $T$ increases the size of these pairs as well as their number increases and diverges if $T = T_{KT}$: it then appears free vortices. ii) Above $T_{KT}$ there is a high-temperature phase uniquely composed of free vortices (the correlation function decays exponentially). These vortices are stable topological spin formations which lower the correlation of the system: they are characterized by a vorticity $q$ defined as $\int d\Phi(r) = 2\pi q$ where $\Phi(r)$ is the angle involved in the description of spin vortices ($q$ is a winding number). As a result, in this phase, there are no symmetry relationships between the orientation of first-nearest neighbor spins even if the crystallographic lattice as well as the associated decorated lattice show symmetry elements. It means that *it is impossible to apply the duality principle in the case of a classical xy model*. In conclusion the duality principle can only be used with systems showing *no vorticity* and a regular distribution of exchange energies which must be independent of the distance between involved spins.

### 2.2.6.3 Conditions of Application of The Duality Principle

If the conditions of validity of the duality principle are fulfilled, it becomes possible to begin the determination of quantum numbers $l$ and $m$ (respectively, $l'$ and $m'$) inside a single corre-lation domain of the $\mathcal{D}$ space associated with the decorated lattice. Indeed, all the spins show the same orientation inside their correlation domain in $\mathcal{S}$ because they are strongly correlated. By construction the image of this correlation domain in $\mathcal{D}$ contains the same number of spins and bonds as in the corresponding initial Kadanoff blocks in $\mathcal{S}$ but it is necessary to determine the set of quantum numbers characterizing all the bonds in $\mathcal{D}$. Thus, by extension, we can also speak of Kadanoff blocks in $\mathcal{D}$. In the simplest case the initial domain in $\mathcal{S}$ (respectively, its image in $\mathcal{D}$) can be reduced to a single lattice bond.

Technically we have to examine integrals $F_{i,j}$ (cf (2.16)) appearing in the characteristic polynomial associated with $Z_N(0)$ (cf (2.17)). Each of these integrals gives the spatial average of spin orientations in $\mathcal{S}$, for a given lattice site $(i, j)$ surrounded by four first-nearest neighbors. *The aim of the work then consists in: i) finding local relationships between quantum numbers l and l' (respectively, m and m') characterizing the bonds inside the correlation domain, ii) and inside the first nearest-neighbor correlation domains in $\mathcal{D}$, iii) extending the work to the whole lattice for deriving general selection rules.* Due to the fact that there exist mirror symmetries in $\mathcal{S}$ because the spin orientations are similar inside a correlation domain, it becomes reasonable to examine the mirror symmetries in $\mathcal{D}$ which concern the state of the same spins.

In the *first step*, if this domain of $\mathcal{D}$ contains no symmetry element characterizing the decorated lattice (Fig. 1a, case *a*1) it is necessary to consider a similar domain chosen so that the couple of domains is characterized by at least one symmetry element. Both domains of this couple in $\mathcal{D}$ are not necessarily characterized by the same set of quantum numbers because the corresponding spin orientations in $\mathcal{S}$ can be different. When the initial domain in $\mathcal{D}$ contains a single symmetry element we have the following possibilities which are interesting to examine: i) considering the domain itself and each of the 7 remaining domains, separately; ii) taking into account the 8 domains simultaneously (Fig. 1a, case *a*2). If the initial correlation domain contains 4 symmetry elements the determination of the relationship between quantum numbers $l$ or $l'$ (respectively, $m$ or $m'$) can directly begin so that all the bonds inside each Kadanoff block in $\mathcal{D}$ are characterized by a couple $(l, m)$ or $(l', m')$ (Fig. 1a, case *a*3). Finally, due to the imbricate character of the universal selection rules, we can also find relationships between quantum numbers characterizing the bonds located on both sides of the common frontier between first-nearest neighbor domains (not necessarily containing a lattice symmetry element).

In a *second step* it then becomes possible to extend the reasoning to all the first-nearest neighbor Kadanoff blocks in $\mathcal{D}$. For a given temperature all these blocks have the same size, as explained above. This property is also fulfilled in the $\mathcal{D}$ space by construction due to the conservation of the number of spins. i) As a result, if considering a first-nearest neighbor block containing at least one $\mathcal{D}$-symmetry element, the reasoning can be extended to all the first-nearest blocks showing the same symmetry element (see Fig. 1b, case *b*1). ii) If the first-nearest neighbor block does not contain $\mathcal{D}$-symmetry elements, one can consider a couple of such blocks showing at least one symmetry element (Fig. 1b, case *b*2). It is then possible to derive relationships between the $l$'s (respectively, the $m$'s) characterizing the bonds belonging to this couple of blocks and relate them to the corresponding quantum numbers of the initial block.

Proceeding likewise for all the first-nearest blocks in the $\mathcal{D}$ space we determine all the quantum numbers $l$ and $m$ (respectively, $l'$ and $m'$) and thus obtain a new domain surrounding the initial block and composed of several independent domains. Finally we can proceed in a similar way for all the blocks which are first-nearest neighbors of this domain (second-near-est neighbors with respect to the initial block, Fig. 1b, case *b*3). It is also possible to consider the first-nearest neighbor domains along a horizontal median (double arrow 1) or a vertical one (double arrow 2). A last possibility consists in examining new correlation domains by turning around the final domain obtained at the end of step 1 (arrow 3). As a result we can derive all the coefficients $l$ and $m$ (respectively, $l'$ and $m'$) for the whole lattice i.e., *general selection rules*. The reasoning being independent of the number of lattice sites it can be ex-tended to the case of infinite lattices (thermodynamic limit) due to the imbricate character of selection rule characterizing each site.

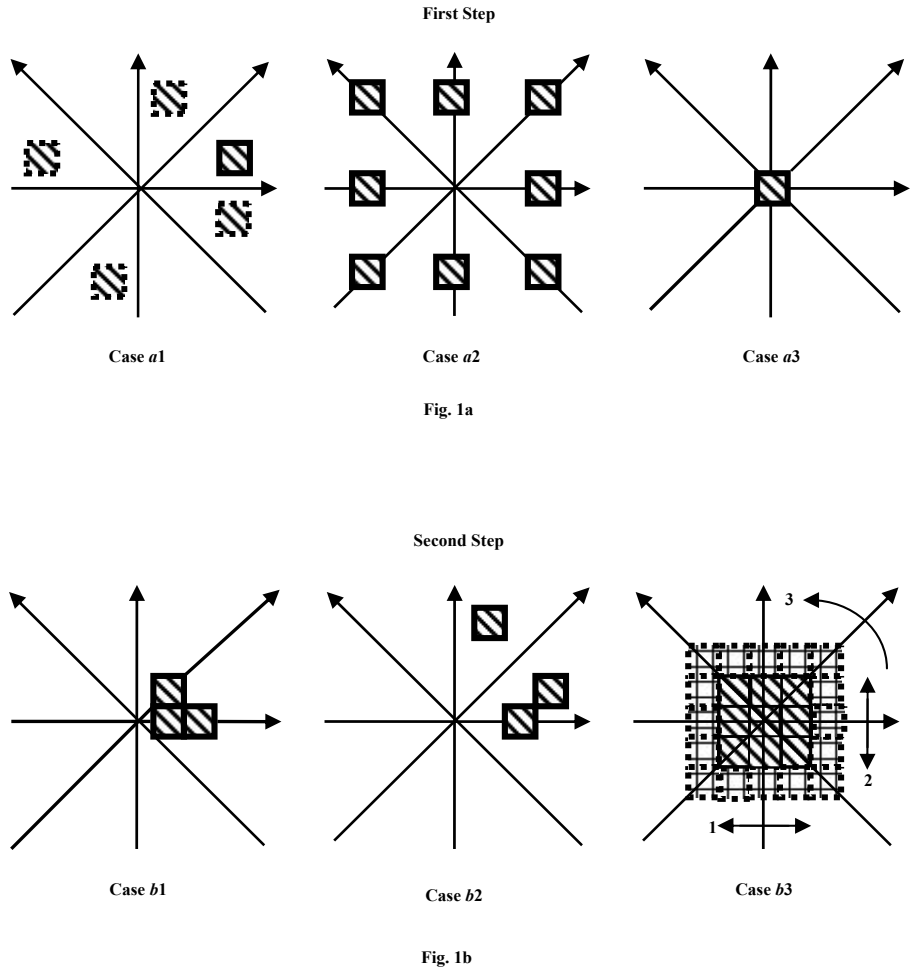

**Fig. 1** The initial correlation domain of the $\mathcal{D}$ space is represented by a solid-contoured diagonally hatched square; for the other possible positions the domains are dashed-contoured; for sake of simplicity the decorated lattice lines and rows have been omitted but only the decorated lattice symmetry axes are represented. **a** In the first step the initial correlation domain shows zero (case *a*1), one (case *a*2) or four symmetry elements (case *a*3). **b** In the second step it is possible to consider a new similar first-nearest neighbor correlation domain (cases *b*1 and *b*2) by using lattice symmetry elements; when all the first-nearest neighbor correlation domains have been examined there is a new correlation domain (oblique lines, case *b*3): it can be composed of a horizontal strip (double arrow 1), a vertical one (double arrow 2) or a square domain (arrow 3); the initial process can be re-iterated for the new second-nearest neighbor correlation domains (dashed-contoured); the arrows respectively give the sense of evolution for choosing a new correlation domain.

In addition, this reasoning is temperature-independent: as previously explained the size of each Kadanoff block diminishes with increasing $T$ in $\mathcal{S}$ space. By construction the size of the corresponding domain in $\mathcal{D}$ also diminishes. Thus the procedure described above can be renewed for each new $T$-value in spite of the fact that the number of independent domains increases. It means that *the set of quantum numbers l and m (respectively, l' and m') is unique because the $\mathcal{D}$-symmetry elements are temperature-independent*.

It has an important consequence: if the existence of correlation domains is necessary for justifying the procedure of establishing selection rules, it means that, finally, these rules are independent of the domain size in $\mathcal{S}$ and in its corresponding image in $\mathcal{D}$. In other words, *the establishment of selection rules can be achieved by just taking into account the symmetry elements of each decorated sublattice*.

### 2.2.6.4 The Physical Case: Thermodynamic Limit

The case of thermodynamic limit is less restrictive than that of a finite lattice because we deal with a purely mathematical problem. Indeed, irrespectively of the existence of lattice symmetry elements in the $\mathcal{D}$ space, $Z_N(0)$ is always expressed as a characteristic polynomial whose structure has been previously examined (*cf* (2.17)).

This polynomial consists of two parts. The first part contains the general term $[F_{i,j}\lambda_l(-\beta J)^2]^{2N(2N+1)}$ (for instance in case *a*) with one integral $F_{i,j}$ per site (giving a spatial average of the spin orientation) and one eigenvalue $\lambda_l(-\beta J)$ per bond (*cf* (2.8)) i.e., all the bonds are characterized by the same integer $l$. As a result it is not necessary to apply the duality principle. But, that supposes that the separate numerical study of integrals $F_{i,j}$ allows one to select a unique $m$-value so that $F_{i,j}$ is maximum. The second part is a product of terms such as $[F_{i,j}\lambda_l(-\beta J)^2]^n$ with $n < 2N(2N+1)$ so that only $n$ bonds are characterized by the same integer $l$; the other ones i.e., $n_1, n_2, ... n_k$ bonds are characterized by different integers $l_1, l_2, ... l_k$, respectively and we must have $n + n_1 + n_2 + ... + n_k = 4N(2N+1)$ (for instance in case *a*). For finding all the bonds characterized by the same integer $l_k$ it becomes necessary to use the duality principle in order to have a closed-form expression.

In the infinite lattice limit the highest eigenvalue naturally arises in the first-rank term $[F_{i,j}\lambda_l(-\beta J)^2]^{2N(2N+1)}$: the corresponding contribution $u_{max} = F_{i,j}\lambda_L(-\beta J)^2$ dominates all the other ones inside this term as well as all the terms composing the second-rank one (see Appendix 1). This can occur in the whole temperature range or in a smaller temperature range if there exist thermal crossover phenomena among the set of eigenvalues. In the latter case the predominant eigenvalue over a temperature range becomes subdominant when temperature is outside this range and a new eigenvalue previously subdominant becomes dominant and so on. In the case of 1$d$ spin chains, we always have the same highest eigenvalue (within the framework described previously). But, we shall see below that, for 2$d$ isotropic couplings, there exist thermal crossover phenomena within the set of eigenvalues. This behavior is detailed and explained below (*cf* Sec. 2.3.1). Of course the corresponding work is purely numerical.

As a result the corresponding contribution of the whole set of subdominant eigenvalues to the characteristic polynomial becomes negligible in the first-rank term as well as in the second-rank one. In other words we show in Appendix 1 that, in the thermodynamic limit, it is not necessary to know the exact structure of all these second-rank terms so that the duality principle becomes irrelevant as well as its restrictive conditions of applicability. Thus, if $u_{max}$ $= F_{i,j}\lambda_L(-\beta J)^2$ is the highest degree term in a given temperature range, it means that the characteristic polynomial (i.e., $Z_N(0)$) reduces to $[F_{i,j}\lambda_L(-\beta J)^2]^{2N(2N+1)}$ (in case *a*).

Under these conditions all the discarded cases detailed above (notably anisotropic couplings leading to spin arrangements characterized by a vorticity *q* and random couplings describing spin glasses) can formally be treated for infinite lattices composed of classical spins due to the fact that (2.6) involving the commutation of all the Hamiltonian operators is fulfilled. The unique difficulty then consists in expanding the local operator $\exp(-\beta H^{ex})$ on a basis of new adequate eigenfunctions $\mathscr{F}_{l,m}(\theta,\varphi)$. They are analogous to the spherical harmonics appearing in (2.7), for isotropic couplings, associated with the new eigenvalues $\lambda'_l(-\beta J_i)$ ($i = 1, 2$) which are analogous to the Bessel functions of (2.8). If the eigenfunctions and the eigenvalues respectively show a closed-form expression, then the new equation (2.7) becomes analytical.

A numerical study of the thermal behavior of the general term $F_{i,j}\lambda'_{l_{i,j}}(-\beta J)^2$ allowing to construct the characteristic polynomial giving $Z_N(0)$ is then necessary for determining the dominant term characterized by $l_{i,j} = l'_{i,j} = L$ over a given temperature range. $F_{ij}$ is always given by (2.11) where, now, each spherical harmonics $Y_{l,m}(\theta,\varphi)$ is replaced by the new eigenfunction $\mathscr{F}_{l,m}(\theta,\varphi)$.

### 2.2.7 Selections Rules Inside a Correlation Domain

We now detail the effects of mirror symmetries on quantum numbers *l* and *m* (respectively, *l'* and *m'*) in the $\mathscr{D}$ space associated with the decorated spin lattice. We recall that, in the case of a plane lattice showing edges (case *a*), the associated decorated lattice is composed of two plane sublattices, each one showing a square unit cell: each bond of the first sublattice is characterized by a quantum number *l* (or *l'*) (subspace $\mathscr{L}$) while each bond of the second one is characterized by *m* (or *m'*) (subspace $\mathscr{M}$). In addition, for each sublattice, there are two couples of basic symmetry elements: i) mirrors $\mathscr{M}_1$ and $\mathscr{M}_2$ perpendicular to the lattice plane whose respective traces are the two medians; ii) through a rotation of $\pm \pi/4$ around an axis perpendicular to the lattice plane and located at the intersection of $\mathscr{M}_1$ and $\mathscr{M}_2$ we obtain a second couple of mirrors $\mathscr{M}_1'$ and $\mathscr{M}_2'$ whose respective traces are the two diagonals.

In the case of a lattice wrapped on a torus (case *b*), the associated decorated lattice is also composed of two sublattices: each sublattice is obtained by wrapping on a torus the corresponding $\mathscr{L}$-sublattice (respectively, $\mathscr{M}$-sublattice) of case *a*. As a result each sublattice is characterized by a square unit cell and each bond is described by a quantum number *l* or *l'* in space $\mathscr{L}$ (respectively, *m* or *m'* in space $\mathscr{M}$). Mirrors $\mathscr{M}_1$ and $\mathscr{M}_2$ also exist in $\mathscr{D}$ space: the previous $\mathscr{M}_1$ mirror of case *a* becomes an equatorial plane and $\mathscr{M}_2$ mirror is always perpendicular to $\mathscr{M}_1$ but now can rotate around the vertical axis of revolution $\Delta$ (two other symmetry elements also exist and are examined in Sec. 2.2.8.2 in which there are related to mirrors $\mathscr{M}_1'$ and $\mathscr{M}_2'$ of case *a*). Due to the similarity of mirrors characterizing cases *a* and *b* the reasoning can indifferently be applied to each case.

We must then find a relationship between the effects induced by these mirrors while applying the duality principle whose conditions of validity have been detailed above. For instance, in case *a*, the symmetry with respect to the first diagonal $i = j$ called $D_1$ (mirror $\mathscr{M}_1'$) and to the second diagonal $i = -j$ called $D_2$ (mirror $\mathscr{M}_2'$) leads to:

$$m'_{i+1,j} = m_{j,i},\; m_{i,j-1} = m'_{j,i},\; m'_{i,j} = m_{j,i-1},\; m_{i,j} = m'_{j+1,i},\; (\mathscr{M} \text{ space})$$

$$l'_{i+1,j} = l_{j,i},\; l_{i,j-1} = l'_{j,i},\; l'_{i,j} = l_{j,i-1},\; l_{i,j} = l'_{j+1,i},\; (\mathscr{L} \text{ space}), \text{ site } (i, j)\; (D_1),$$

$$m'_{j+1,-i} = m_{i,-(j+1)} \, , \, m_{j,-(i+1)} = m'_{i+1,-j} \, , \, m'_{j,-i} = m_{i,-j}, \, m_{j,-i} = m'_{i,-j}, \, (\mathcal{M} \text{ space}) \qquad (2.24)$$

$$l'_{j+1,-i} = l_{i,-(j+1)} \, , \, l_{j,-(i+1)} = l'_{i+1,-j} \, , \, l'_{j,-i} = l_{i,-j}, \, l_{j,-i} = l'_{i,-j}, \, (\mathcal{L} \text{ space}) \text{ site } (i, -j) \, (D_2),$$

and similar relationships for sites $(-i, -j)$ and $(-i, j)$.

Thus, considering the effects of mirrors $\mathcal{M}_1'$ and $\mathcal{M}_2'$ in the expression of $Z_N(0)$ given by (2.17) allows one to write that $Z_N(0)$ *is independent of the bond orientation*. Indeed, for a lattice showing edges (case *a*), the bonds can be orientated from the upper horizontal edge to the lower one and from the left vertical edge to the right one. This situation also prevails in the case of a torus (case *b*) due to the fact that the upper and lower horizontal edges of the initial square lattice are linked as well as the left and right vertical edges. Thus, each site carrying a classical spin $S_{i,j}$ is characterized by: (i) two horizontal ''right'' arrows, with one arrow coming from its first ''left'' horizontal neighbor (couple $(l_{i,j-1}, m_{i,j-1})$) and one arrow going to its first ''right'' horizontal neighbor (couple $(l_{i,j}, m_{i,j})$); the corresponding contributions to the integral $F_{i,j}$ in term of spherical harmonics are given by $Y_{l_{i,j-1},m_{i,j-1}}(\theta_{i,j}, \varphi_{i,j})$ and $Y^*_{l_{i,j},m_{i,j}}(\theta_{i,j}, \varphi_{i,j})$, respectively; (ii) two vertical ''downwards'' arrows, with one arrow coming from its first upper vertical neighbor (couple $(l'_{i+1,j}, m'_{i+1,j})$) and one arrow going down to its first lower vertical neighbor (couple $(l'_{i,j}, m'_{i,j})$), with the corresponding harmonics $Y_{l'_{i+1,j},m'_{i+1,j}}(\theta_{i,j}, \varphi_{i,j})$ and $Y^*_{l'_{i,j},m'_{i,j}}(\theta_{i,j}, \varphi_{i,j})$.

As a result we can derive the action of mirrors $\mathcal{M}_1$ and $\mathcal{M}_2$ which are obtained from mirrors $\mathcal{M}_1'$ and $\mathcal{M}_2'$ by a rotation of $\pi/4$ around the axis located at their intersection. In terms of symmetry it means that, in case *a* (lattice showing edges) we deal with the horizontal and vertical medians of respective equations $i = 0$ and $j = 0$ which are traces of $\mathcal{M}_1$ and $\mathcal{M}_2$ mirrors. In case *b* (lattice wrapped on a torus) we have to consider the equatorial plane mirror $\mathcal{M}_1$ and the perpendicular rotating mirror $\mathcal{M}_2$. *The effect of such mirrors consists in transforming a spherical harmonics into its conjugated form and vice versa*, as for the symmetries with respect to diagonals. As a result we must have $F_{i,j} = F_{-i,j}^*$ ($\mathcal{M}_1$ mirror) and $F_{i,j} = F_{i,-j}^*$ ($\mathcal{M}_2$ mirror). As all the integrals $F_{i,j}$ are real due the *SRm* condition (*cf* (2.19)) we finally have (see Fig. 2a):

$$\mathcal{M}_1(F_{i,j}) = F_{-i,j} \, , \qquad (2.25)$$

so that:

$$m'_{i+1,j} = m'_{-i,j}, \; l'_{i+1,j} = l'_{-i,j} \text{ (bonds 1)}, \, m_{i,j-1} = m_{-i,j-1}, \, l_{i,j-1} = l_{-i,j-1} \text{ (bonds 2)},$$
$$m'_{i,j} = m'_{-(i-1),j}, \, l'_{i,j} = l'_{-(i-1),j} \text{ (bonds 3)}, \, m_{i,j} = m_{-i,j} \, , \, l_{i,j} = l_{-i,j} \text{ (bonds 4)}, \qquad (2.26)$$

and

$$\mathcal{M}_2(F_{i,j}) = F_{i,-j}, \qquad (2.27)$$

with now:

$$m'_{i+1,-j} = m'_{i+1,j}, \, l'_{i+1,-j} = l'_{i+1,j} \text{ (bonds 1)}, \, m_{i,-(j+1)} = m_{i,j} \, , \, l_{i,-(j+1)} = l_{i,j} \text{ (bonds 2)},$$
$$m'_{i,-j} = m'_{i,j} \, , \, l'_{i,-j} = l'_{i,j} \text{ (bonds 3)}, \, m_{i,-j} = m_{i,j-1}, \, l_{i,-j} = l_{i,j-1} \text{ (bonds 4)}. \qquad (2.28)$$

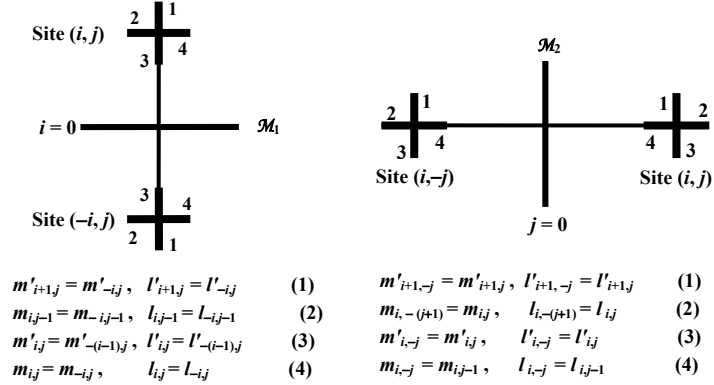

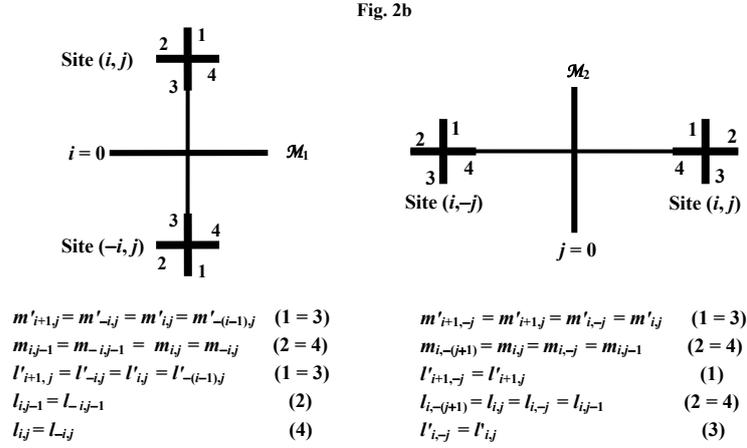

**Fig. 2** **a** Relationships between coefficients $m$ and $m'$ (respectively, $l$ and $l'$) derived from mirror symmetries. **b** Transformation of these relationships after taking into account the local selection rules $SRm$ and $SRl$.

If combining previous results with $SRm$'s (*cf* (2.19)) characterizing $(i, j)$ and $(-i, j)$ sites in the case of a horizontal mirror $\mathcal{M}_1$, we get by adding both relations:

$$m_{i,j-1} = m_{-i,j-1} = m_{i,j} = m_{-i,j}, \tag{2.29}$$

whereas by subtracting them

$$m'_{i+1,j} = m'_{-i,j} = m'_{i,j} = m'_{-(i-1),j}. \tag{2.30}$$

Similarly, for $(i, j)$ and $(i, -j)$ sites, in the case of a vertical mirror $\mathcal{M}_2$, we have by adding both $SRm$'s:

$$m'_{i+1,-j} = m'_{i+1,j} = m'_{i,-j} = m'_{i,j} \tag{2.31}$$

whereas by substracting them

$$m_{i,-(j+1)} = m_{i,j} = m_{i,-j} = m_{i,j-1}. \tag{2.32}$$

Now, if considering the $l$'s and $l''$s, the situation is a little bit more complicated. It is easily shown that, if we examine equations (2.22) and (2.23) derived from the initial non-vanishing condition of the CG coefficients involved in the expansion of integrals $F_{-i,j}$ and $F_{i,j}$, we have $L_{-i,j} = L_{i,j}$, $A_{-i,j} = A_{i,j}$ and $A'_{-i,j} = A'_{i,j}$ so that $g_{-i,j} = g_{i,j}$, $g'_{-i,j} = g'_{i,j}$. By subtracting $SRl2$ characterizing $(i, j)$ and $(-i, j)$ sites we obtain:

$$l'_{i+1,j} = l'_{-i,j} = l'_{i,j} = l'_{-(i-1),j} \ . \tag{2.33}$$

Equivalently, in the expansion of integrals $F_{i,-j}$ and $F_{i,j}$, we have $L_{i,-j} = L_{i,j}$, $A_{i,-j} = A_{i,j}$ and $A'_{i,-j} = A'_{i,j}$ so that $g_{i,-j} = g_{i,j}$, $g'_{i,-j} = g'_{i,j}$. By substracting $SRl2$ characterizing sites $(i, -j)$ and $(i, j)$ we obtain:

$$l_{i,-(j+1)} = l_{i,j} = l_{i,-j} = l_{i,j-1} \ . \tag{2.34}$$

All the previous results are summarized in Fig. 2b. *Thus, when dealing with a horizontal $\mathcal{M}_1$ mirror ($(i, j)$ and $(-i, j)$ sites), we have $m_{i,j-1} = m_{-i,j-1} = m_{i,j} = m_{-i,j} = m_1^0$ for the horizontal bonds and $m'_{i+1,j} = m'_{-i,j} = m'_{i,j} = m'_{-(i-1),j} = m_1'^0$ for the vertical bonds (with $m_1^0 \neq m_1'^0$ or $m_1^0 = m_1'^0$). However, only the coefficients $l'_{i+1,j}$, $l'_{-i,j}$, $l'_{i,j}$ and $l'_{-(i-1),j}$ characterizing the vertical bonds are equal. Similarly, when dealing with a vertical $\mathcal{M}_2$ mirror ($(i, j)$ and $(i, -j)$ sites), we have $m_{i,-(j+1)} = m_{i,j} = m_{i,-j} = m_{i,j-1} = m_2^0$ for the horizontal bonds and $m'_{i+1,-j} = m'_{i+1,j} = m'_{i,-j} = m'_{i,j} = m_2'^0$ for the vertical bonds (with $m_2^0 \neq m_2'^0$ or $m_2^0 = m_2'^0$). But now, only the coefficients $l_{i,-(j+1)}$, $l_{i,j}$, $l_{i,-j}$ and $l_{i,j-1}$ characterizing the horizontal bonds are equal.*

### 2.2.8 Generalization of Selections Rules to the Whole Lattice

#### 2.2.8.1 Case *a* : Lattice Showing Edges

In Step 1 we begin by using the basic symmetry elements i.e., mirrors $\mathcal{M}_1$ and $\mathcal{M}_2$ (see Fig. 2). Under these conditions let us examine all the sites of a horizontal line $i$. At site $(i, 0)$ we respectively have $m_{i,-1} = m_{i,0}$ on the one hand and $l_{i,-1} = l_{i,0}$ on the other one (with $-N \leq i \leq N$) due to the symmetry with respect to the vertical median ($\mathcal{M}_2$ mirror). Considering $(i, -1)$ and $(i, 1)$ sites we have $m_{i,-2} = m_{i,1}$ so that, due to the result obtained at site $(i, 0)$, we have $m_{i,-2} = m_{i,-1} = m_{i,0} = m_{i,1}$; by analogy we can directly write $m'_{i+1,-1} = m'_{i+1,1} = m'_{i,-1} = m'_{i,1}$ and $l_{i,-2} = l_{i,-1} = l_{i,0} = l_{i,1}$, $l'_{i+1,-1} = l'_{i+1,1}$, $l'_{i,-1} = l'_{i,1}$ ($-N \leq i \leq N$) through $\mathcal{M}_2$-symmetry (see Fig. 2b). Finally, due to the imbricate character of lattice sites a similar reasoning applied to the remaining sites of line $i$ allows one to show that all the corresponding horizontal bonds are characterized by the same couple of values $m_{i,j} = m_i$ and $l_{i,j} = l_i$ for any $j$, with $-N \leq j \leq N$ and $-N \leq i \leq N$.

Owing to $\mathcal{M}_1$-symmetry this property may be extended to each vertical line $j$ so that $m'_{i,j} = m'_j$ and $l'_{i,j} = l'_j$ for any $i$, with $-N \leq i \leq N$ and $-N \leq j \leq N$. Without considering further symmetries for different horizontal lines $i \neq i'$ (respectively, different vertical lines $j \neq j'$), we have $m_i \neq m_{i'}$ and $l_i \neq l_{i'}$ (respectively, $m'_j \neq m'_{j'}$ and $l'_j \neq l'_{j'}$). The symmetry with respect to the two mirrors $\mathcal{M}_1$ and $\mathcal{M}_2$ permits to derive that, more generally, the horizontal lines of ranks $+i$ and $-i$, with $-N \leq i \leq N$ (respectively the vertical lines of rank $+j$ and $-j$, with $-N \leq j \leq N$) are constituted of bonds characterized by the same couple of values $(l_i, m_i)$ (respectively, $(l'_j, m'_j)$).

The situation becomes particular if we consider the horizontal edges $i = -N$ and $i = N$ (respectively, the vertical ones $j = -N$ and $j = N$). For instance let us consider the upper horizontal edge $i = N$. For site $(N, 0)$ we have $m_{N,-1} = m_{N,0}$ due the mirror symmetry $\mathcal{M}_2$ and the *SRm* condition (*cf* (2.19)) gives $m'_{N+1,0} + m_{N,-1} = m_{N,0} + m'_{N,0}$, with now $m'_{N+1,0} = 0$ due to the fact that there is no vertical bond beyond the edge. We immediately derive that $m'_{N,0} = 0$ and more generally, if extending this reasoning to the remaining sites $(N, j)$, we have $m'_{N,j} = 0$ with $-N \leq j \leq N$. This procedure can be re-iterated for all the horizontal lines $i$ from lines $i = N - 1$ to $i = 0$ (due to the mirror $\mathcal{M}_1$ it is not necessary to consider the lines between $i = -1$ and $i = -N$). As a result all the vertical in-bonds are characterized by $m'_{i,j} = 0$, with $-N < i < N$ and $-N < j < N$, whereas now $m_{N,j} = 0$ or $m_{N,j} \neq 0$.

A similar reasoning using the mirror symmetry $\mathcal{M}_1$ allows one to find that $m_{i,j} = 0$ for all the horizontal lines $i$. Consequently, we derive $m_{ed} = 0$ for the four lattice edges so that:

$$m_{i,j} = m'_{i,j} = m_a = 0 \quad \forall\ (i,j) \in \text{Square Lattice (case } a). \tag{2.35}$$

This particular *m*-value is independent of the number of lattice bonds so that it remains valid in the thermodynamic limit.

In Step 2 the symmetry with respect to the diagonal $D$ ($D$-symmetry, exchange of indices 1 and 2) allows one to derive $l_i = l'_j$ if $i = j$ and $i = -j$ but two consecutive horizontal (respectively, vertical) lines do not have the same coefficient $l_i$ (respectively, $l'_j$).

### 2.2.8.2 Case *b* : Lattice Wrapped on a Torus

We examine the case of a lattice showing edges but now wrapped on a torus (case *b*). We have previously seen that there are 3 possibilities for building a torus, respectively labeled *b*1, *b*2 and *b*3. We define the outer curvature radius $R_>$ and the inner radius $R_<$. $R$ is the radius of the circle obtained by rotating the centre of the vertical circle around the axis of revolution $\Delta$ (see Fig. 3a). We have $r = (R_> - R_<)/2$, $R = (R_> + R_<)/2$ and $R_< = R_> - 2r$ so that $R - r = R_<$. Thus, $R - r > 0$ if $R_< > 0$ (ring torus), $R - r = 0$ if $R_< = 0$ (horn torus) and $R - r < 0$ if $R_< < 0$ (spindle torus).

We recall that, in case *b*1, the horizontal lines $i = N$ and $i = -N$ (respectively, the vertical rows $j = N$ and $j = -N$) are linked to each other owing to $2N + 1$ vertical (respectively, horizontal) extra bonds. It means that $r = R_> = (2N+1)a/2\pi$ where $a$ is the lattice spacing. We then have $R_< = -R_>$ and $R = 0$, $R - r < 0$ (spindle torus). In case *b*2 the horizontal lines $i = N$ and $i = -N$ (respectively, the vertical rows $j = N$ and $j = -N$) coincide. In this case we have $r = R_> = 2Na/2\pi$ and $R = 0$, $R - r < 0$ (spindle torus). Case *b*3 is a mix of cases *b*1 and *b*2: for instance the horizontal lines $i = N$ and $i = -N$ coincide (there are $2N$ vertical bonds) and the vertical rows $j = N$ and $j = -N$ are linked by means of $N_e$ horizontal extra bonds on line $i = 0$ ($2NN_e$ for the $2N$ horizontal lines $i$). We then deal with a rectangular lattice with square unit cells and wrapped on a torus. In this case we have $R_> = (2N + N_e)a/2\pi$, $r = 2Na/2\pi$. If $R - r = R_< > 0$ (ring torus) we must have $R_< = R_> - 2r > 0$ i.e., $N_e > 2N$ (at least $4N^2$ bonds for the $2N$ horizontal lines $i$). For sake of simplicity we choose this latter case. In the thermodynamic limit $N \to +\infty$, all the curvature radii diverge: so we always have $R - r \geq 0$ if the condition $N_e \geq 2N$ is fulfilled, so that $R_>$ diverges more rapidly than $r$.

Now we examine the symmetry elements of a ring torus ($R - r \geq 0$). The Villarceau theorem states that four circles can be drawn from any point located at the torus surface [19]. The first one labeled $C_1$ (see Fig. 3b) is in the plane $\mathcal{M}'_1$ (containing this point) parallel to the equatorial plane of the torus ($\mathcal{M}_1$ mirror). The second one labeled $C_2$ is perpendicular to $C_1$ ($\mathcal{M}_2$ mirror). Due to the symmetry of revolution the $\mathcal{M}_2$ mirror can rotate around the axis of revolution $\Delta$. The two other circles are called Villarceau circles, respectively labeled $V_1$ and $V_2$. These circles are tangent to the torus in two points: the initial point P and the point located at the extremity of the circle diameter (see Fig. 3b). In addition Garnier has shown that, in fact, the Villarceau circles are the intersection between the torus and a sphere of radius $R$ and center $(r, 0)$ for $V_1$ (respectively, $(-r, 0)$ for $V_2$) where the couple $(y, z)$ is defined by the origin O and the axes O$y$ and O$z$ appearing in Fig. 3a [19]. The corresponding Villarceau planes respectively containing the Villarceau circles $V_1$ and $V_2$ cut the torus obliquely with respect to mirror $\mathcal{M}_1$ (see Fig. 3c). In this case too each of these planes can rotate around $\Delta$ while conserving the corresponding inclination angle $\psi$. Thus, if cutting obliquely the torus by a Villarceau plane, the two parts obtained are symmetrical with respect to this plane like those obtained by perpendicularly cutting the torus by a plane containing circle $C_1$ or $C_2$ (mirror $\mathcal{M}_1$ or $\mathcal{M}_2$). As a result each Villarceau plane becomes itself a mirror.

The correspondence with the symmetry elements of case *a* is straightforward. If opening the torus for retrieving a planar lattice showing edges, the medians of case *a* are the respective traces of mirrors $\mathcal{M}_1$ and $\mathcal{M}_2$ which are shared by cases *a* and *b*. The respective traces of Villarceau circles $V_1$ and $V_2$ at the torus surface locally appear as the diagonals $D_1$ and $D_2$ of case *a* (see Fig. 3c).

Thus, Step 1 is shared by cases *a* and *b* because they have the same basic symmetry elements i.e., mirrors $\mathcal{M}_1$ and $\mathcal{M}_2$. Consequently we can use all the corresponding results obtained in case *a*. i) Owing to $\mathcal{M}_2$-symmetry and due to the imbricate character of lattice sites all the horizontal bonds are characterized by the same couple of values $m_{i,j} = m_i$ and $l_{i,j} = l_i$ for any $j$, with $-N \leq j \leq N$ and $-N \leq i \leq N$. ii) Owing to $\mathcal{M}_1$-symmetry this property may be extended to each vertical line $j$ so that $m'_{i,j} = m'_j$ and $l'_{i,j} = l'_j$ for any $i$. Without considering further symmetries for different horizontal lines $i \neq i'$ (respectively, different vertical lines $j \neq j'$), we have $m_i \neq m_{i'}$ and $l_i \neq l_{i'}$ (respectively, $m'_j \neq m'_{j'}$ and $l'_j \neq l'_{j'}$). However, due to the fact that there are no edges for a torus, we have indifferently $m_i = 0$ or $m_i \neq 0$ (respectively, $m'_j = 0$ or $m'_j \neq 0$). Finally, the symmetry with respect to the two mirrors $\mathcal{M}_1$ and $\mathcal{M}_2$ permits to derive that, more generally, the horizontal lines of ranks $+i$ and $-i$, with $-N \leq i \leq N$ (respectively the vertical lines of rank $+j$ and $-j$, with $-N \leq j \leq N$) are constituted of bonds characterized by the same couple of values $(l_i, m_i)$ (respectively, $(l'_j, m'_j)$).

In Step 2 we use the fact that the torus is characterized by an axis of revolution $\Delta$. As a result the $\mathcal{M}_2$ mirror containing this axis can rotate around it (see Fig. 3a). Thus, two consecutive vertical lines are constituted of bonds characterized by the same couple of values $(l'_j, m'_j)$. This property also prevails for all the remaining vertical lines, indifferently (see Fig. 2).

Step 3 concerns the *D*-symmetry operation (exchange of horizontal and vertical lattice lines i.e., indices 1 and 2 for $J_i$): it is conditioned by the existence of bitangent planes to the torus, each one containing a Villarceau circle. We must have $R - r > 0$ (ring torus) or $R = r$ (horn torus). This symmetry permits to derive that $m_i = m'_j$ and $l_i = l'_j$ if $i = j$ (or $i = -j$). Due to the fact that the planes containing the Villarceau circles are inclined, this result cannot be obtained in a single step for all the horizontal and vertical torus bonds. But, due to the revolution axis $\Delta$, this operation can be extended to all the torus points and allows to characterize all the torus bonds. As we have just shown that all the bonds of the vertical lines of the torus are characterized by the same couple $(l'_j, m'_j)$, we derive that all the horizontal lines as well as all the vertical ones are characterized by the same couple $(l, m)$ or equivalently

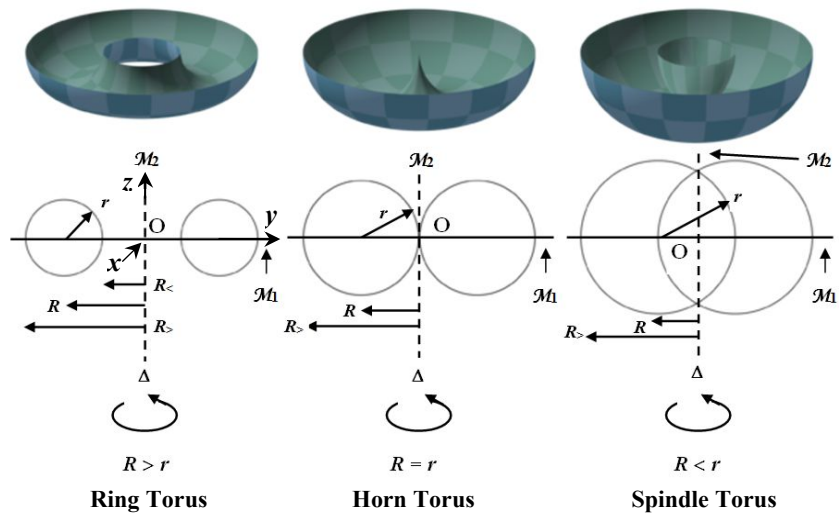

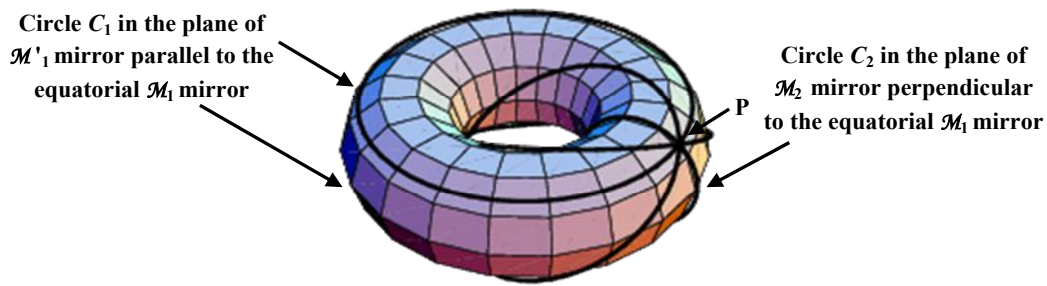

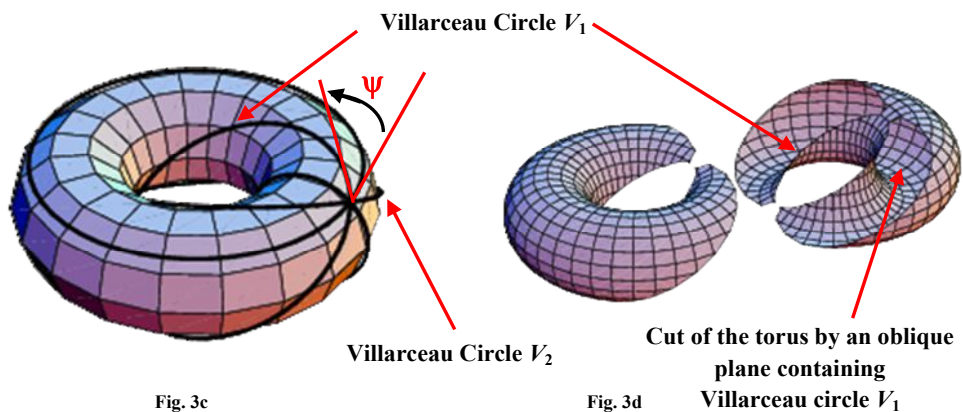

**Fig. 3** **a** Bottom-halves and cross-sections of the three torus classes. **b** The four symmetry elements of the torus. **c** Villarceau circles; ψ characterizes the cut inclination: it represents the angle between the plane parallel to the equatorial mirror $\mathcal{M}_1$ and the plane containing the Villarceau circle. **d** Cut of the torus by an oblique plane containing the Villarceau circle $V_1$.

$$m_{i,j} = m'_{i,j} = m_b \ , \ l_{i,j} = l'_{i,j} = l_b \ \ \forall \ (i,j) \in \text{Torus (case } b\text{)}. \tag{2.36}$$

This result is in perfect agreement with the geometrical properties of the toroidal environment, for each lattice site. In summary the property followed by coefficients $l$ and $m$ is similar to that encountered in the case of a finite chain showing cyclic conditions, as previously recalled (see Sec. 2.2.4).

In the case of a spindle torus ($R - r < 0$) we always have mirrors $\mathcal{M}_1$ and $\mathcal{M}_2$ but there are no more Villarceau circles: each of them becomes a hyperbola (see Fig. 3a). In other words the couple of Villarceau circles $V_1$ and $V_2$ is now a couple of hyperboles $H_1$ and $H_2$ for any point P located at the torus surface. The corresponding planes containing $H_1$ and $H_2$ also allow to characterize a $D$-symmetry. As a result this symmetry permits to derive that $m_i = m'_j$ and $l_i = l'_j$ if $i = j$ (or $i = -j$) and (2.36) is verified. More generally we derive that (2.36) is valid for any type of torus ($R > r$, ring torus, $R = r$, horn torus or $R < r$, spindle torus).

## 2.2.9 Expression of the Zero-Field Partition Function of a Finite Square Lattice (case *a*) and a Finite Torus (case *b*)

Due to the previously detailed selection rules we are now able to write the zero-field partition function for a lattice showing edges and for a torus. In the case of a square lattice (SL) showing edges (case *a*) the general form of the zero-field partition function previously given by (2.17) can be written as:

$$Z_N^{SL}(0) = (4\pi)^{4N(2N+1)} \left[ \sum_{l=0}^{+\infty} [\lambda_l(-\beta J_1)\lambda_l(-\beta J_2)]^{2N(2N+1)} \prod_{i=-N}^{N} \prod_{j=-N}^{N} F_{i,j}(l,l,0) \right.$$

$$\left. + \prod_{i=-N}^{N} \prod_{j=-N}^{N} \sum_{l_i=0}^{+\infty}{'} \sum_{l_j=0}^{+\infty}{'} F_{i,j}(l_i,l_j,0) \lambda_{l_i}(-\beta J_1) \lambda_{l_j}(-\beta J_2) \right]. \tag{2.37}$$

The notation $\sum_{l_i=0}^{+\infty}{'} \sum_{l_j=0}^{+\infty}{'}$ means that $l_i$ and $l_j$ are chosen so that the corresponding current second-rank term cannot be the first rank one. In addition, for in-sites, the integral $F_{i,j}(l_i,l_j,0)$ is given by:

$$F_{i,j}(l_i,l_j,0) = \frac{(2l_i+1)(2l_j+1)}{4\pi} \sum_{L=|l_i-l_j|}^{l_i+l_j} \frac{1}{2L+1} \left[ C_{l_i\ 0\ l_j\ 0}^{L\ 0} \right]^4 \quad i \neq N, j \neq N, \tag{2.38}$$

(with $F_{i,j}(l_i, l_j, 0) = F_{i,j}(l, l, 0)$ when $l_i = l_j = l$) and for lattice edges and corners:

$$F_{i,j}(l_{ed},l_j,0) = F_{i,j}(l_i,l_{ed},0) = \sqrt{\frac{2l+1}{4\pi}} \left[ C_{l_{ed}\ 0\ l\ 0}^{l_{ed}\ 0} \right]^2, \ l = l_i \text{ or } l_j, \ F_{\pm N,\pm N}(l_{ed},l_{ed},0) = 1. \tag{2.39}$$

The first term in the right-hand side of (2.37) represents the first-rank terms of the characteristic polynomial, as previously defined in Sec. 2.2.4.[5] It can also be obtained by simply imposing $l_{i,j} = l'_{i,j} = l$ and $m_{i,j} = m'_{i,j} = 0$, independently of the application of the duality principle (which leads itself to $m_{i,j} = m'_{i,j} = 0$). It means that its expression is valid for any spin arrangement (showing a random or a regular distribution of exchange energies) and for any vorticity $q$ if the new eigenvalue $\lambda_l(-\beta J)$ and the new integral $F_{i,j}$ are expressed via new eigenfunctions corresponding to the new local exchange Hamiltonian $H_{i,j}^{ex}$.

---

[5] A further numerical study confirms this property (see Sec. 2.3.1).

The second term appearing in the right-hand side (2.37) represents the second-rank terms. Their expression can only be detailed owing to the duality principle. It means that it is strictly valid for 2$d$ isotropic couplings.

Indeed, the application of the duality principle has allowed to derive that $m_{i,j} = m'_{i,j} = 0$ for the whole lattice. As explained in Sec. 2.2.6.4 the second-rank term structure appears as a product of terms such as $[F_{i,j}\lambda_l(-\beta J)^2]^n$ with $n < 2N(2N+1)$ so that only $n$ bonds are characterized by the same integer $l$; the remaining ones i.e., $n_1, n_2, \ldots n_k$ bonds are characterized by different integers $l_1, l_2, \ldots l_k$, respectively and we must have $n + n_1 + n_2 + \ldots + n_k = 4N(2N+1)$. The symmetry with respect to the two mirrors $\mathcal{M}_1$ and $\mathcal{M}_2$ permits to derive that, more generally, the horizontal lines of ranks $+i$ and $-i$, with $-N \leq i \leq N$ (respectively the vertical lines of rank $+j$ and $-j$, with $-N \leq j \leq N$) are constituted of bonds characterized by the same couple of values $(l_i, m_i)$ (respectively, $(l'_j, m'_j)$). The $D$-symmetry (exchange of 1 and 2 indices) allows one to derive $l_i = l'_j$ if $i = j$ and $i = -j$ but two consecutive horizontal (respectively, vertical) lines do not show the same coefficient $l_i$ (respectively, $l'_j$).

In the case of a torus (case $b$, with here $N_e = 2N + 1$ by sake of simplicity) the general form of the zero-field partition function reduces to:

$$Z_N^{\text{Torus}}(0) = (4\pi)^{2(2N+1)^2} \sum_{l=0}^{+\infty} \left[\lambda_l(-\beta J_1)\lambda_l(-\beta J_2)\right]^{(2N+1)^2} \sum_{m=-l}^{+l} \left[F_{in}(l,l,m)\right]^{(2N+1)^2} \quad (2.40)$$

where the function $\lambda_l(-\beta J_i)$, with $i = 1, 2$, is given by (2.8). The general integral $F_{in}(l, l, m)$ can directly be written from (2.16) and, as all the torus sites are in-sites, we have $F_{in}(l, l, m) = F(l, l, m)$ with

$$F(l,l,m) = \frac{(2l+1)^2}{4\pi} \sum_{L=0}^{2l} \frac{1}{2L+1}\left[C_{l\ 0\ l\ 0}^{L\ 0} C_{l\ m\ l\ m}^{L\ 2m}\right]^2. \quad (2.41)$$

We can formally rewrite (2.40) as follows:

$$Z_N^{\text{Torus}}(0) = (4\pi)^{2(2N+1)^2} \left[\sum_{l=0}^{+\infty} \left[F(l,l,0)\lambda_l(-\beta J_1)\lambda_l(-\beta J_2)\right]^{(2N+1)^2} + \right.$$

$$\left. + \sum_{l=1}^{+\infty} \left[\lambda_l(-\beta J_1)\lambda_l(-\beta J_2)\right]^{(2N+1)^2} \sum_{\substack{m=-l,\\m\neq 0}}^{+l} \left[F(l,l,m)\right]^{(2N+1)^2}\right] \quad (2.42)$$

As in (2.37) the first term in the right-hand side of (2.42) represents the first-rank term of the characteristic polynomial. It can also be obtained by simply imposing $l_{i,j} = l'_{i,j} = l$ and $m_{i,j} = m'_{i,j} = 0$, independently of the application of the duality principle. As in case $a$ (lattice showing edges) it means that its expression is stricty valid for any spin arrangement and for any vorticity $q$ on condition to express the new eigenvalue $\lambda_l(-\beta J)$ and the new integral $F_{i,j}$ via the new eigenfunctions corresponding to the new local exchange Hamiltonian $H_{i,j}^{ex}$.

The second term appearing in the right-hand side of (2.42) is the second-rank term. This term can be exclusively obtained owing to the duality principle. As in case $a$ it means that its expression is strictly valid for $2d$ isotropic. The symmetry with respect to the two mirrors $\mathcal{M}_1$ and $\mathcal{M}_2$ permits to derive that, more generally, the horizontal lines of ranks $+i$ and $-i$, with $-N \leq i \leq N$ (respectively the vertical lines of rank $+j$ and $-j$, with $-N \leq j \leq N$) are constituted of bonds characterized by the same couple of values $(l_i, m_i)$ (respectively, $(l'_j, m'_j)$). Due to the revolution symmetry the Villarceau plane containing the Villarceau circle (cf Fig. 3) which characterizes the $D$-symmetry (exchange of indices 1 and 2) can turn around the revolution axis $\Delta$ : it then allows one to derive that $l_i = l'_j$ if $i = j$ and $i = -j$ for all the torus sites in contrast to the case of the lattice showing edges (case $a$) so that all the horizontal and vertical torus lines are characterized by the same coefficients $l$ and $m$, respectively.

In the thermodynamic limit ($N \to +\infty$) we must have $Z_N^{SL}(0) \approx Z_N^{Torus}(0)$. Under these conditions we expect that only the first-rank terms of (2.37) and (2.42) remain, the second-rank terms being respectively negligible, in the whole temperature range (see Appendix 1). In addition, in the same limit, this means that, if comparing these first-rank terms, we must always have $F(l, l, 0) > F(l, l, m)$ for any $m \neq 0$ so that the ratio $[F(l,l,m)/F(l,l,0)]^{(2N+1)^2}$ becomes negligible if $N \to +\infty$. As a result a numerical study is unavoidable for verifying these assertions.

Finally, as the equation $Z_N^{SL}(0) \approx Z_N^{Torus}(0)$ is valid independently of the application of the duality principle, if $N \to +\infty$, we have no restriction regarding the nature of the spin arrangement and the corresponding vorticity $q$ on condition to express the new eigenvalue $\lambda_l(-\beta J)$ and the new integral $F_{i,j}$ via the new eigenfunctions corresponding to the new local exchange Hamiltonian $H_{i,j}^{ex}$.

## 2.3 Zero-Field Partition Function in the Thermodynamic Limit

### 2.3.1 Case of Finite Temperatures

We have just seen that the value $m = m_{i,j} = m'_{i,j} = 0$ is selected for all the bonds of a $2d$ lattice showing edges (case $a$) and a finite or an infinite size. When the two radii of curvature of the torus (case $b$) become infinite, we must have $Z_N^{SL}(0) \approx Z_N^{Torus}(0)$ as previously noted. In other words, as for the chain closed at its extremities (cyclic boundary conditions) and becoming infinite, the value $m = m_{i,j} = m'_{i,j} = 0$ must also be selected for all the bonds of the torus (case $b$), in the thermodynamic limit ($N \to +\infty$).

In this respect we have studied the integral $F(l, l, m)$ given by (2.41). In Fig. 4a we have reported the ratio $F(l, l, m)/F(0,0,0)$ vs $l$ for various $m$-values such as $|m| \leq l$ (with $F(0,0,0) = 1/4\pi$). We immediately observe that this ratio rapidly decreases for increasing $|m|$-values, for any $l$. However, we have zoomed the beginning of each curve corresponding to the case $|m| = l$. This trend is not followed but we always have $F(l, l, m) < F(l, l, 0)$ (see Fig. 4b). In addition, in Fig. 4c, for $m = 0$, we observe that $F(l_i, l_j, 0)$ decreases for $l_j < l_i$. As a result, when $N \to +\infty$, the integral $F(l, l, 0)$ appears as the dominant term in the generic one $[F(l,l,m)]^{(2N+1)^2} \approx [F(l,l,m)]^{4N^2}$ so that the value $m = 0$ is selected.

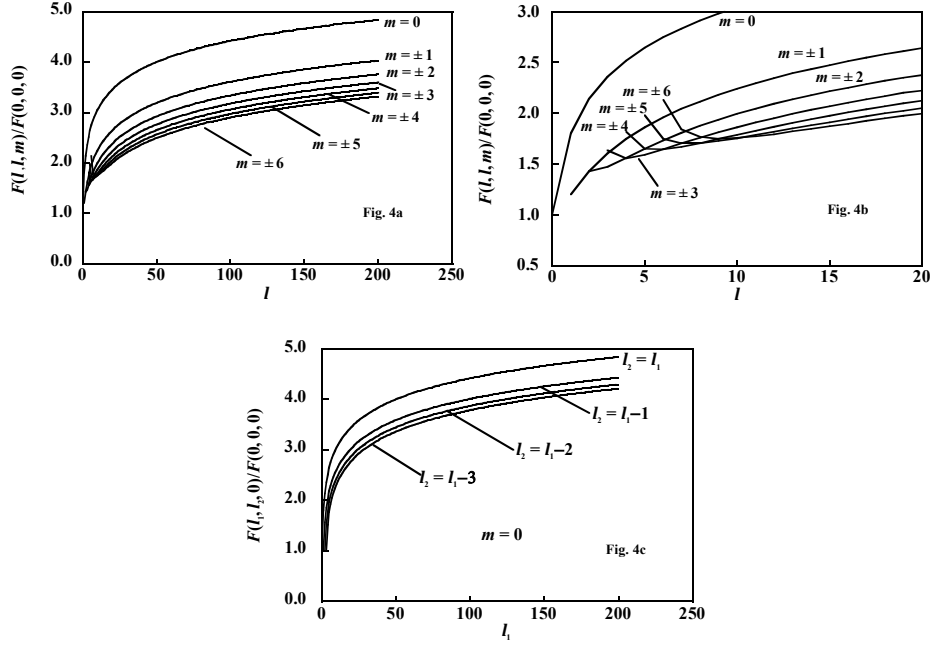

**Fig. 4 a** Numerical study of the ratio $F(l, l, m)/F(0, 0, 0)$ vs $l$ for various values of $m$ ($F(l, l, m)$ is given by (2.41) and $F(0, 0, 0) = 1/4\pi$). **b** Zoom of the previous study. **c** Numerical study of the ratio $F(l_1, l_2, 0)/F(0, 0, 0)$ for various values of $l_2 \leq l_1$.

In a first step we focus on the zero-field partition function of a lattice showing edges (case *a*). Independently of the application of the duality principle, we can directly start from (2.17). We note that, if the value $m = 0$ is selected when $N \to +\infty$, (2.17) (which gives the characteristic polynomial associated with $Z_N(0)$ for a lattice showing edges (case *a*) or a lattice wrapped on a torus (case *b*)) exactly gives (2.37). Indeed, for a lattice showing edges both equations can commonly be rewritten as

$$Z_N^{\text{SL}}(0) = (4\pi)^{8N^2}\left[\sum_{l=0}^{+\infty}\left[F(l,l,0)\lambda_l(-\beta J_1)\lambda_l(-\beta J_2)\right]^{4N^2}\right.$$

$$\left.+\prod_{i=-N}^{N}\prod_{j=-N}^{N}\sum_{l_i=0}^{+\infty}{'}\sum_{l_j=0}^{+\infty}{'}F(l_i,l_j,0)\lambda_{l_i}(-\beta J_1)\lambda_{l_j}(-\beta J_2)\right],\text{ as } N\to+\infty \text{ (case } a\text{)} \quad (2.43a)$$

As previously noted the first term in the right-hand side of (2.43a) represents the first-rank terms of the characteristic polynomial whereas the second one gives all the second-rank terms. The notation $\sum_{l_i=0}^{+\infty}{'}\sum_{l_j=0}^{+\infty}{'}$ has been defined after (2.37). In Appendix 1 we show that, in the thermodynamic limit, the second-rank terms become negligible with respect to the first-rank ones, for any form of their mathematical expression. As a result (2.43a) reduces to

$$Z_N^{\text{SL}}(0) = (4\pi)^{8N^2}\sum_{l=0}^{+\infty}\left[F(l,l,0)\lambda_l(-\beta J_1)\lambda_l(-\beta J_2)\right]^{4N^2},\text{ as } N\to+\infty \text{ (case } a\text{)}. \quad (2.43b)$$

In a second step, when the lattice wrapped on a torus becomes infinite, the edge effects become negligible. As a result the zero-field partition function associated with the torus showing two infinite radii of curvature (*cf* (2.40) and (2.42)) reduces to:

$$Z_N^{\text{Torus}}(0) = (4\pi)^{8N^2} \sum_{l=0}^{+\infty} \left[ F(l,l,0) \lambda_l(-\beta J_1) \lambda_l(-\beta J_2) \right]^{4N^2}, \text{ as } N \to +\infty \quad \text{(case } b) \quad (2.43c)$$

because the value $m = 0$ is selected. The integral $F(l, l, 0)$ is now given by (2.41) with $l = l_i = l_j$. This result brings a strong validation to the application of duality principle: in case *a*, for a finite lattice showing edges, we have shown that $m = 0$ owing to this principle; as the reasoning can be extended to infinite lattices we retrieve this result by a separate numerical study in case *b* so that (2.43b) and (2.43c) coincide.

Now we must wonder if all the current terms of the previous *l*-series must be kept i.e., if the series must be truncated, for a given range of temperature. If we define the ratio $r_l$

$$r_l = \frac{F(l,l,0)}{F(0,0,0)} \frac{\lambda_l(-\beta J_1) \lambda_l(-\beta J_2)}{\lambda_0(-\beta J_1) \lambda_0(-\beta J_2)} \quad (2.44)$$

(2.43c) can be rewritten as:

$$Z_N^{\text{Torus}}(0) = (4\pi)^{8N^2} \left[ F(0,0,0) \lambda_0(-\beta J_1) \lambda_0(-\beta J_2) \right]^{4N^2} \sum_{l=0}^{+\infty} (r_l)^{4N^2}, \text{ as } N \to +\infty \quad (2.45)$$

where $r_0 = 1$.

The numerical study is restricted to the case $J = J_1 = J_2$ for sake of simplicity but the reasoning can easily be extended to the general case $J_1 \neq J_2$. In addition it is achieved for temperatures $T > 0$ K in the whole range $[0+\varepsilon, +\infty[$, with $\varepsilon \ll 1$. Let us consider the following ratio:

$$\frac{r_{l\pm 1}}{r_l} = \frac{F(l \pm 1, l \pm 1, 0)}{F(l,l,0)} \left( \frac{\lambda_{l\pm 1}(-\beta J)}{\lambda_l(-\beta J)} \right)^2, J = J_1 = J_2. \quad (2.46)$$

We have studied the thermal behavior of the ratio $r_{l+1}/r_l$ for various finite *l*-values. It is reported Fig. 5a. We observe that $\log_{10}(r_{l+1}/r_l)$ shows a decreasing linear behavior with respect to $k_B T / |J|$. We have zoomed Fig. 5a in the very low-temperature domain (see Fig. 5b). If $r_{l+1}/r_l < 1$ $\log_{10}(r_{l+1}/r_l) < 0$ and $\log_{10}(r_{l+1}/r_l) > 0$ if $r_{l+1}/r_l > 1$. We can then observe a succession of *crossovers*, each crossover being characterized by a specific temperature called *crossover temperature* $T_{\text{CO}}$. $T_{\text{CO}}$ is the solution of the following equation:

$$r_l(T_{\text{CO}}) = r_{l+1}(T_{\text{CO}}) \quad (2.47)$$

i.e., owing to (2.46):

$$\frac{\lambda_{l+1}(-|J|/k_B T_{\text{CO}})}{\lambda_l(-|J|/k_B T_{\text{CO}})} = \left[ \frac{F(l,l,0)}{F(l+1,l+1,0)} \right]^{1/2}. \quad (2.48)$$

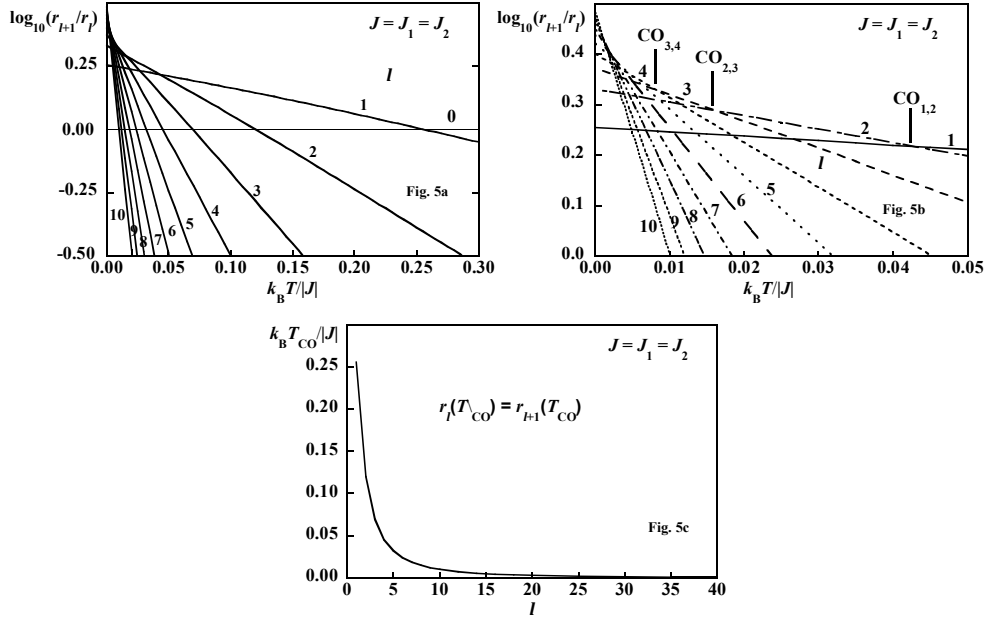

**Fig. 5 a** Thermal variations of the ratio $\log_{10}(r_{l+1}/r_l)$ for various values of $l$ where $r_{l+1}/r_l$ is defined by (2.46). **b** Zoom of the plot allowing to have a better insight of the crossover phenomena between various $l$-regimes. **c** Plot of the crossover temperature $T_{CO}$ vs $l$.

For instance, for the reduced temperatures such as $k_B T/|J| \geq 0.255$, the value $l = 0$ is dominant i.e., $\lambda_0(-\beta J)$ represents the dominant term of the characteristic polynomial. All the other terms $\lambda_l(-\beta J)$ with $l > 0$ are subdominant. When $0.255 \geq k_B T/|J| \geq 0.043$ $l = 1$ is dominant so that $\lambda_1(-\beta J)$ is now the dominant term of the characteristic polynomial whereas $\lambda_0(-\beta J)$ has become the subdominant one as well as all the other terms $\lambda_l(-\beta J)$ with $l > 1$ etc... In that case the crossover temperature corresponding to the transition between the regimes respectively characterized by $l = 0$ and $l = 1$ is labeled $T_{CO_{0,1}}$.

We have reported $k_B T_{CO}/|J|$ vs $l$ in Fig. 5c. As expected we observe that $T_{CO}$ rapidly decreases when $l$ increases. It means that, when the temperature is cooling down, near absolute zero, it appears a succession of closer and closer crossovers so that all the eigenvalues, characterized by an increasing $l$-value, successively play a role. But, when $T \approx 0$ K, all these eigenvalues intervene due to the fact that the crossover temperatures are closer and closer. As a result we can say that $T = 0$ K *plays the role of critical temperature*. This aspect is more detailed in the next section.

How interpreting this phenomena? In the 1$d$-case (infinite spin chain) we always have $\lambda_0(-\beta J)$ as dominant eigenvalue in the whole range of temperature, the integral $F(l, 0)$ being always equal to unity. In the 2$d$-case the situation is more complicated. The appearance of successive predominant eigenvalues is due to the presence of $F(l, l, 0) \neq 1$, for any $l > 0$. A numerical fit shows that the ratio $F(l, l, 0)/F(0, 0, 0)$ appearing in Fig. 3a increases with $l$ *according to a logarithmic law*, more rapidly than the ratio $|\lambda_l(-\beta J)/\lambda_0(-\beta J)|^2$ decreases with $l$, for a given temperature. The particular case $l \to +\infty$ is examined in the next section.

Owing to the numerical study summarized in Fig. 5b, with $J_1 = J_2$ or $J_1 \neq J_2$, we are thus led to consider the respective lower and upper limits $T_{i,<}$ and $T_{i,>}$ of a temperature range $[T_{i,<}, T_{i,>}]$ (where $T_{i,<}$ and $T_{i,>}$ are the crossover temperatures characterizing two consecutive regimes). In this regime one of the current second-rank terms (in infinite number) of the characteristic polynomial labeled $u_{l_i,l_j}$ and defined as

$$u_{l_i,l_j}(T) = F(l_i, l_j, 0)\lambda_{l_i}(-\beta J_1)\lambda_{l_j}(-\beta J_2), \quad T \in [T_{i,<}, T_{i,>}] \tag{2.49}$$

behaves as the dominant term called $u_{max}$ if $l_i = l_j$; the corresponding $l$-value is called $l_{max}$. The remaining terms can be classified in the decreasing modulus order from $u_{l_i} = u_{max}$:

$$u_{max}(T) > u_1(T) > u_2(T) > \ldots > u_\infty(T), \quad T \in [T_{i,<}, T_{i,>}]. \tag{2.50}$$

This property remains valid in the whole range of temperature $[0+\varepsilon,+\infty[$, $\varepsilon \ll 1$, for each new $u_{l_i} = u_{max}$ changing with the range $[T_{i,<}, T_{i,>}]$. In Appendix 1, for each range $[T_{i,<}, T_{i,>}]$, we rigorously show that: i) the second-rank terms of (2.43a) become negligible with respect to the first-rank ones; ii) the corresponding closed-form expression of the zero-field partition function of a 2$d$ square lattice showing edges (case $a$), labeled $Z_N^{SL}(0)$, and that of a torus (case $b$), labeled $Z_N^{Torus}(0)$, coincide in the thermodynamic case ($N \to +\infty$), the edge effects becoming negligible. As the whole range of temperature $[0+\varepsilon,+\infty[$, with $\varepsilon \ll 1$, is covered by adding all the consecutive ranges $[T_{i,<}, T_{i,>}]$, the result can be extended to $[0+\varepsilon,+\infty[$. Thus, as for a chain closed at its extremities (cyclic boundary conditions) and becoming infinite, the value $m = m_{i,j} = m'_{i,j} = 0$ is selected for all the bonds and we can justify again that

$$Z_N^{SL}(0) \approx Z_N^{Torus}(0) \approx (4\pi)^{8N^2} \sum_{l=0}^{+\infty} [F(l,l,0)\lambda_l(-\beta J_1)\lambda_l(-\beta J_2)]^{4N^2}, \text{ as } N \to +\infty. \tag{2.51}$$

This expression is valid independently of the application of the duality principle.

### 2.3.2 Study near $T = 0$ K

Now we thoroughly examine the case $T \to 0$ K (i.e., $\beta|J| \to +\infty$) as $l \to +\infty$. As previously we restrict the study to the case $J = J_1 = J_2$ for sake of simplicity (the reasoning can easily be extended to the general case $J_1 \neq J_2$). In addition the forthcoming results are going to allow the comparison with low-temperature results previously published in this case [4,5]. Let us begin with the integral $F(l, l, 0)$ composed of four spherical harmonics $Y_{l,0}(\theta,\varphi)$. In the infinite $l$-limit we have [12]:

$$Y_{l,0}(\theta,\varphi) \approx \frac{1}{\pi\sqrt{\sin\theta}}\left\{\left(1-\frac{3}{8l}\right)\cos\left((2l+1)\frac{\theta}{2}-\frac{\pi}{4}\right) - \frac{1}{8l\sin\theta}\cos\left((2l+3)\frac{\theta}{2}-\frac{3\pi}{4}\right)\right\},$$

$$+ O(l^{-2}), \text{ as } l \to +\infty, \varepsilon \leq \theta \leq \pi - \varepsilon, 0 < \varepsilon \ll 1/l, 0 \leq \varphi \leq 2\pi. \tag{2.52}$$

Then, after reporting this asymptotic behavior in the integral form of $F(l, l, 0)$ given by (2.11), a tedious but not complicated calculation allows one to show that:

$$F(l,l,0) \approx \frac{1}{\pi^3}\ln(l), \text{ as } l \to +\infty \tag{2.53}$$

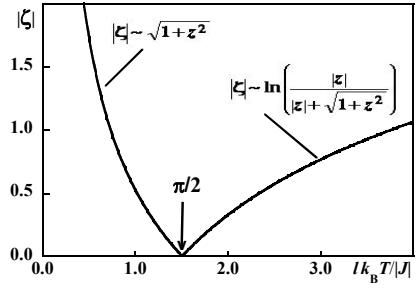

**Fig. 6** Thermal variations of $|\zeta|$ for various values of $lk_BT/|J| = 1/|z|$.

so that:

$$\frac{F(l\pm 1, l\pm 1, 0)}{F(l,l,0)} \to 1 \pm \frac{1}{l\ln(l)}, \text{ as } l \to +\infty. \tag{2.54}$$

This ratio is always slightly greater (sign +) or slightly lower (sign −) than unity but tends to unity by increasing $l$-values. Now we must examine the ratios $\lambda_{l\pm 1}(-\beta|J|)/\lambda_l(-\beta|J|)$ i.e., $I_{l\pm 1}(-\beta|J|)/I_l(-\beta|J|)$, as $l \to +\infty$. Intuitively, when $T \to 0$, we guess that the comparisons $\beta|J| \gg l$, $\beta|J| \sim l$ and $\beta|J| \ll l$ are essential. This study is detailed further.

The behavior of Bessel function $I_l(-\beta|J|)$ as $l \to +\infty$ and $\beta|J| \to +\infty$ has been established by Olver [20]. In Appendix 2 we have extended this work to a large order $l$ (but not necessarily infinite) and to any real argument $z$ varying from a finite value to infinity. Thus, the study of the Bessel differential equation in the large $l$-limit necessitates the introduction of the dimensionless auxiliary variable:

$$\zeta = -\frac{J}{|J|}\left[\sqrt{1+z^2} + \ln\left(\frac{|z|}{1+\sqrt{1+z^2}}\right)\right], |z| = \frac{\beta|J|}{l}. \tag{2.55}$$

The numerical study of $|\zeta|$ is reported in Fig. 6. As expected we observe that there are two branches. $|\zeta|$ vanishes for a numerical value of $|z_0|^{-1}$ very close to $\pi/2$ so that there are 3 domains which are interpreted later. Let $T_0$ be the corresponding temperature; we find:

$$l\frac{k_B T_0}{|J|} = \frac{\pi}{2}. \tag{2.56}$$

In the formalism of renormalization group $T_0$ is called a *fixed point*. In the present 2$d$ case we have $l \to +\infty$. Thus we derive that $T_0 \to T_c = 0$ K as $l \to +\infty$ so that the critical temperature can be seen as a fixed point, as expected. As a result all the thermodynamic functions can be expanded as series of current term $|T - T_0|/T_0$ near $T_0 = T_c = 0$ K. In Fig. 6 we observe that, below $T_c$, we have $|z| \gg 1$ (i.e., $\beta|J| \gg l$) and $|\zeta| \sim (1+z^2)^{1/2}$ (i.e., $|\zeta| \sim |z|$). Above $T_c$, we have $|z| \ll 1$ (i.e., $\beta|J| \ll l$) and $|\zeta| \sim 1 + \ln(|z|/(1+(1+z^2)^{1/2})$ (i.e., $|\zeta| \sim 1 + \ln(|z|/2)$). Finally, if $T \approx T_0 = T_c$, $|z| \approx z_0$ i.e., $\beta|J| \approx l$ and $|\zeta| \approx 0$.

For convenience we now introduce the coupling constant $g$ at temperature $T$ as well as its reduced value $g^*$:

$$g = \frac{k_B T}{|J|}, \quad g^* = \frac{T}{T_c}. \tag{2.57}$$

$g$ measures the strength of quantum fluctuations. $g^*$ is a universal parameter and is $l$-independent. At the critical point $T_0 = T_c$ we have $g^* = 1$. Owing to (2.56), the critical coupling $g_c$ can be written as:

$$g_c = \frac{k_B T_c}{|J|}, \quad g_c = \frac{\pi}{2l}. \tag{2.58}$$

Chubukov *et al.* have found that, at the critical temperature $T_c$, the critical coupling is $g_c = 4\pi/\Lambda$ where $\Lambda = 2\pi/a$ is a relativistic cutoff parameter (*a* being the lattice spacing) [5]. Haldane has evaluated $g_c$ in the case of a classical spin lattice [6]. He proposed $g_c = 2\sqrt{d}\,a/S$ or equivalently $g_c = 2a/S$ if referring to the vertical rows or horizontal lines of the 2$d$-lattice characterized by the same exchange energy $J = J_1 = J_2$. In our case $S = 1$ so that $g_c = 2a$ or $g_c = 4\pi/\Lambda$, with $\Lambda = 2\pi/a$.

Independently of the introduction of the cutoff parameter $\Lambda$ we have to explain the difference between the value of $g_c$ given by (2.58) and the factor 4 necessary for obtaining $g_c = 4\pi/\Lambda$. Chakravarty *et al.* have evaluated $g_c$ for a $d$-dimensional spin lattice. They found that (in $\Lambda$-unit) [4]

$$g_c(d) = \frac{2(d-1)}{K_d}, \quad K_d^{-1} = 2^{d-1}\pi^{d/2}\Gamma\!\left(\frac{d}{2}\right). \tag{2.59}$$

Let us recall that the hypersurface $S_d$ of a $d$-dimensional unit sphere is $S_d = 2\pi^{d/2}/\Gamma(d/2)$ so that $K_d = S_d/(2\pi)^d$ is the integration measure. It means that $g_c$ is proportional to a geometrical factor through a dimensional one. This is explained in the next section while giving a physical interpretation to $g_c$ (see after (3.15)). When $d = 2$, $K_d = 1/2\pi$. Taking into account (2.59) with $g_c = 4\pi/\Lambda$ or $g_c = 2(2\pi)/\Lambda$, (2.58) can be rewritten as $g_c = 2(4K_d)^{-1}/2l$. In the case of Chakravarty *et al.* [4] or Chubukov *et al* [5] the spin lattice has a spacing *a* and similar Landé factors $G$ (the spin density is $S/a$). In our case the spin lattice is composed of two imbricate sublattices characterized by alternating Landé factors $G$ and $G'$ so that each sublattice described by $G$ or $G'$ has a spacing $2a$ (the spin density is $S/2a$) As a result, the surface of integration is 4 times greater in our case than in the case of Chakravarty *et al.* or Chubukov *et al.* [4,5]. Thus, for the cutoff parameter $\Lambda$, the correspondence with the notation of Chubukov is $\Lambda = 2l$. Now a first physical interpretation can be given from (2.58).

We introduce: i) the thermal de Broglie wavelength $\lambda_{DB}$, ii) the low-temperature spin wave celerity $c = 2\sqrt{2}|J|Sa/\hbar$ along the diagonal of the lattice (i.e., $c = 2|J|a/\hbar$ along the vertical rows or horizontal lines of the lattice characterized by the same exchange energy $J = J_1 = J_2$ and spacing *a*, with $\sqrt{S(S+1)} = 1$) and iii) the slab thickness $L_\tau$ along the $i\tau$-imaginary axis of the $D$-space time ($D = 3$):

$$\lambda_{DB} = 2\pi \frac{\hbar c}{k_B T}, \quad L_\tau = \frac{\hbar c}{k_B T}. \tag{2.60}$$

In other words we have $L_\tau = \lambda_{DB}/2\pi$. By definition we must have

$$\lambda_{DB} \gg a \tag{2.61}$$

i.e., $\Lambda = 2\pi/a \gg 2\pi/\lambda_{DB}$ or equivalently

$$\Lambda \gg L_\tau^{-1}. \tag{2.62}$$

At the critical point $T_c = 0$ K $\lambda_{DB} \to +\infty$ (as well as $L_\tau$). It means that spins are strongly correlated. For finite temperatures $\lambda_{DB}$ and $L_\tau$ become finite. The adequate tool for estimating the correlation between any couple of spins is the correlation length $\xi$ (that we define below after (2.71) from the decay of the spin-spin correlation). As a result $\lambda_{DB}$ (or $L_\tau$) appears as the adequate unit length for measuring the correlation length. From (2.62) we see that $L_\tau$ is more convenient.

Under these conditions we generalize the application of the cutoff parameter $\Lambda$. $|z|$ defined by the second of (2.55) can be rewritten

$$|\tilde{z}| = \frac{\beta|J|}{\Lambda}. \tag{2.63}$$

Thus $|\tilde{z}|$ is a scaling parameter and $|\tilde{z}|\Lambda/2 = |z|l = \beta|J|$ is nothing but the inverse of a temperature scale. $|\zeta|\Lambda/2 = l|\zeta|$ which is a function of $z$ (cf (2.55)) can be rewritten vs $|\tilde{z}|$ and is denoted $|\tilde{\zeta}|\Lambda$. So $|\tilde{\zeta}|$ is another scaling parameter. It vanishes at $T = T_c$ like $|\zeta|$ (see Fig. 6), as shown by the renormalization approach [4].

In Appendix 2 we have established the $l$-polynomial expansion of the ratio $\lambda_{l\pm1}(lz)/\lambda_l(lz)$ (notably containing the dimensionless variable $l|\zeta|$) (see (6.20) and (6.21)) as $T \to 0$. Using a similar reasoning as above we can directly express the ratios $\lambda_{\Lambda\pm1}(\tilde{z}\Lambda)/\lambda_\Lambda(\tilde{z}\Lambda)$

$$\frac{\lambda_{\Lambda\pm1}(\tilde{z}\Lambda)}{\lambda_\Lambda(\tilde{z}\Lambda)} \approx -\frac{J}{|J|}\left\{\mp\left(\frac{1}{|\tilde{z}|} + \frac{1}{|\tilde{z}|\Lambda}\right) + \frac{I'_\Lambda(|\tilde{z}|\Lambda)}{I_\Lambda(|\tilde{z}|\Lambda)}\right\}, \text{ as } T \to 0,$$

$$\frac{I'_\Lambda(|\tilde{z}|\Lambda)}{I_\Lambda(|\tilde{z}|\Lambda)} \approx \frac{\varepsilon}{\tilde{u}|\tilde{z}|}\left[1 - \frac{\tilde{u}}{\Lambda} - \frac{\tilde{u}^2}{2\Lambda^2} - 2\exp\left(-|\tilde{\zeta}|\Lambda\left(1 - \frac{\tilde{u}}{2\Lambda} - \frac{3\tilde{u}^2}{8\Lambda^2}\right)\right) + O(\Lambda^{-3})\right],$$

$$\varepsilon = \begin{cases} +1 \text{ if } T > T_c \\ -1 \text{ if } T < T_c \end{cases}, \text{ as } T \to 0, \tag{2.64}$$

where $\varepsilon$ is the sign of the slope of $d|\tilde{\zeta}|/dT$ and where $\tilde{u}$ is given by:

$$\tilde{u} = \frac{1}{\sqrt{1+\tilde{z}^2}}. \tag{2.65}$$

As a result, in the low-temperature domain, it becomes possible to expand the ratios $\lambda_{\Lambda\pm1}(\widetilde{z}\Lambda)/\lambda_\Lambda(\widetilde{z}\Lambda)$ vs a general scaling parameter that we now need to define.

Chakravarty *et al.*[4] have introduced the physical parameters $\rho_s$ and $\Delta$ defined as:

$$\rho_s = |J|(1 - g^*), \quad \Delta = |J|(g^* - 1). \tag{2.66}$$

In the 2*d*-case $\rho_s$ and $\Delta$ have the dimension of an energy $JS^2$ (in our case $J$). $\rho_s$ is the *spin stiffness* of the ordered ground state (Néel state for an antiferromagnet) and $\Delta$ is the *energy gap between the ground state and a first excited state*. We justify these interpretations below. The writing of $\rho_s$ and $\Delta$ is motivated by the fact that these parameters must obey Josephson's scaling law near $T_c$ [22]:

$$\rho_s \approx (g_c - g)^{(D-2)\nu}, \quad \Delta \approx (g - g_c)^{(D-2)\nu} \tag{2.67}$$

where $D = d + 1$ is the space time dimension (here $D = 2 + 1$) and $\nu$ is the usual correlation length critical exponent ($\nu = 1$ as evaluated by Chakravarty *et al.* [4]). At the critical point $g^* = 1$ $\rho_s$ and $\Delta$ vanish at $T_c = 0$ K and, near critically, we have $\rho_s \ll |J|$ and $\Delta \ll |J|$ where $|J|$ finally appears as the bare value of $\rho_s$ and $\Delta$ i.e., their value at 0 K. For all the previous reasons we are led to introduce the following parameters:

$$\frac{\rho_s}{k_B T} = \frac{1}{g} - \frac{1}{g_c}, \quad \frac{\Delta}{k_B T} = 4\pi\left(\frac{1}{g_c} - \frac{1}{g}\right) \tag{2.68}$$

where the factor $4\pi$ appears in $\Delta$ for notational convenience (the correspondence with the notation of Chakravarty *et al.* is assumed in (2.67) by multiplying $\Delta$ by the factor $4\pi$).

The expression giving $g^*$ must be handled with care because we have $T_0 \to T_c = 0$ K as $\Lambda \to +\infty$. Thus, when $g^* < 1$, it means that this ratio remains finite by always imposing $T < T_0$ i.e., owing to (2.63) which can be rewritten as $|\widetilde{z}|/\widetilde{z}_c = 1/g^*$, $\beta|J| > \Lambda$. When $g^* > 1$ we always have $T > T_0$ and $\beta|J| < \Lambda$. Finally $\beta|J| \sim \Lambda$ means that we have $T \sim T_0 = T_c$ in the vanishing temperature limit.

$\lambda_{\Lambda\pm1}(\widetilde{z}\Lambda)/\lambda_\Lambda(\widetilde{z}\Lambda)$ with $|\widetilde{z}|\Lambda = \beta|J|$, jointly appears with the ratio of integrals $F(\Lambda \pm 1, \Lambda \pm 1, 0)/F(\Lambda, \Lambda, 0)$ (*cf* (2.54)) in the ratios $r_{\Lambda\pm1}/r_\Lambda$ given by (2.46). If $T_0 \to T_c = 0$ K ($\Lambda \to +\infty$) we have $F(\Lambda \pm 1, \Lambda \pm 1, 0)/F(\Lambda, \Lambda, 0) \to 1$ by upper values (sign +), respectively lower values (sign −) of $\Lambda$ at the order $(\Lambda\ln(\Lambda))^{-1}$. As a result it means that the low-temperature behavior of the ratios $r_{\Lambda\pm1}/r_\Lambda$ is mainly given by that of the ratios of two consecutive eigenvalues $\lambda_{\Lambda\pm1}(\widetilde{z}\Lambda)/\lambda_\Lambda(\widetilde{z}\Lambda)$ expressed by (2.64). This is a reasonable result because all the physics is derived from the thermal behavior of this ratio. Up to first order in $\Lambda^{-1}$ we can write:

$$\frac{r_{\Lambda\pm1}}{r_\Lambda} \approx \left(\frac{\lambda_{\Lambda\pm1}(\widetilde{z}\Lambda)}{\lambda_\Lambda(\widetilde{z}\Lambda)}\right)^2 + O((\Lambda\ln(\Lambda))^{-1}), \text{ as } T \to 0. \tag{2.69}$$

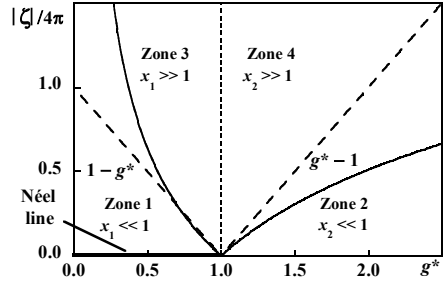

**Fig. 7** Thermal variations of $|\zeta|/4\pi$ *vs* $g^*$ and domains of predominance *vs* dimensionless parameters $x_1$ and $x_2$ defined by (2.75) and (2.76); $x_1$ and $x_2$ control the scaling properties of the magnetic system.

We always have $\lambda_{\Lambda+1}(\tilde{z}\Lambda) < \lambda_{\Lambda}(\tilde{z}\Lambda)$ for any value of $|\tilde{z}|\Lambda = \beta|J|$. As this property remains valid in the whole range of temperature we always have $r_{\Lambda+1}/r_{\Lambda} < 1$. However, it is interesting to known the *T*-decreasing law of $r_{\Lambda+1}/r_{\Lambda}$ when $T \to 0$. In Appendix 3, we have derived the closed-form expression of the spin-spin correlation $|<\boldsymbol{S}_{0,0}.\boldsymbol{S}_{0,1}>|$ between first-nearest neighbors valid for any temperature. The general closed-form expression of the spin-spin correlation between any couple of spins (i.e., for any couple of lattice site) will be detailed in another article. Thus we show that, in the low-temperature limit, near $T_c = 0$ K:

$$\left(\frac{r_{\Lambda+1}}{r_{\Lambda}}\right)^{1/2} \approx |<\boldsymbol{S}_{0,0}.\boldsymbol{S}_{0,1}>|, \text{ as } \Lambda \to +\infty, T \to 0, \qquad (2.70)$$

In addition (*cf* (7.7)-(7.13)) we have

$$\left(\frac{r_{\Lambda+1}}{r_{\Lambda}}\right)^{1/2} \approx \left|\frac{I'_{\Lambda}(|\tilde{z}|\Lambda)}{I_{\Lambda}(|\tilde{z}|\Lambda)}\right|, \text{ as } \Lambda \to +\infty, T \to 0. \qquad (2.71)$$

The Ornstein-Zernike theorem [23] describes the decreasing law of the spin-spin correlation at long distance *r* where *r* is the distance between any couple of sites $(i, j)$ and $(i+k, j+k')$, expressed *vs* the lattice parameter $2a$. Near the critical temperature $T_c = 0$ K the spin-spin correlation $|<\boldsymbol{S}_{i,j}.\boldsymbol{S}_{i+k,j+k'}>|$ decays as $\exp(-2a/\xi)$, where $\xi$ is the correlation length (this asymptotic form has also powers of *r* as prefactor when $r \gg 1$). When temperature strongly approaches $T_c$ the correlation length becomes quasi infinite and we have $|<\boldsymbol{S}_{0,0}.\boldsymbol{S}_{0,1}>| \to 1 - 2a/\xi$. If taking into account previous remarks done after (2.62) we can introduce the renormalized correlation length $\tilde{\xi}$ expressed in $L_\tau$-unit. Due to its universal behavior near $T_c$ we have automatically

$$\frac{\xi}{2a} \approx \frac{\tilde{\xi}}{L_\tau}, \text{ as } T \to 0 \qquad (2.72)$$

so that the decay of the spin-spin correlation between first nearest neighbors behaves as $|<S_{0,0}.S_{0,1}>| \to 1 - L_\tau/\widetilde{\xi}$ near $T_c = 0$ K. As a result the ratio of convergence $r_{\Lambda+1}/r_\Lambda$ appears equivalently as

$$\left(\frac{r_{\Lambda+1}}{r_\Lambda}\right)^{1/2} \approx 1 - 2a/\xi + \ldots, \quad \left(\frac{r_{\Lambda+1}}{r_\Lambda}\right)^{1/2} \approx 1 - L_\tau/\widetilde{\xi} + \ldots, \text{ as } T \to 0. \quad (2.73)$$

It means that we can have an estimate of the low-temperature behavior of the correlation length from the study of the ratio $r_{\Lambda+1}/r_\Lambda$. In other words, it becomes possible to identify the various zones appearing in Fig. 7, each of them being characterized by a specific regime of the correlation length.

Now we express all the scaling parameters previously encountered. In a first step, we expand $|\widetilde{\zeta}|$ in the low-temperature limit. We have the following asymptotic behaviors:

$$\frac{|\widetilde{\zeta}|}{4\pi} \approx |\zeta_F|_<, \quad |\zeta_F|_< = 1 - g^*, \quad T < T_c, \quad \frac{|\widetilde{\zeta}|}{4\pi} \approx |\zeta_F|_>, \quad |\zeta_F|_> = g^* - 1, \quad T > T_c, \quad (2.74)$$

so that the thermal study of $|\widetilde{\zeta}|$ is reduced to two domains: $g^* < 1$ i.e., $g < g_c$ ($T < T_c$) and $g^* > 1$ i.e., $g > g_c$ ($T > T_c$). Each of these domains can be divided itself into two subdomains according to as $|\widetilde{\zeta}| > |\zeta_F|$ or $|\widetilde{\zeta}| < |\zeta_F|$. In other words, in Fig. 7, if considering a vertical line $g^* = K$ where $K$ is any positive constant, with $g^* < 1$ or $g^* > 1$, as well as two different temperatures $T_1$ and $T_2$ such as $T_1 > T_2$ for instance, we have $T_1\Lambda_1 = T_2\Lambda_2$ due to the second of (2.57) so that $\Lambda_1 < \Lambda_2$ with $\Lambda_1 \gg 1$ and $\Lambda_2 \gg 1$. Recalling the remark done after (2.68) and concerning the behavior of $g^*$ it simply means that, in (2.58), $T_c$ decreases more rapidly towards 0 K when considering $\Lambda_2$. Physically it means that, in the diagram of Fig. 7, temperature increases along a vertical line $g^* = K$ when starting from 0 along the $|\zeta|$-axis. As a result we are led to introduce the following parameters:

$$x_1 = \frac{k_B T}{2\pi\rho_s}, \quad x_2 = \frac{k_B T}{\Delta} \quad (2.75)$$

where the factor $2\pi = g_c\Lambda/2$ also appears for notational convenience (as noted for $\Delta$ after (2.68)). Thus, if using the definition of $\rho_s$ and $\Delta$ (cf (2.69)), we verify that $x_1$ and $x_2$ are increasing functions of temperature when starting from 0 along a vertical line $g^* = K$ parallel to the $|\zeta|$-axis. As $\rho_s$ and $\Delta$ vanish at $T_0 = T_c$, $x_1$ and $x_2$ become infinite at this fixed point. Finally $x_1$ and $x_2$ can also be written as:

$$x_1 = \frac{2g^*}{1-g^*}\Lambda^{-1}, \quad x_2 = \frac{g^*}{g^*-1}\Lambda^{-1}. \quad (2.76)$$

As expected the scaling parameters $x_1$ and $x_2$ appear as functions of the universal parameter $g^*$. From a physical point of view and as noted by Chakravarty et al. [4] as well as by Chubu-kov *et al.* [5], $x_1$ and $x_2$ control the scaling properties of the magnetic system. $x_1\Lambda$ and $x_2\Lambda$ have been plotted in Fig. 8.

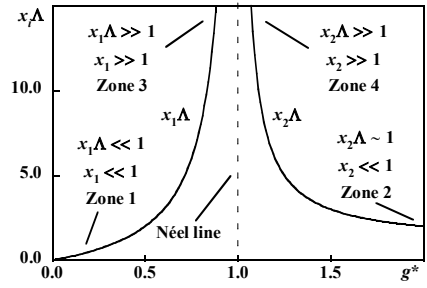

**Fig. 8** Plot of $x_1\Lambda$ and $x_2\Lambda$ defined by (2.76) *vs* $g^* = T/T_c$ and domains of predominance derived from the asymptotic behaviors of $x_1\Lambda$ and $x_2\Lambda$.

In summary, the comparison between Figs. 7 and 8 allows one to write that for $g^* < 1$ ($T < T_c$) we have two possibilities: if $|\tilde{\zeta}| > |\zeta_F|_<$ then $x_1 > 1$ and if $|\tilde{\zeta}| < |\zeta_F|_<$ then $x_1 < 1$. Similarly, for $g^* > 1$ ($T > T_c$) we have: $x_2 > 1$ if $|\tilde{\zeta}| > |\zeta_F|_>$ and $x_2 < 1$ if $|\tilde{\zeta}| < |\zeta_F|_>$.

There is an analytical continuity between $x_1$ and $x_2$ while passing through $T_0 = T_c$. We then guess that there are only 3 domains of predominance: $x_1 \ll 1$ ($T < T_c$ and $|\tilde{\zeta}| < |\zeta_F|_<$, Zone 1) i.e., $\rho_s \gg k_B T$, $x_2 \ll 1$ ($T > T_c$ and $|\tilde{\zeta}| < |\zeta_F|_>$, Zone 2) i.e., $\Delta \gg k_B T$; finally $x_1 \gg 1$ ($T < T_c$ and $|\tilde{\zeta}| > |\zeta_F|_<$, Zone 3) i.e., $\rho_s \ll k_B T$ and $x_2 \gg 1$ ($T > T_c$ and $|\tilde{\zeta}| > |\zeta_F|_>$, Zone 4) i.e., $\Delta \ll k_B T$. In this latter domain, along the line $T = T_c$ ($g^* = 1$), we directly reach the *Néel line* (see Fig. 8). Each of these domains previously described corresponds to a particular magnetic regime. The physical meaning of each regime is going to arise from the low-temperature study of the ratio $\lambda_{\Lambda+1}(\tilde{z}\Lambda)/\lambda_\Lambda(\tilde{z}\Lambda)$ *vs* $x_1$ or $x_2$.

However it remains a last step i.e., expressing the scaling parameter $|\tilde{\zeta}|$ *vs* $x_1$ or $x_2$. It is achieved in Appendix 4. We have rigorously shown:

$$|\tilde{\zeta}|\Lambda \approx 2\,\mathrm{argsh}\left(\frac{\exp(-1/x_1)}{2}\right), \text{(Zones 1 and 3)}, \quad (2.77)$$

$$|\tilde{\zeta}|\Lambda \approx 2\,\mathrm{argsh}\left(\frac{\exp(1/2x_2)}{2}\right), \text{(Zones 2 and 4)}. \quad (2.78)$$

As a result behaviors can be derived for the four zones of the magnetic diagram given by Fig. 6. In Zone 1 ($x_1\Lambda \ll 1$, $x_1 \ll 1$) we have

$$|\tilde{\zeta}|\Lambda \approx \exp(-1/x_1), x_1 \ll 1 \text{ (Zone 1)}. \quad (2.79)$$

In Zone 2 ($x_2\Lambda \sim 1$, $x_2 \ll 1$)

$$|\tilde{\zeta}|\Lambda \approx \frac{1}{x_2} + 2\exp(-1/x_2),\ x_2 \ll 1\ \text{(Zone 2)}. \qquad (2.80)$$

In Zones 3 ($x_1\Lambda \gg 1$, $x_1 \gg 1$) and 4 ($x_2\Lambda \gg 1$, $x_2 \gg 1$) we have

$$|\tilde{\zeta}|\Lambda \approx 2\ln\left(\frac{1+\sqrt{5}}{2}\right) - \frac{2}{\sqrt{5}x_1},\ x_1 \gg 1\ \text{(Zone 3)}. \qquad (2.81)$$

$$|\tilde{\zeta}|\Lambda \approx 2\ln\left(\frac{1+\sqrt{5}}{2}\right) + \frac{1}{\sqrt{5}x_2},\ x_2 \gg 1\ \text{(Zone 4)}. \qquad (2.82)$$

At the common frontier between Zones 3 and 4, when directly reaching $T_c$, $x_1$ and $x_2$ become infinite (*cf* Fig. 7) and $|\tilde{\zeta}|\Lambda$ shows the common limit:

$$|\tilde{\zeta}|\Lambda \approx 2\ln\left(\frac{1+\sqrt{5}}{2}\right),\ x_1 \to +\infty,\ x_2 \to +\infty. \qquad (2.83)$$

The ratio $\alpha = (1+\sqrt{5})/2$ is the golden mean. As a result, starting from a closed expression of $|\tilde{\zeta}|$ given by (2.55) for $|\zeta|$, we directly obtained for $|\tilde{\zeta}|\Lambda$ the result of Chubukov *et al.* derived from a renormalization technique [5]. Adopting their notation we define:

$$|\tilde{\zeta}|\Lambda \approx X_i(x_i),\ i = 1,\ 2. \qquad (2.84)$$

As a result, if comparing with (2.71)-(2.73), we immediately derive that, near $T_c$, $\xi/2a$ (or equivalently $\tilde{\xi}/L_\tau$) scales as

$$\frac{\xi}{2a} \approx \left(|\tilde{\zeta}|\Lambda\right)^{-1},\ \text{as}\ T \to 0. \qquad (2.85)$$

The low-temperature behaviors of the ratio $I'_\Lambda(|\tilde{z}|\Lambda)/I_\Lambda(|\tilde{z}|\Lambda)$ have been detailed up to order $\Lambda^{-1}$ in Appendix 4. This study allows one to obtain all the thermal behaviors of the correlation length $\xi$ expressed in *a*-unit (and its renormalized expression $\tilde{\xi}$ expressed in $L_\tau$-unit) for each of the various zones of the phase diagram appearing in Fig. 7. Below we directly report these results. In Zone 1 ($x_1\Lambda \ll 1$, $x_1 \ll 1$) we have:

$$\xi = \frac{e}{8}\frac{\hbar c}{2\pi\rho_s}\exp\left(\frac{2\pi\rho_s}{k_B T}\right)\left(1 + \frac{k_B T}{4\pi\rho_s}\right),\ x_1 \ll 1\ \text{(Zone 1, RCR)}. \qquad (2.86)$$

We exactly retrieve the result first obtained by Hasenfratz and Niedermayer [24] and confirmed by Chubukov *et al.* [5]. This is the *Renormalized Classical Regime* (RCR) characterized by $\rho_s \gg k_B T$: the divergence of $\xi$ describes a *long-range order* when $T$ approaches $T_c = 0$ K. In Zone 2 ($x_2\Lambda \sim 1$, $x_2 \ll 1$), we find:

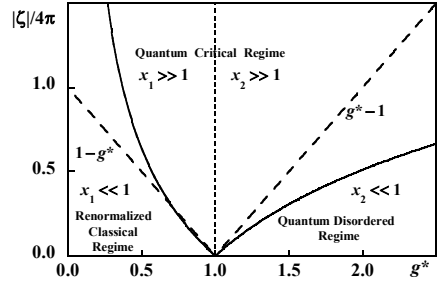

**Fig. 9** Regimes characterizing each magnetic phase ($g^* = T/T_c$).

$$\tilde{\xi} \approx \frac{\hbar c}{\Delta}, \; x_2 \ll 1 \; (\text{Zone 2, QDR}). \tag{2.87}$$

We deal with the *Quantum Disordered Regime* (QDR) characterized by $\Delta \gg k_B T$. Owing to (2.75) we have $\Delta = k_B T/x_2$ so that $\tilde{\xi} \approx L_\tau x_2 \ll L_\tau$ as $x_2 \ll 1$. Equivalently, due to (2.72) we have $\xi \approx 2ax_2 \ll 2a$: we then deal with a *short-range order*. In Zones 3 ($x_1 \gg 1$) and 4 ($x_2 \gg 1$) we finally show:

$$\tilde{\xi} \approx C^{-1} \frac{\hbar c}{k_B T}\left(1 + \frac{2}{C\sqrt{5}x_1}\right), \; x_1 \gg 1 \; (\text{Zone 3, QCR}), \tag{2.88}$$

$$\tilde{\xi} \approx C^{-1} \frac{\hbar c}{k_B T}\left(1 - \frac{1}{C\sqrt{5}x_2}\right), \; x_2 \gg 1 \; (\text{Zone 4, QCR}) \tag{2.89}$$

where we have set:

$$C = 2\ln\left(\frac{1+\sqrt{5}}{2}\right) = 0.962 \; 424, \; C^{-1} = 1.039 \; 043. \tag{2.90}$$

We now deal with the *Quantum Critical Regime* (QCR). At the frontier between Zones 3 ($\rho_s \ll k_B T$) and 4 ($\Delta \ll k_B T$) i.e., along the vertical line reaching the Néel line at $T_c$, $x_1$ and $x_2$ become infinite so that:

$$\tilde{\xi}_c \approx C^{-1} \frac{\hbar c}{k_B T}, \; T = T_c \tag{2.91}$$

i.e., $\tilde{\xi} \approx L_\tau$ as $C^{-1}$ is close to unity, as predicted by the renormalization group analysis [4,5]. As a result the corresponding critical exponent is:

$$\nu = 1. \tag{2.92}$$

As for Zone 1, when $T$ approaches $T_c = 0$ K, the divergence of $\xi$ in Zones 3 and 4 as well when reaching the Néel line, characterizes a long-range order. Thus, in the low-temperature limit, if studying the convergence ratio of the $l$-series giving the zero-field partition function $Z_N(0)$ valid for any temperature (cf (2.51)), we retrieve all the exact results derived by a renormalization technique, exclusively valid near $T_c = 0$ K, for the correlation length [4,5]. These results are gathered in Fig. 9.

In summary this low-temperature study has allowed one to characterize each magnetic phase by its specific regime. From a physical point of view, in Zone 3 ($x_1 \gg 1$), the energy scale $k_B T$ is the largest energy which kills the critical spin fluctuations i.e, it represents the upper limit above which the magnetic system becomes paramagnetic. The ground state is ordered: the spin fluctuations are quantum critical at the shortest scales lower than $\xi$ and are progressively quenched by an increasing temperature. For small $x_1$ the antiferromagnet (respectively, the ferromagnet) is in the renormalized-classical region labeled Zone 1 ($x_1 \ll 1$). The system behaves as if it has long-range order. In other words the lattice is composed of independent blocks of length $\xi$. When thermal fluctuations become stronger and stronger the magnetic order is finally destroyed. In Zone 4 ($x_2 \gg 1$) the temperature predominates the small zero-temperature gap $\Delta$. The system is ordered and behaves as in Zone 3. In Zone 2 ($x_2 \ll 1$) the ground state gap $\Delta$ quenches the spin fluctuations, thus putting the system in the quantum disordered region.

## 3 Free Energy Density

### 3.1 Definitions

The free energy density for the two-dimensional lattice is defined as

$$\mathcal{F}(T) = -\frac{k_B T}{S} \ln Z_N(0) \tag{3.1}$$

where $Z_N(0)$ is the zero-field partition function given by (2.51) and $S$ the lattice surface, in the thermodynamic limit. In the vicinity of the critical point we have seen that the lattice is composed of independent blocks of spins. When $J = J_1 = J_2$ we deal with square blocks of side length $\xi$ where $\xi$ is the correlation length (cf (2.91)). Thus we have $S = \xi^2$ i.e., $S = C^{-2}(\hbar c/k_B T)^2$ where $C$ is given by (2.90). The free energy density can be rewritten

$$\mathcal{F}(T) = -\hbar c C^{-1} \xi^{-D} \ln Z_N(0), \; D = 3. \tag{3.2}$$

As a result, at $T = 0$ K, $\mathcal{F}(T)$ has the form $\hbar c \Upsilon \xi^{-D}$ where $\Upsilon$ is an universal number defined by

$$\Upsilon = C^{-1} = 1.039\;043 \tag{3.3}$$

so that the hyperscaling hypothesis is verified [25]. Finally $\xi^D$ represents the volume of the slab of space time which is infinite in $D - 1$ dimensions but of finite length $L_\tau$ in the remaining direction along the $i\tau$-axis. In our case $D = 3$.

## 3.2 Low Temperature Behaviors

Starting from the expression of the free energy density in the vicinity of the critical point (*cf* (3.2)) we define the free energy density per lattice bond:

$$\frac{\mathcal{F}(T)}{8N^2} = -\frac{\hbar c C^2}{8N^2}\left(\frac{k_B T}{\hbar c}\right)^3 \ln Z_N(0) \tag{3.4}$$

where $Z_N(0)$ is the zero-field partition function. In the thermodynamic limit, if $J = J_1 = J_2$, (2.43) and (2.51) reduce to

$$Z_N(0) \approx (4\pi)^{8N^2} \sum_{l=0}^{+\infty} \left[F(l,l,0)\lambda_l(-\beta J)^2\right]^{4N^2}, \text{ as } N \to +\infty. \tag{3.5}$$

When *T* is cooling down we have previously seen that the predominance of each eigenvalue changes. Let $\lambda_{l_i}(-\beta J)$ be the dominant eigenvalue in the range of temperature $[T_{i,<}, T_{i,>}]$. In this range the *l*-summation giving $Z_N(0)$ reduces to a single term $F(l,l,0)\lambda_l(-\beta J)^2$ so that we have $\ln Z_N(0) \approx \ln(F(l,l,0)\lambda_l(-\beta J)^2)$. Near the critical point $T_c = 0$ K all the eigenvalues have a close thermal behavior and the dominant ones are characterized by an index $l \gg 1$. In addition the argument $\beta|J|$ can be replaced by $|\tilde{z}|\Lambda$ (*cf* (2.63)) with $\Lambda = 2l \gg 1$. Thus, in the low-temperature limit, for any lattice size and notably in the thermodynamic limit ($N \to +\infty$), we always have $F(l_i,l_i,0) \ll |\lambda_{l_i}(-\beta J)|$ so that the contribution of $F(l_i,l_i,0)$ to $\mathcal{F}(T)$ can be neglected. In addition it is cancelled when calculating $\mathcal{F}(T) - \mathcal{F}(0)$. This property remains valid for decreasing *l*-values. As a result we have per lattice bond

$$\frac{1}{8N^2}\ln Z_N(0) \approx \ln\left|\lambda_{l_i}(-\beta J)\right|, T_i \in [T_{i,<}, T_{i,>}], \text{ as } N \to +\infty. \tag{3.6}$$

For a given temperature *T*, each eigenvalue $\lambda_l(-\beta J)$ is dominant within a range $\delta T_{l_i} = T_{l_i,>} - T_{l_i,<}$ with $T = \sum_{i=0}^{i_{max}}(T_{l_i,>} - T_{l_i,<})$, $T_{l_0,<} = 0$, $T_{l_{i-1},>} = T_{l_i,<}$ ($i \neq 0$) and $T_{l_{i_{max}},>} = T$ or equivalently $\sum_{i=0}^{+\infty}\delta T_{l_i}/T = 1$. Thus $\delta T_{l_i}/T$ appears as the corresponding weight of the eigenvalue $\lambda_{l_i}(-\beta J)$ at temperature *T*. As a result the contribution per bond to $Z_N(0)$ is $\delta Z_N(0)_{l_i} = \lambda_{l_i}(\beta|J|)\delta T_{l_i}/T$, as $Z_N(0)$ is an even function of $\beta J$. Similarly we can define the elementary contribution to $\ln Z_N(0)$ by $\delta \ln Z_N(0)_{l_i} = \ln(\lambda_{l_i}(\beta|J|))\delta T_{l_i}/T$.

While approaching $T_c = 0$ K the ranges $T_{l_i,>} - T_{l_i,<}$ are tighter and tighter and the corresponding weight is very small. We tend towards a continuum: there is an increasing number of eigenvalues characterized by a higher and higher index *l* (replaced by $\Lambda$ near $T_c = 0$ K). In other words we deal with an infinity of eigenvalues showing an infinite argument so that their total contribution dominates the ones characterized by a lower value of *l* in spite of the fact that they have a larger weight (see Fig. 5). Under these conditions, if $\lambda_\Lambda(|\tilde{z}|\Lambda)$ is the dominant eigenvalue in the range $\delta T_\Lambda = T_{\Lambda,>} - T_{\Lambda,<}$ and $\delta T_\Lambda/T$ the corresponding weight at temperature $T$, $\delta Z_N(0)_\Lambda = \lambda_\Lambda(|\tilde{z}|\Lambda)\delta T_\Lambda/T$ is the $\Lambda$-contribution per bond to $Z_N(0)$. Similarly the $\Lambda$-contribution per bond to $\ln Z_N(0)$ is $\delta \ln Z_N(0)_\Lambda = \ln(\lambda_\Lambda(|\tilde{z}|\Lambda))\delta T_\Lambda/T$. As the dominant eigenvalue changes between two consecutive ranges $\delta T_\Lambda$ and $\delta T_{\Lambda+1}$, with $\delta T_\Lambda \approx \delta T_{\Lambda+1}$, if $\Lambda \to +\infty$, we can formally write for a given temperature $T$ close to $T_c = 0$ K

$$\delta \ln(Z_N(0)) \approx \lim_{M \to +\infty} \sum_{\Lambda=0}^{M} \ln(\lambda_\Lambda(|\tilde{z}|\Lambda))\frac{\delta T_\Lambda}{T}, \quad \lim_{M \to +\infty} \sum_{\Lambda=0}^{M} \frac{\delta T_\Lambda}{T} = 1 \qquad (3.7)$$

where $\delta T_\Lambda = T_{\Lambda,>} - T_{\Lambda,<}$ is the temperature range for which the eigenvalue $\lambda_\Lambda(|\tilde{z}|\Lambda)$ given by

$$\lambda_\Lambda(|\tilde{z}|\Lambda) \approx \frac{1}{|\tilde{z}|\Lambda}\left(\frac{|\tilde{z}|}{\sqrt{1+\tilde{z}^2}}\right)^{1/2}\left\{\exp(|\tilde{\zeta}|\Lambda)\sum_{s=0}^{+\infty}\frac{U_s}{(\Lambda)^s} + \exp(-|\tilde{\zeta}|\Lambda)\sum_{s=0}^{+\infty}\frac{U_s}{(-\Lambda)^s}\right\}, \text{ as } \Lambda \to +\infty \quad (3.8)$$

is dominant. The coefficients $U_s$ are defined by (6.12) and $|\tilde{\zeta}|\Lambda$ is given by (8.7).

Thus $\delta \ln Z_N(0)$ appears as a continuous suite of different continuous convergent functions $\ln \lambda_\Lambda(|\tilde{z}|\Lambda)$, each of them being exclusively valid in the range $\delta T_\Lambda$, with all the $\delta T_\Lambda$'s of unequal widths. As a result we are dealing with a *Riemann sum*.

In the thermodynamic limit, the elementary free energy density per lattice bond can be expressed as

$$\frac{\delta \mathcal{F}(T)}{8N^2} = -\hbar c C^2 \left(\frac{k_B T}{\hbar c}\right)^3 \lim_{M \to +\infty} \sum_{\Lambda=0}^{M} \ln(\lambda_\Lambda(|\tilde{z}|\Lambda))\frac{\delta T_\Lambda}{T}, \text{ as } N \to +\infty \qquad (3.9)$$

where the dimensionless weight $\delta T_\Lambda/T$ must be linked to a surface element that we identify in Appendix 5 (*cf* (9.27)). In this appendix we thoroughly explain how to transform this discrete sum as a *Kurzweil-Henstock integral*, in the infinite $M$-limit (*cf* (9.20)). Its value essentially comes from the larger and larger value of $\Lambda$. In other words it means that the integral must be calculated owing to a steepest descent method. As a result, $\delta \mathcal{F}(T)$ is evaluated near its extremum (i.e., at the saddle point). Thus, due to the constraint $S^2 = 1$ we are led to introduce a Lagrange multiplier field $\gamma$ (see Appendix 5) [27]. The search of the integral extremum allows one to show

$$\|\gamma\| = \min\left(\frac{|\tilde{\zeta}|\Lambda}{L_\tau}\right), \quad \frac{|\tilde{\zeta}|\Lambda}{L_\tau} = \frac{m_L c^2}{\hbar c}, \text{ as } T \to 0. \qquad (3.10)$$

As $(|\tilde{\zeta}|\Lambda)_{min}$ is given by the common limit of (8.19) and (8.20) i.e., (8.21) we have

$$\|\gamma\| = \frac{2\ln\alpha}{L_\tau}, \quad \|\gamma\| = \frac{m_0 c^2}{\hbar c} \tag{3.11}$$

where $\alpha = (1+\sqrt{5})/2$ is the golden mean and $C = \ln\alpha$ is given by (2.90). This is the saddle-point value of the auxiliary field $\gamma$. In the infinite volume system ($L_\tau \to +\infty$ i.e., at $T = 0$ K rigorously) $\|\gamma\|_\infty = 0$ i.e., $m_0 = 0$. Finally, in Appendix 5, we show that, near the saddle point, the free energy density per lattice bond can be written in the thermodynamic limit as

$$\frac{\mathcal{F}(T)}{8N^2} = \hbar c C \left[ \frac{1}{L_\tau} \int \frac{d^2 k}{4\pi^2} \ln(1-\exp(-u_k)) + \frac{1}{2\hbar} \int \frac{d^3 p}{8\pi^3} \ln\left( \frac{p^2 + (m_0 c)^2}{\hbar^2 \Lambda^{-2}} \right) - \frac{(m_0 c/\hbar)^2}{2g} \right],$$

$$\text{as } N \to +\infty. \tag{3.12}$$

with

$$\frac{d^3 p}{(2\pi)^3 \hbar} = \frac{d^2 k}{(2\pi)^2} \frac{d(p/\hbar)}{2\pi}, \quad u_k = L_\tau\sqrt{k^2+\gamma^2}, \quad u_k = \frac{\varepsilon_k}{k_B T}, \quad \varepsilon_k = \sqrt{(pc)^2 + (m_0 c^2)^2}, \tag{3.13}$$

where $\varepsilon_k$ is the relativistic energy and $\boldsymbol{p}$ the relativistic momentum. In the previous equation we can now introduce the critical coupling $g_c$ for $D = 3$

$$\frac{1}{g_c} = \frac{1}{\hbar} \int \frac{d^3 p}{8\pi^3} \frac{1}{p^2}. \tag{3.14}$$

At this step we can note that this result can be generalized to a $D$-space time. In that case $g_c$ is given by:

$$\frac{1}{g_c} = \frac{1}{\hbar} \int \frac{d^D p}{(2\pi)^D} \frac{1}{p^2} \tag{3.15}$$

with $d^D p = S_D\, p^{D-1} dp$, $D = d + 1$, $S_D = 2\pi^{D/2}/\Gamma(D/2)$ so that finally, if integrating (3.15) we retrieve the result of Chakravarty *et al.*[4] (expressed in $\Lambda$-unit)

$$\widetilde{g}_c(d) = \frac{2(d-1)(2\pi)^d}{S_d} \Lambda^{1-d}, \tag{3.16}$$

thus finding *a posteriori* the geometrical factor that we suspected. If expressing the ratio $S_{d+1}/S_d$ i.e., $S_D/S_d = 2^{D-2}(D-1)\Gamma((D-1)/2)^2/\Gamma(D)$ we can obtain the surface of space time $S_D$ swept by the relativistic momentum $\boldsymbol{p}$ vs $S_d$. $S_D/(2\pi)^D$ represents the integration measure $K_D$. In other words, in the $\Lambda$-scale, $\widetilde{g}_c(D)$ is expressed per unit of swept surface of space time

$$\widetilde{g}_c(D) = A(D) \frac{(2\pi)^D}{S_D} \Lambda^{2-D}, \quad A(D) = \frac{2^{D-2}(D-1)(D-2)\Gamma\left(\frac{D-1}{2}\right)^2}{\pi\Gamma(D)}. \tag{3.17}$$

The second integral in (3.12) is badly divergent when $p$ becomes infinite (ultraviolet domain). However all divergences disappear if introducing the infinite volume contribution $\mathcal{F}_\infty(T) = \mathcal{F}(0)$ for which $m_0 = 0$ ($L_\tau \to +\infty$). As a result we have:

$$\frac{\mathcal{F}(T)-\mathcal{F}(0)}{8N^2} = \hbar c C \left[ \frac{1}{L_\tau} \int \frac{d^2k}{4\pi^2} \ln(1-\exp(-u_k)) + \frac{1}{2\hbar} \int \frac{d^3p}{8\pi^3} \ln\left(\frac{p^2+(m_0 c/\hbar)^2}{p^2}\right) \right.$$
$$\left. - \frac{(m_0 c/\hbar)^2}{2g} \right], \text{ as } N \to +\infty. \quad (3.18)$$

This is the exact relationship that Chubukov *et al.* have derived from a renormalization approach in the vicinity of the critical point [5]. Consequently *it brings an important verification to our result derived from $Z_N(0)$ whose expression is valid for any temperature and considered in the low-temperature limit*.

If $g < g_c$, owing to the definition of the spin stiffness $\rho_s$ given by (2.68) and after a first integration in (3.18), we get:

$$\frac{\mathcal{F}(T)-\mathcal{F}(0)}{8N^2} = \hbar c C \left[ \frac{1}{L_\tau} \int_{X_1(x_1)}^{+\infty} \frac{\varepsilon d\varepsilon}{2\pi} \ln(1-\exp(-\varepsilon)) + \frac{1}{2\hbar} \int \frac{d^3p}{8\pi^3} \left\{ \ln\left(\frac{p^2+(m_0 c/\hbar)^2}{p^2}\right) \right. \right.$$
$$\left. \left. - \frac{(m_0 c/\hbar)^2}{p^2} \right\} - \frac{(m_0 c/\hbar)^2}{2} \frac{\rho_s}{k_B T} \right]. \quad (3.19)$$

The first integral can be expressed owing to polylogarithms (or Jonquière functions) [28]:

$$\operatorname{Li}_s(z) = \sum_{k=1}^{+\infty} \frac{z^k}{k^s}, \quad (3.20)$$

and the Riemann zeta function $\zeta(s) = \operatorname{Li}_s(1)$ ($s > 1$). We have:

$$\frac{\mathcal{F}(T)-\mathcal{F}(0)}{8N^2} = -\frac{\hbar c C}{2\pi}\left(\frac{k_B T}{\hbar c}\right)^3 \left[ X_1(x_1)\operatorname{Li}_2\!\left(e^{-X_1(x_1)}\right) + \operatorname{Li}_3\!\left(e^{-X_1(x_1)}\right) \right.$$
$$\left. + \frac{X_1^3(x_1)}{6} + \frac{X_1^2(x_1)}{2x_1} + \ldots \right], \text{ (Zones 1 and 3, } T<T_c\text{)}, \quad (3.21)$$

with

$$X_1(x_1) = \exp(-1/x_1), \quad x_1 \ll 1 \text{ (Zone 1, } T<T_c\text{)},$$
$$X_1(x_1) = 2\ln\!\left(\frac{\sqrt{5}+1}{2}\right) - \frac{2}{\sqrt{5}\,x_1}, \quad x_1 \gg 1 \text{ (Zone 3, } T<T_c\text{)}. \quad (3.22)$$

As a result, if $x_1 \ll 1$ (Zone 1, $T < T_c$), (3.21) becomes:

$$\frac{\mathcal{F}(T)-\mathcal{F}(0)}{8N^2} = -\frac{\hbar c C}{2\pi}\left(\frac{k_B T}{\hbar c}\right)^3\left[\zeta(3)+\zeta(2)e^{-1/x_1}+\frac{e^{-2/x_1}}{2x_1}+\frac{e^{-3/x_1}}{6}+\ldots\right], x_1 \ll 1$$

$$\text{(Zone 1, } T < T_c\text{).} \quad (3.23)$$

When $x_1 \gg 1$ (Zone 3, $T < T_c$) the argument of Jonquière function can be expanded vs $1/x_1 \ll 1$ as well as the function itself. We find:

$$\frac{\mathcal{F}(T)-\mathcal{F}(0)}{8N^2} = -\frac{\hbar c C}{2\pi}\left(\frac{k_B T}{\hbar c}\right)^3\left[C_0+\frac{C_1}{x_1}+\frac{C_2}{x_1^2}+\frac{C_3}{x_1^3}+\ldots\right], x_1 \gg 1 \text{ (Zone 3, } T < T_c\text{),} \quad (3.24)$$

with:

$$C_0 = C\mathrm{Li}_2\left(e^{-C}\right)+\mathrm{Li}_3\left(e^{-C}\right)+\frac{C^3}{6},\ C_0 = 0.961\ 646,$$

$$C_1 = C\left[C'\mathrm{Li}_1\left(e^{-C}\right)+\frac{C}{2}(1-C')\right],\ C_1 = 0.463\ 130,$$

$$C_2 = \frac{C'}{2}\left[CC'\mathrm{Li}_0\left(e^{-C}\right)-C'\mathrm{Li}_1\left(e^{-C}\right)+C(C'-2)\right],\ C_2 = -0.430\ 409,$$

$$C_3 = \frac{C'^2}{6}\left[CC'\mathrm{Li}_{-1}\left(e^{-C}\right)-2C'\mathrm{Li}_0\left(e^{-C}\right)-C'+3\right],\ C_3 = 0.248\ 109,$$

$$C = 2\ln\left(\frac{1+\sqrt{5}}{2}\right), C = 0.962\ 424,\ C' = \frac{2}{\sqrt{5}}, C' = 0.894\ 427. \quad (3.25)$$

If $g > g_c$, the following relation between the $T = 0$ gap, $\Delta$, and the coupling $g$ is useful:

$$\frac{1}{g} = \frac{1}{\hbar}\int\frac{d^3 p}{8\pi^3}\frac{1}{p^2+(\Delta/\hbar c)^2} \quad (3.26)$$

As a result and after a first integration in (3.18), we get:

$$\frac{\mathcal{F}(T)-\mathcal{F}(0)}{8N^2} = \hbar c C\left(\frac{k_B T}{\hbar c}\right)^3\left[\int_{X_2(x_2)}^{+\infty}\frac{\varepsilon d\varepsilon}{2\pi}\ln(1-\exp(-\varepsilon))\right.$$

$$\left.+\frac{1}{2\hbar}\int\frac{d^3 p}{8\pi^3}\left\{\ln\left(\frac{p^2+(m_0 c/\hbar)^2}{p^2+(\Delta/\hbar c)^2}\right)-\frac{(m_0 c/\hbar)^2-(\Delta/\hbar c)^2}{p^2+(\Delta/\hbar c)^2}\right\}\right]. \quad (3.27)$$

In that case too this is the exact relation that Chubukov *et al.* have derived from a renormalization approach in the vicinity of the critical point [5]. As in the case $g < g_c$, if expressing the first integral with polylogarithms, we get:

$$\frac{\mathcal{F}(T)-\mathcal{F}(0)}{8N^2} = -\frac{\hbar c C}{2\pi}\left(\frac{k_B T}{\hbar c}\right)^3 \left[X_2(x_2)\mathrm{Li}_2\left(e^{-X_2(x_2)}\right)+\mathrm{Li}_3\left(e^{-X_2(x_2)}\right)\right.$$
$$\left. + \frac{X_2^3(x_2)}{6} - \frac{X_2^2(x_2)}{4x_2} + \frac{1}{12x_2^3} + ...\right] \quad \text{(Zones 2 and 4, } T > T_c\text{)} \quad (3.28)$$

with

$$X_2(x_2) \approx \frac{1}{x_2} + 2\exp(-1/x_2),\ x_2 \ll 1 \text{ (Zone 2, } T > T_c\text{)},$$

$$X_2(x_2) \approx 2\ln\left(\frac{1+\sqrt{5}}{2}\right) + \frac{1}{\sqrt{5}x_2},\ x_2 \gg 1 \text{ (Zone 4)}. \quad (3.29)$$

When $x_2 \ll 1$ (Zone 2, $T > T_c$) (3.28) becomes:

$$\frac{\mathcal{F}(T)-\mathcal{F}(0)}{8N^2} = -\frac{\hbar c C}{2\pi}\left(\frac{k_B T}{\hbar c}\right)^3\left[\left(1+\frac{1}{x_2}\right)e^{-1/x_2} + \frac{1}{4}\left(\frac{17}{2}+\frac{5}{x_2}\right)e^{-2/x_2}\right.$$
$$\left. + \frac{1}{9}\left(\frac{101}{6}+\frac{1}{x_2}\right)e^{-3/x_2} + ...\right],\ x_2 \ll 1 \text{ (Zone 2, } T > T_c\text{)}. \quad (3.30)$$

When $x_2 \gg 1$ (Zone 4, $T > T_c$) the argument of Jonquière function can be expanded *vs* $1/x_2 \ll 1$ as well as the function itself. We find:

$$\frac{\mathcal{F}(T)-\mathcal{F}(0)}{8N^2} = -\frac{\hbar c C}{2\pi}\left(\frac{k_B T}{\hbar c}\right)^3\left[C_0' + \frac{C_1'}{x_2} + \frac{C_2'}{x_2^2} + \frac{C_3'}{x_2^3} + ...\right], x_2 \gg 1 \text{ (Zone 4, } T > T_c\text{)}, \quad (3.31)$$

with:

$$C_0' = C_0, C_0' = 0.961\ 646, C_1' = -\frac{C_1}{2}, C_1' = -0.231\ 565,$$

$$C_2' = \frac{C_2}{4}, C_2' = -0.107\ 602, C_3' = -\frac{C_3}{8} + \frac{1}{12}, C_3' = 0.052\ 320 \quad (3.32)$$

where $C_0$, $C_1$, $C_2$ and $C_3$ are given by (3.25). If $x_1 \ll 1$ (Zone 1, $T < T_c$) the leader term of the free energy density is

$$\frac{\mathcal{F}(T)-\mathcal{F}_\infty(T)}{8N^2} = -\frac{\hbar c C}{2\pi}\left(\frac{k_B T}{\hbar c}\right)^3 \zeta(3) + ...,\ x_1 \ll 1 \text{ (Zone 1, } T < T_c\text{)}, \quad (3.33)$$

with $\mathcal{F}_\infty(T) = \mathcal{F}(0)$. As a result, coming simultaneously from Zone 3 ($T < T_c$, $x_1 \gg 1$) and Zone 4 ($T > T_c$, $x_2 \gg 1$) towards the common vertical frontier directly leading to the Néel line ($T = T_c$) for which $x_1 \approx x_2 \to +\infty$, the common leader term of the free energy density can be written as:

$$\frac{\mathcal{F}(T) - \mathcal{F}_\infty(T)}{8N^2} = -\hbar c C \frac{\zeta(3)}{2\pi}\left(\frac{k_B T}{\hbar c}\right)^3 \widetilde{C} + ... \quad (3.34)$$

with

$$\widetilde{C} = \frac{C_0}{\zeta(3)} = \frac{4}{5}. \quad (3.35)$$

We retrieve the complete result of Fisher and de Gennes which states that, if hyperscaling is valid, the free energy density satisfies [25]

$$\frac{\mathcal{F}(T) - \mathcal{F}_\infty(T)}{8N^2} = -\hbar c C \frac{\Gamma(D/2)\zeta(D)}{\pi^{D/2}} \frac{\widetilde{C}}{L_\tau^D} + ... \quad (3.36)$$

with here $D = 3$ and $L_\tau = \hbar c / k_B T$. In addition we retrieve the exact result of Sachdev for the universal number $\widetilde{C}$ [26]. All these results will be interpreted in a forthcoming paper while examining the low-temperature behaviors of the specific heat.

## 4 Conclusion

We here conclude this article by recalling the main highlights of our results. We have developed a general method for deriving the characteristic polynomial associated with the zero-field partition function $Z_N(0)$ of a finite lattice showing edges or wrapped on a torus. In a first step we have reconsidered the spin space by defining a new decorated lattice. It is similar to the crystallographic one characterized by a square unit cell and composed of two sublattices. Each bond of the first square sublattice is characterized by an integer $l$ or $l'$ and the whole square sublattice by the set (…, $l_{i,j}$,…, $l'_{i,j}$, …). Similarly each bond of the second square sublattice is described by a relative integer $m$ or $m'$ (with $m \in [-l, +l]$, respectively $m' \in [-l', +l']$) and the whole square sublattice by the set (…, $m_{i,j}$,…, $m'_{i,j}$,…). Under these conditions the aim of the study has finally consisted in examining the effects of sublattice symmetries over coefficients $m_{i,j}$ and $m'_{i,j}$ (respectively, coefficients $l_{i,j}$ and $l'_{i,j}$) for obtaining a univocal set of selection rules for coefficients $l$ and $m$.

For doing this work we have used a *duality principle* and we have defined the conditions of validity which must be fulfilled, thus leading to list the nature of the allowed spin couplings and the structure of corresponding spin arrangements. Finally we have shown that, in the thermodynamic limit, this principle becomes irrelevant, thus allowing to reconsider all the discarded cases when applying the duality principle to finite lattices. We conclude that: i) *only the 2d classical Heisenberg model is integrable in the case of finite lattices*; ii) *all the 2d classical spin systems are fully integrable in the case of infinite lattices*. More particularly, it means that the 2d classical *xy* model, which is nonintegrable on a finite lattice, becomes integrable in the thermodynamic limit.

In case *a* (lattice showing edges) the application of duality principle has led to the following selections rules: i) the horizontal lines of ranks $+i$ and $-i$, with $-N \leq i \leq N$ (respectively the vertical lines of rank $+j$ and $-j$, with $-N \leq j \leq N$) are constituted of bonds characterized by the same couple of values $(l_i, m_i)$ or $(l'_j, m'_j)$ but two consecutive lines do not show the same couple $(l_i, m_i)$ (respectively, $(l'_j, m'_j)$) ; ii) all the lines are characterized by $m = 0$, independently of the lattice size. In case *b* (lattice wrapped on a torus) all the lines are characterized by the same coefficient $l$ (respectively $m$ but $m = 0$ or $m \neq 0$).

In the thermodynamic limit the value $m = 0$ is numerically selected for the torus characterized by two infinite radii of curvature (case *b*) as for the lattice showing edges (case *a*) whereas all the *l*'s are equal to a common value $l_0$. *This result brings a strong validation to the use of duality principle*: in case *a* we have shown that $m = 0$ owing to this principle; as the reasoning can be extended to infinite lattices, we also have $m = 0$. We retrieved this result by a separate numerical study in case *b*.

As a result a very simple closed-form expression can be obtained for the characteristic polynomial giving the zero-field partition function $Z_N(0)$ *valid for any temperature T*.

We have shown that, if $J = J_1 = J_2$, it appears *crossovers* between two consecutive terms of the characteristic polynomial. Coming from high temperatures where the eigenvalue characterized by $l = 0$ is dominant, *l*-eigenvalues, with increasing $l > 0$, are successively dominant when $T$ is cooling down. In the vicinity of absolute zero the dominant ones are characterized by $l \to +\infty$. As they are composed of Bessel functions they show a close behavior when $T \to 0$ so that all the *l*-eigenvalues become equivalent, confirming the fact that the critical temperature is $T_c = 0$ K.

From this low-temperature study we obtain a diagram exhibiting three magnetic phases: it is exactly similar to the one derived from the renormalization group technique [4,5].Then, owing to the Ornstein-Zernike theorem we obtain three low-temperature behaviors of the correlation length i.e., one per magnetic phase of the diagram, in agreement with previous results obtained from a renormalization technique [5]. At $T_c = 0$ K the critical exponent is $\nu = 1$, as expected. As a result we have identified three regimes: the Renormalized Classical Regime (RCR), the Quantum Disordered Regime (QDR) and the Quantum Critical Regime (QCR).

Finally, from the exact closed-form expression of $Z_N(0)$, we have expressed the free energy density $\mathcal{F}$. Near $T_c = 0$ K and for each magnetic phase, we have respectively obtained the same expression than the corresponding one derived from the renormalization group approach [5]. These results will be interpreted in a forthcoming paper while examining the low-temperature behaviors of the specific heat. Closed-form expressions valid for any temperature will be also derived for the spin-spin correlations, the correlation length and the susceptibility. In the vicinity of the critical point we shall retrieve the respective expressions obtained from a renormalization technique [5].

**Acknowledgments** I particularly thank Janis Kliava (LOMA, University of Bordeaux) for fruitful discussions and his constant support during the writing of the present article.

## Appendix 1: Expression of the Zero-Field Partition Function of a Square Lattice in the Thermodynamic Limit

When $T$ approaches $T_c = 0$ K all the eigenvalues are very close due to the fact that the crossover temperatures are closer and closer. This characterizes the transition at the critical point $T_c$. A special study for $T \approx 0$ K is made in Appendix 5. For $T > 0$ K, between two consecutive crossover temperatures $T_{i,<}$ and $T_{i,>}$, we must show in a first step that, in the thermodynamic limit, $Z_N^{\text{SL}}(0) \approx Z_N^{\text{Torus}}(0)$ with

$$Z_N^{\text{Torus}}(0) \approx (4\pi)^{8N^2} [u_{\max}(T)]^{4N^2}, \quad T \in [T_{l_i,<}, T_{l_i,>}],$$

$$u_{\max}(T) = F(l_i, l_i, 0)\lambda_{l_i}(-\beta J_1)\lambda_{l_i}(-\beta J_2), \text{ as } N \to +\infty, \quad (5.1)$$

if $\lambda_{l_i}(-\beta J_k)$, $k = 1, 2$, is the dominant eigenvalue characterizing the range $[T_{l_i,<}, T_{l_i,>}]$, with $l_{\max} = l_i$; the integral $F(l_i, l_j, 0)$ is given by (2.41) in which $m = 0$. In addition we have $T = \sum_{i=0}^{i_{\max}} (T_{l_i,>} - T_{l_i,<})$, $T_{l_0,<} = 0$, $T_{l_{i-1},>} = T_{l_i,<}$ ($i \neq 0$) and $T_{l_{i_{\max}},>} = T$. But, if we wish to consider a closed-form expression of the zero-field partition function $Z_N^{\text{Torus}}(0)$ valid in the whole range $[0+\varepsilon, +\infty[$, with $\varepsilon \to 0$, we must show that all the terms of the $l$-series given by (2.42) must be kept. If neglecting the edge contributions in the thermodynamic limit ($N \to +\infty$) the zero-field partition function of the finite square lattice can be more generally rewritten as:

$$Z_N^{\text{SL}}(0) = (4\pi)^{8N^2} \prod_{i=-(N-1)}^{N-1} \prod_{j=-(N-1)}^{N-1} \sum_{l_i=0}^{+\infty} \sum_{l_j=0}^{+\infty} u_{l_i,l_j}(T) \quad (5.2)$$

with:

$$u_{l_i,l_j}(T) = F(l_i, l_j, 0)\lambda_{l_i}(-\beta J_1)\lambda_{l_j}(-\beta J_2), \quad l_i = l_j \text{ or } l_i \neq l_j, \quad T \in [T_{l_i,<}, T_{l_i,>}]. \quad (5.3)$$

Under these conditions we must finally show that $Z_N^{\text{SL}}(0) \approx Z_N^{\text{Torus}}(0)$ if $N \to +\infty$, for any $T$.

A numerical study leads to a classification of terms $u_{l_i,l_j}(T)$, with $l_i \neq l_j$. If factorizing the major term $u_{\max}(T)$ of the polynomial expression giving $Z_N(0)$ (5.2) becomes when $N \to +\infty$:

$$Z_N^{\text{SL}}(0) = (4\pi)^{8N^2} [u_{\max}(T)]^{4N^2} \{1 + S_1(N,T) + S_2(N,T)\}, \quad T \in [T_{l_i,<}, T_{l_i,>}] \quad (5.4)$$

with:

$$S_1(N,T) = \sum_{\substack{l=0, \\ l \neq l_{\max}}}^{+\infty} \left[\frac{u_l(T)}{u_{\max}(T)}\right]^{4N^2}, \quad S_2(N,T) = \prod_{i=-(N-1)}^{N-1} \prod_{j=-(N-1)}^{N-1} \sum_{l_i=0}^{+\infty} \sum_{\substack{l_j=0, \\ l_j \neq l_i}}^{+\infty} \frac{u_{l_i,l_j}(T)}{u_{\max}(T)}. \quad (5.5)$$

$S_1(N,T)$ and $S_2(N,T)$ are absolutely convergent series characterized by positive current terms lower than unity. Each term of $S_2(N,T)$ is composed of a product of terms $u_{l_i,l_j}(T)/u_{\max}(T)$ carrying an exponent lower than $4N^2$ but the sum of all these exponents is equal to $4N^2$. Due to the numerical property of $u_{l_i}(T)$ and $u_{l_i,l_j}(T)$, for a given $T \in [T_{l_i,<}, T_{l_i,>}]$, a classification of the various terms of $S_1(N,T)$ and $S_2(N,T)$ can be achieved due to (2.50). As a result the series $S(N,T) = S_1(N,T) + S_2(N,T)$ is also an absolutely convergent series characterized by an infinite sum of positive vanishing current terms symbolically labeled $X_k(N,T)$. These terms can be written in the decreasing modulus order so that

$$S(N,T) = \sum_{k=0}^{+\infty} X_k(N,T), \; 0 < X_k(N,T) < 1, \tag{5.6}$$

with:

$$X_1(N,T) > X_2(N,T) > \ldots > X_\infty(N,T), \quad T \in [T_{l_i,<}, T_{l_i,>}]. \tag{5.7}$$

Now we artificially share the infinite series $S(N,T)$ into two parts:

$$S(N,T) = S^B_{k_i}(N,T) + S^E_{k_i}(N,T), \; T \in [T_{l_i,<}, T_{l_i,>}], \tag{5.8}$$

with:

$$S^B_{k_i}(N,T) = \sum_{k=0}^{k_i} X_k(N,T), \; S^E_{k_i}(N,T) = \sum_{k=k_i}^{+\infty} X_k(N,T) \tag{5.9}$$

where $S^B_{k_i}(N,T)$ and $S^E_{k_i}(N,T)$ are the beginning and the end of $S(N,T)$, respectively. We have the natural inequalities $S^B_{k_i}(N,T) < S(N,T)$ and $S^E_{k_i}(N,T) < S(N,T)$. As we deal with an infinite (absolutely convergent) series made of positive vanishing current terms $X_k(N,T)$ it is always possible to find a particular value $k_i = k_1$ of the general index $k$ such as:

$$S^B_{k_1}(N,T) = S^E_{k_1}(N,T) = \frac{\varepsilon}{2}, \; S(N,T) = \varepsilon, \; 0 < \varepsilon < 1, \; T \in [T_{l_i,<}, T_{l_i,>}]. \tag{5.10}$$

If increasing $N \gg 1$ of $n > 0$ we automatically have

$$S^K_{k_1}(N+n,T) < S^K_{k_1}(N,T) = \frac{\varepsilon}{2}, \; K = B, E, \; T \in [T_{l_i,<}, T_{l_i,>}] \tag{5.11}$$

because the inequality $0 < X_k(N,T) < 1$ imposes $0 < X_k(N+n,T) < X_k(N,T) < 1$. Finally, if calling $S(N+n,T)$ the sum $S^B_{k_1}(N+n,T) + S^E_{k_1}(N+n,T)$ we have

$$S(N+n,T) < S(N,T) = \varepsilon, \; T \in [T_{l_i,<}, T_{l_i,>}]. \tag{5.12}$$

As a result we derive

$$S(N,T) = S_1(N,T) + S_2(N,T) \to 0, \text{ as } N \to +\infty, \text{ for } T \in [T_{l_i,<}, T_{l_i,>}], \tag{5.13}$$

and

$$Z^{SL}_N(0) \approx (4\pi)^{8N^2} [u_{\max}(T)]^{4N^2}, \text{ as } N \to +\infty, \text{ for } T \in [T_{l_i,<}, T_{l_i,>}]. \tag{5.14}$$

This reasoning can be repeated for each new range of temperature $[T_{l_j,<}, T_{l_j,>}]$, with $j \neq i$. Consequently, if summing (5.14) over all the ranges $[T_{l_i,<}, T_{l_i,>}]$ so that $T = \sum_{i=0}^{i_{max}} (T_{l_i,>} - T_{l_i,<})$, $T_{l_0,<} = 0$, $T_{l_{i-1},>} = T_{l_i,<}$ ($i \neq 0$) and $T_{l_{i_{max}},>} = T$ and comparing with (5.4) and (5.5) we always have

$$\sum_{l=0}^{+\infty} u_l(T)^{4N^2} \gg \prod_{i=-(N-1)}^{N-1} \prod_{j=-(N-1)}^{N-1} \sum_{l_i=0}^{+\infty} \sum_{\substack{l_j=0, \\ l_j \neq l_i}}^{+\infty} u_{l_i,l_j}(T), \text{ as } N \to +\infty \qquad (5.15)$$

so that finally

$$Z_N^{SL}(0) \approx (4\pi)^{8N^2} \sum_{l=0}^{+\infty} \left[ F(l,l,0) \lambda_l(-\beta J_1) \lambda_l(-\beta J_2) \right]^{4N^2}, \text{ as } N \to +\infty. \qquad (5.16)$$

In other words this expression is valid for any temperature $T = \sum_{i=0}^{i_{max}} (T_{l_i,>} - T_{l_i,<})$. As $Z_N^{SL}(0) \approx Z_N^{Torus}(0)$ due to the fact that edge contributions are negligible when $N \to +\infty$ (2.43b) and (2.43c) are then proved: both equations are strictly similar.

## Appendix 2: Asymptotic Expansion of Modified Bessel Functions of Large Order, for Any Real Argument

In this appendix we wish to derive the behavior of the modified Bessel function of the first kind of large order $l$ i.e., $I_{l+1/2}(z) \approx I_l(z)$, where $l$ is not necessarily infinite, for any real argument $z$ varying from a finite value to infinity. In other words we generalize a work previous published by Olver [20] but restricted to very large orders and positive arguments. We set $f_l(lz) = \lambda_l(lz)$ i.e., owing to (2.8)

$$f_l(lz) = \left(\frac{\pi}{2lz}\right)^{1/2} I_{l+1/2}(lz), \ z = -\frac{\beta J}{l}. \qquad (6.1)$$

The discussion can be restricted to the study of positive argument owing to:

$$f_l(lz) = \left(-\frac{J}{|J|}\right)^l f_l(l|z|). \qquad (6.2)$$

For mathematical reasons explained in the main text of this article (see Sec. 2.3.2) we also need to know the behavior of the ratio $f_{l+1}(lz)/f_l(lz)$ (respectively, $f_{l-1}(lz)/f_l(lz)$).

Owing to the well-known recurrence relationships between modified Bessel functions we have exactly:

$$\frac{f_{l\pm1}(l|z|)}{f_l(l|z|)} = \mp\left(\frac{1}{|z|} + \frac{1}{2l|z|}\right) + \frac{I'_{l+1/2}(l|z|)}{I_{l+1/2}(l|z|)}, \qquad (6.3)$$

with:

$$I'_{l+1/2}(l|z|) = \frac{dI_{l+1/2}(l|z|)}{d(l|z|)}. \qquad (6.4)$$

Thus, only the behavior of $I_{l+1/2}(|Z|) \approx I_l(|Z|)$ and $I'_{l+1/2}(|Z|) \approx I'_l(|Z|)$, with $|Z| = l|z| = \beta|J|$, must be known for $|Z|$ and $l$ large but not necessarily infinite.

In this respect we here adapt the work of Olver initially achieved for positive values of $Z$ [20]. It is based on the fact that the searched asymptotic expansion of $I_l(Z)$ must tend towards the well-known asymptotic one when $|Z| \to +\infty$. This latter one is easily derived from the Kummer function $M(a, b, Z)$ for $Z$ real, large but not necessarily infinite, so that we have [20]

$$I_{l+1/2}(Z) = \frac{1}{\sqrt{2\pi Z}} \left\{ \exp(Z) \sum_{k=0}^{R-1} \frac{C_k}{k!} (2Z)^{-k} + O(|Z|^{-R}) \right.$$

$$\left. + \exp(i(l+1)\pi) \exp(-Z) \sum_{k=0}^{S-1} \frac{C_k}{k!} (-2Z)^{-k} + O(|Z|^{-S}) \right\},$$

for large real $|Z|$ (6.5)

where $C_k = (l+1)_k(-l)_k$, the $(c)_k$'s being Pochhammer's symbols [21]. The factor $\exp(i(l+1)\pi)$ is associated with the change of $Z$ sign. It must be supressed if $Z > 0$ and when using $|Z|$. When $|Z| \to +\infty$ the second term in the right-hand side of (6.5) disappears and we retrieve the well-known asymptotic expansion of $I_{l+1/2}(Z)$ if expanding the $C_k$'s [21].

Olver has shown that the function $w = |z|^{1/2} I_l(l|z|)$ satisfies the differential equation [20]:

$$\frac{d^2 w}{dz^2} = \left( \frac{1+z^2}{z^2} l^2 - \frac{1}{4z^2} \right) w, \quad w = |z|^{1/2} I_l(l|z|). \qquad (6.6)$$

We briefly summarize his results. Setting

$$\left( \frac{d\zeta}{d|z|} \right)^2 = \frac{1+z^2}{z^2}, \quad W = \left( \frac{d|z|}{d\zeta} \right)^{-1/2} w, \quad W = \left( \frac{1+z^2}{z^2} \right)^{1/4} w \qquad (6.7)$$

then $W$ satisfies the equation

$$\frac{d^2 W}{d\zeta^2} = \left\{ l^2 + f(\zeta) \right\} W \qquad (6.8)$$

where

$$f(\zeta) = -\frac{z'^2}{4z^2} + z'^{1/2} \frac{d^2}{d\zeta^2} \left( z'^{-1/2} \right), \quad z' = \frac{dz}{d\zeta}. \qquad (6.9)$$

As $z'$ must be positive it means that $\zeta$ and $z$ must show the same sign. The first equation in (6.7) can be integrated to obtain

$$\zeta = -\frac{J}{|J|}\left[\sqrt{1+z^2} + \ln\left(\frac{|z|}{1+\sqrt{1+z^2}}\right)\right]. \tag{6.10}$$

The numerical study of $|\zeta|$ is done in the main text (see Fig. 6). Reporting this result in the first part of (6.9) allows one to derive $f(\zeta)$ but its expression is irrelevant here. The solutions of (6.8) are given by the following series

$$W_1(\zeta) \approx \exp(l\zeta)\sum_{s=0}^{+\infty}\frac{U_s(\zeta)}{l^s}, \quad W_2(\zeta) \approx \exp(-l\zeta)\sum_{s=0}^{+\infty}\frac{U_s(\zeta)}{(-l)^s} \tag{6.11}$$

where the coefficients $U_s(\zeta)$ have been expressed by Olver [20]. We just recall below the sequence of the first coefficients $U_s(u)$:

$$U_0(u) = 1, \quad U_1(u) = \frac{u}{8} - \frac{5u^3}{24}, \quad U_2(u) = \frac{9u^2}{128} - \frac{77u^4}{192} + \frac{385u^6}{1152}, \ldots, \tag{6.12}$$

with :

$$u = \frac{1}{\sqrt{1+z^2}}, \quad |z| = \frac{\beta|J|}{l}. \tag{6.13}$$

Because of the uniform property of the expansions given by (6.11), the function $I_l(l|z|)$ can be expressed vs $W_1(|\zeta|)$ and $W_2(|\zeta|)$. We skip intermediate steps so that, when $|Z|$ becoming large, we have:

$$I_l(l|z|) \approx \frac{(1+z^2)^{-1/4}}{\sqrt{2\pi l}}\left\{\exp(l|\zeta|)\sum_{s=0}^{+\infty}\frac{U_s(u)}{l^s} + \exp(-l|\zeta|)\sum_{s=0}^{+\infty}\frac{U_s(u)}{(-l)^s}\right\}. \tag{6.14}$$

The factor $\exp(i(l+1)\pi)$ associated with the change of argument sign (cf (6.5)) has disappeared because the property $I_l(l|z|) = I_{-l}(l|z|)$ must hold for the respective expansions of $I_l(l|z|)$ and $I_{-l}(l|z|)$. But we still have $I_l(-lz) = (\text{sign}(z))^l I_l(lz)$. The derivative $I'_{l+1/2}(l|z|) \approx I'_l(l|z|)$ defined in (6.4), with $l \gg 1$, can directly be obtained by differentiating the previous equation with respect to $|z|$. Using the expression of $d|\zeta|/d|z|$ where $\zeta$ is given by (6.10) we obtain $du/d|\zeta| = -u^2(1-u^2)$ so that finally

$$I'_l(l|z|) \approx \varepsilon\frac{(1+z^2)^{1/4}}{\sqrt{2\pi l}|z|}\left\{\exp(l|\zeta|)\sum_{s=0}^{+\infty}\frac{V_s(u)}{l^s} - \exp(-l|\zeta|)\sum_{k=0}^{+\infty}\frac{V_s(u)}{(-l)^s}\right\}, \quad \varepsilon = \begin{cases}+1 \text{ if } T > T_c \\ -1 \text{ if } T < T_c\end{cases} \tag{6.15}$$

where $\varepsilon$ is the sign of the slope of $d|\zeta|/dT$. The coefficients $V_s(u)$ are polynomials in $u$ given by:

$$V_s = U_s - u(1-u^2)\left(\frac{1}{2}U_{s-1} + u\frac{dU_{s-1}}{du}\right), \quad s \geq 1. \tag{6.16}$$

The first three are:

$$V_0(u) = 1, \quad V_1(u) = -\frac{3u}{8} + \frac{7u^3}{24}, V_2(u) = -\frac{15u^2}{128} + \frac{99u^4}{192} - \frac{455u^6}{1152}, \ldots \quad (6.17)$$

Regarding the parity we have $\Gamma'_l(-lz) = (\text{sign}(z))^{l+1}\Gamma'_l(lz)$. Then, reporting all the previous results allows one to derive both ratios $f_{l+1}(lz)/f_l(lz)$ and $f_{l-1}(lz)/f_l(lz)$, when $l$ becomes large but not necessarily infinite, for a $|z|$-argument varying from a finite value to infinity. If setting:

$$X_\pm = 1 + \widetilde{X}_\pm, \quad \widetilde{X}_\pm = \sum_{s=1}^{+\infty} \frac{X_s}{(\pm l)^s}, \quad X_\pm = U_\pm \text{ or } V_\pm, X_s = U_s \text{ or } V_s \quad (6.18)$$

where the coefficients $U_s$ and $V_s$ are respectively given by (6.12), (6.16) and (6.17) the polynomial expansion of the ratios $f_{l+1}(lz)/f_l(lz)$ and $f_{l-1}(lz)/f_l(lz)$ can symbolically be written as:

$$\frac{f_{l\pm 1}(lz)}{f_l(lz)} \approx -\frac{J}{|J|}\left\{\mp\left(\frac{1}{|z|} + \frac{1}{2l|z|}\right) + \frac{1}{u|z|}\frac{1+\widetilde{V}_+}{1+\widetilde{U}_+}\left[1 - \exp(-2l|\varsigma|)\left(\frac{1+\widetilde{V}_-}{1+\widetilde{V}_+} + \frac{1+\widetilde{U}_-}{1+\widetilde{U}_+}\right)\right]\right\}$$

$$+ O(\exp(-4l|\varsigma|)), \text{ as } T \to 0. \quad (6.19)$$

The parity of the ratio $f_{l\pm 1}(lz)/f_l(lz)$ only depends on the sign of $-J$. Taking into account all the previous informations we can finally write if $l \gg 1$:

$$\frac{\lambda_{l+1}(lz)}{\lambda_l(lz)} \approx -\frac{J}{|J|}\left\{\mp\left(\frac{1}{|z|} + \frac{1}{2l|z|}\right) + \frac{I'_l(l|z|)}{I_l(l|z|)}\right\}, \text{ as } T \to 0 \quad (6.20)$$

with

$$\frac{I'_l(l|z|)}{I_l(l|z|)} \approx \frac{\varepsilon}{u|z|}\left[1 - \frac{u}{2l} - \frac{u^2}{8l^2} - 2\exp(-2l|\varsigma|)\left(1 - \frac{u}{4l} - \frac{3u^2}{32l^2}\right)\right] + O(\exp(-4l|\varsigma|)), \text{ as } T \to 0, (6.21)$$

where $\varepsilon = \pm 1$ is sign of the slope of $d|\varsigma|/dT$.

## Appendix 3: Spin-Spin Correlation Along a Lattice Line, in the Thermodynamic Limit

In the simplest case of a lattice characterized by a single type of exchange energy $J = J_1 = J_2$ the physical meaning of the ratio of convergence $r_{\Lambda+1}/r_\Lambda$, in the low-temperature limit i.e., near the fixed point $\widetilde{z}_c = 1/4\pi$, imposes to calculate the spin-spin correlation between any couple of spins located on the *same horizontal lattice line* (*cf* (2.70)), for instance, at sites $(i, j)$ and $(i, j+k)$. However the following reasoning is valid if $J_1 \neq J_2$.

The corresponding spin-spin correlation labeled $<\boldsymbol{S}_{i,j}.\boldsymbol{S}_{i,j+k}>$ can be defined as

$$<\boldsymbol{S}_{i,j}.\boldsymbol{S}_{i,j+k}> = \frac{1}{Z_N(0)}\int d\boldsymbol{S}_{-N,-N}\ldots\int \boldsymbol{S}_{i,j}d\boldsymbol{S}_{i,j} \times$$

$$\times \ldots\int \boldsymbol{S}_{i,j+k}d\boldsymbol{S}_{i,j+k}\ldots\int d\boldsymbol{S}_{N,N}\exp\left(-\beta\sum_{i=-N}^{N}\sum_{j=-N}^{N}H_{i,j}^{ex}\right) \quad (7.1)$$

and, because of isotropic couplings, the *z-z* spin-spin correlation is such as $<S_{i,j}^z.S_{i,j+k}^z> = <\boldsymbol{S}_{i,j}.\boldsymbol{S}_{i,j+k}>/3$. For any lattice size the numerator of (7.1) can be expanded on the infinite basis of spherical harmonics as the zero-field partition function $Z_N(0)$:

$$<S_{i,j}^z.S_{i,j+k'}^z> = \frac{(4\pi)^{8N^2}}{3Z_N(0)}\prod_{i=-N}^{N}\prod_{j=-N}^{N}\sum_{l_{i,j}=0}^{+\infty}\lambda_{l_{i,j}}(-\beta J)\sum_{l'_{i,j}=0}^{+\infty}\lambda_{l'_{i,j}}(-\beta J)\sum_{m_{i,j}=-l_{i,j}}^{+l_{i,j}}\sum_{m'_{i,j}=-l'_{i,j}}^{+l'_{i,j}}F'_{i,j} \quad (7.2)$$

where the integral $F'_{i,j}$ is now

$$F'_{i,j} = \int d\boldsymbol{S}_{i,j}X_{i,j}Y_{l'_{i+1,j},m'_{i+1,j}}(\boldsymbol{S}_{i,j})Y_{l_{i,j-1},m_{i,j-1}}(\boldsymbol{S}_{i,j})Y^*_{l_{i,j},m_{i,j}}(\boldsymbol{S}_{i,j})Y^*_{l'_{i,j},m'_{i,j}}(\boldsymbol{S}_{i,j}) \quad (7.3)$$

with

$$X_{K,K'} = \cos\theta_{K,K'} \text{ if } K = i,\ K' = j \text{ or } K' = j + k, X_{K,K'} = 1 \text{ if } K \neq i, K' \neq j \text{ and } K' \neq j + k. \quad (7.4)$$

When $X_{K,K'} = 1$, we have $F'_{i,j} = F_{i,j}$. For calculating integral $F'_{i,j}$ we use the following transform:

$$\cos\theta_{K,K'}Y_{l_{K,K'},m_{K,K'}}(\boldsymbol{S}_{K,K'}) = C_{l+1}Y_{l_{K,K'}+1,m_{K,K'}}(\boldsymbol{S}_{K,K'}) + C_{l-1}Y_{l_{K,K'}-1,m_{K,K'}}(\boldsymbol{S}_{K,K'}) \quad (7.5)$$

with:

$$C_{l+1} = \left[\frac{(l_{K,K'}+1+m_{K,K'})(l_{K,K'}+1-m_{K,K'})}{(2l_{K,K'}+1)(2l_{K,K'}+3)}\right]^{1/2},\ C_{l-1} = \left[\frac{(l_{K,K'}+m_{K,K'})(l_{K,K'}-m_{K,K'})}{(2l_{K,K'}+1)(2l_{K,K'}-1)}\right]^{1/2}. \quad (7.6)$$

The numerator of (7.2) can be calculated by the same method used for the zero-field partition function $Z_N(0)$. Due to the imbricate character of integrals $F_{i,j}$ we recall that the process of integration can be achieved through three methods: i) integrating from horizontal edge line $i = N$ to $i = -N$ between vertical edge line $j = -N$ and $j = N$ or ii) *vice versa*; iii) integrating simultaneously from the four lattice edges $i = N, i = -N, j = -N$ and $j = N$ in the direction of the lattice heart characterized by site (0,0).

In the thermodynamic limit ($N \to +\infty$) we have $m_{K,K'} = m'_{K,K'} = 0$. In addition, we have seen in Appendix 1 that the *l*-polynomial expansion giving $Z_N(0)$ reduces to the dominant term for which $l_{K,K'} = l'_{K,K'} = l$ for all the horizontal and vertical bonds, notably, for all the sites linked to the horizontal line *i* beginning at site $(i, j)$ and ending at site $(i, j + k)$. However, for the horizontal bonds of line *i*, a special care must be brought to the bond between sites $(0, 0)$ and $(0, 1)$ characterized by $l_{0,0}$: this is the *correlation domain*.

At these sites we respectively have $l'_{1,0} = l_{0,-1} = l'_{0,0} = l$ and $l'_{1,1} = l'_{0,1} = l_{0,1} = l$ due to the fact that these coefficients are involved in the highest-degree term of the characteristic polynomial associated with $Z_N(0)$ (in spite of the fact that there are not necessary the symmetry properties of the $\mathcal{D}$ space allow to retrieve the common value of these coefficients): all the current integrals $F'_{k,k'}$ have been evaluated except $F'_{0,0}$ and $F'_{0,1}$. In other words the integrand of current integrals $F'_{k,k'}$, with $(k, k') \neq (0, 0)$ and $(k, k') \neq (0, 1)$ do not have changed in contrast with the integrand of $F'_{0,0}$ and $F'_{0,1}$. Then we have to determine the evolution of $l_{0,0}$ when passing from its value in the highest-degree term of $Z_N(0)$ i.e., $l_{0,0} = l$ to its new value in the numerator of (7.2) because the integrand of integral $F'_{0,0}$ has become $\cos\theta Y_{l_{0,0},0}(\boldsymbol{S}_{0,0})[Y_{l,0}(\boldsymbol{S}_{0,0})]^2$.

The decomposition law only intervenes at sites $(0, 0)$ and $(0, 1)$. Thus, at site $(0, 0)$, for determining $l_{0,0}$ with respect to the other $l$-coefficients appearing in $F'_{0,0}$ i.e., $l'_{1,0} = l_{0,-1} = l'_{0,0} = l$, we have to apply the decomposition law to the spherical harmonics $Y_{l,0}(\boldsymbol{S}_{0,0})$. The corresponding result can be written

$$F'_{0,0} = C_{l+1} F_{l_{0,0}, l+1} + C_{l-1} F_{l_{0,0}, l-1} \qquad (7.7)$$

with:

$$F_{l_{0,0}, l+\varepsilon} = \int d\boldsymbol{S}_{0,0} [Y_{l,0}(\boldsymbol{S}_{0,0})]^2 Y_{l+\varepsilon,0}(\boldsymbol{S}_{0,0}) Y_{l_{0,0},0}(\boldsymbol{S}_{0,0}), \; \varepsilon = \pm 1. \qquad (7.8)$$

We immediately retrieve the calculation of integrals appearing in that one concerning the zero-field partition function (in (7.8) if choosing $\varepsilon = 0$, integral $F_{l,l}$ is nothing but integral $F(l, l, 0)$ given by (2.41) with $m = 0$). In integral $F_{l_{0,0}, l+\varepsilon}$ we begin by expressing the products of pairs of spherical harmonics as CG series (cf (2.15)). For instance we have

$$[Y_{l,0}(\boldsymbol{S}_{0,0})]^2 = \frac{(2l+1)}{\sqrt{4\pi}} \sum_{L=0}^{2l} \frac{1}{\sqrt{2L+1}} \left[ C^{L\;0}_{l\;0\;l\;0} \right]^2 Y_{L,0}(\boldsymbol{S}_{0,0}),$$

$$Y_{l+\varepsilon,0}(\boldsymbol{S}_{0,0}) Y_{l_{0,0},0}(\boldsymbol{S}_{0,0}) = \sum_{L'=|l+\varepsilon-l_{0,0}|}^{l+\varepsilon+l_{0,0}} \left[ \frac{(2(l+\varepsilon)+1)(2l_{0,0}+1)}{4\pi(2L'+1)} \right]^{1/2} \left[ C^{L'\;0}_{l+\varepsilon\;0\;l_{0,0}\;0} \right]^2 Y_{L',0}(\boldsymbol{S}_{0,0}) \quad (7.9)$$

After inserting these expressions in (7.8) the non-vanishing condition of integral $F_{l_{0,0}, l+\varepsilon}$ imposes $L = L'$ i.e., $l_{0,0} = l + \varepsilon$, $\varepsilon = \pm 1$. If differently expanding the product of two spherical harmonics appearing in the integrand of $F'_{0,0}$ by means of other combinations we also find that $l_{0,0} = l + \varepsilon$). As a result all the lattice bonds are characterized by the integer $l$ whereas the coefficient $l_{0,0}$ characterizing the unique bond of the correlation domain is $l \pm 1$. We have:

$$F_{l_{0,0}, l+\varepsilon} = \frac{(2l+1)(2(l+\varepsilon)+1)}{4\pi} \sum_{L=0}^{L_>} \frac{1}{2L+1} \left[ C^{L\;0}_{l\;0\;l\;0} \; C^{L\;0}_{l+\varepsilon\;0\;l+\varepsilon\;0} \right]^2, \; \varepsilon = \pm 1, \qquad (7.10)$$

with $L_> = \min(2l, 2l+2\varepsilon)$. Thus, after applying at site (0, 0) the decomposition law given by (7.5) and (7.6), each contribution of (7.7) can be written owing to the generic term $C_{l+\varepsilon}F_{l+\varepsilon,l+\varepsilon'}$ ($\varepsilon = \pm 1$, $\varepsilon' = \pm 1$). The local contribution including the radial factor is $C_{l+\varepsilon}F_{l+\varepsilon,l+\varepsilon}\lambda_{l+\varepsilon}(-\beta J)$.

We have discarded the case in which we apply the decomposition law to $Y_{l_{0,0},0}(S_{0,0})$. If doing so and imposing that the final value of integral $F'_{0,0}$ must be similar i.e., independent of the application of the decomposition law, a similar work than the previous one leads to write $l_{0,0} = l$ and some of the coefficients $l'_{1,0}$, $l_{0,-1}$ or $l'_{0,0}$ is equal to $l + \varepsilon$, $\varepsilon = \pm 1$. This result is in contradiction with the fact that we must have $l'_{1,0} = l_{0,-1} = l'_{0,0} = l$. In addition the corresponding first-nearest neighbor integral $F'_{0,-1}$, $F'_{1,0}$ or $F'_{-1,0}$ vanishes. *As a result the decomposition law must uniquely be applied on a spherical harmonics $Y_{l,0}(S_{0,0})$ for which l is known.*

Then, we apply the decomposition law at site (0, 1). We have to consider integral $F'_{0,1}$. The work is similar to that one achieved for integral $F'_{0,0}$ given by (7.7) but the $l$-coefficients are now $l'_{1,1}$, $l'_{0,0}$ and $l_{0,1}$. All these coefficients have been previously determined. We have $l'_{1,1} = l'_{0,1} = l_{0,1} = l$ for reasons explained above. We also consider that $l_{0,0}$ is unknown whereas we have a single possibility for applying the decomposition law to spherical harmonics i.e., on $Y_{l'_{1,1},0}(S_{0,1})$, $Y_{l_{0,1},0}(S_{0,1})$ or $Y_{l'_{0,1},0}(S_{0,1})$ (which are nothing but $Y_{l,0}(S_{0,0})$). We verify that $l_{0,0} = l + \varepsilon$, $\varepsilon = \pm 1$, as independently found when calculating $F'_{0,0}$: the corresponding contribution is $F'_{0,1} = C_{l+1}F_{l+1,l+1} + C_{l-1}F_{l+1,l-1}$ if $l_{0,0} = l + 1$ and $F'_{0,1} = C_{l-1}F_{l-1,l+1} + C_{l+1}F_{l-1,l-1}$ if $l_{0,0} = l - 1$. As a result we can finally write

$$<S_{0,0}\cdot S_{0,1}> = \frac{1}{Z_N(0)}\sum_{l=0}^{+\infty}\left[\lambda_l(-\beta J)^2 F_{l,l}\right]^{4N^2}\left[X_{l+1} + (1-\delta_{l,0})X_{l-1}\right], \text{ as } N\to+\infty, \quad (7.11)$$

with

$$X_{l+1} = C_{l+1}\left(C_{l+1}\frac{F_{l+1,l+1}}{F_{l,l}} + C_{l-1}\frac{F_{l+1,l-1}}{F_{l,l}}\right)\frac{F_{l+1,l+1}}{F_{l,l}}\frac{\lambda_{l+1}(-\beta J)}{\lambda_l(-\beta J)},$$

$$X_{l-1} = C_{l-1}\left(C_{l+1}\frac{F_{l-1,l+1}}{F_{l,l}} + C_{l-1}\frac{F_{l-1,l-1}}{F_{l,l}}\right)\frac{F_{l-1,l-1}}{F_{l,l}}\frac{\lambda_{l-1}(-\beta J)}{\lambda_l(-\beta J)}. \quad (7.12)$$

In the infinite $l$-limit (i.e., near $T_c = 0$ K) we have $C_{l+\varepsilon} \to 1/2$ and $F_{l+\varepsilon,l+\varepsilon'}/F_{l,l} \to 1$ so that

$$|<S_{0,0}\cdot S_{0,1}>| \approx \frac{1}{2}\left[\frac{\lambda_{l+1}(l|z|)}{\lambda_l(l|z|)} + \frac{\lambda_{l-1}(l|z|)}{\lambda_l(l|z|)}\right], \text{ as } l \to +\infty, T \to T_c = 0 \text{ K}, \quad (7.13)$$

or owing to (6.20):

$$|<S_{0,0}\cdot S_{0,1}>| \approx \left|\frac{I'_l(l|z|)}{I_l(l|z|)}\right|, \text{ as } l \to +\infty, T \to T_c = 0 \text{ K}. \quad (7.14)$$

Under these conditions, near $T_c = 0$ K, the ratios $r_{l\pm1}/r_l$ defined by (2.46) are linked by the following relationship:

$$\left(\frac{r_{l+1}}{r_l}\right)^{1/2} + \left(\frac{r_{l-1}}{r_l}\right)^{1/2} = 2\left|<S_{0,0}.S_{0,1}>\right|, \text{ as } l \to +\infty, T \to T_c = 0 \text{ K.} \quad (7.15)$$

In the low-temperature domain $r_{l+1}/r_l$ and $r_{l-1}/r_l$ are very close, as $l \to +\infty$. As a result we derive:

$$\left(\frac{r_{l+1}}{r_l}\right)^{1/2} \approx \left|<S_{0,0}.S_{0,1}>\right|, \text{ as } l \to +\infty, T \to 0. \quad (7.16)$$

Owing to the multiplication theorem applied to Bessel functions, near $T_c = 0$ K, it is easy to show as $\Lambda = 2l$ [21]

$$\left(\frac{r_{l+1}}{r_l}\right)^{1/2} \approx \left(\frac{r_{\Lambda+1}}{r_\Lambda}\right)^{1/2}, \left(\frac{r_{\Lambda+1}}{r_\Lambda}\right)^{1/2} \approx \left|\frac{I'_\Lambda(|\widetilde{z}|\Lambda)}{I_\Lambda(|\widetilde{z}|\Lambda)}\right| \text{ as } \Lambda \to +\infty, T \to 0. \quad (7.17)$$

## Appendix 4: Low-Temperature Behaviors of the Correlation Length

From (2.71) and (2.73) we can equivalently write near $T_c = 0$ K:

$$\left|\frac{I'_\Lambda(|\widetilde{z}|\Lambda)}{I_\Lambda(|\widetilde{z}|\Lambda)}\right| \approx 1 - 2a/\xi + \ldots, \quad \left|\frac{I'_\Lambda(|\widetilde{z}|\Lambda)}{I_\Lambda(|\widetilde{z}|\Lambda)}\right| \approx 1 - L_\tau/\widetilde{\xi} + \ldots \quad (8.1)$$

due to the universal relation $\xi/a = \widetilde{\xi}/L_\tau$ exclusively valid near $T_c$ (*cf* (2.72)). $\xi$ is the correlation length expressed in *a*-unit and $\widetilde{\xi}$ is the renormalized correlation length expressed in $L_\tau$-unit. From (6.19)-(6.21) we have:

$$\left|\frac{I'_\Lambda(|\widetilde{z}|\Lambda)}{I_\Lambda(|\widetilde{z}|\Lambda)}\right| = S_1 - S_2, \; S_1 = \frac{1}{\widetilde{u}|\widetilde{z}|}\frac{1+\widetilde{V}_+}{1+\widetilde{U}_+}, \; S_2 = \frac{1}{\widetilde{u}|\widetilde{z}|}\frac{1+\widetilde{V}_+}{1+\widetilde{U}_+}\exp(-\Lambda|\widetilde{\zeta}|)\left(\frac{1+\widetilde{V}_-}{1+\widetilde{V}_+} + \frac{1+\widetilde{U}_-}{1+\widetilde{U}_+}\right),$$
$$\text{as } T \to 0, \quad (8.2)$$

where $\widetilde{z}$ and $\widetilde{u}$ are given by (2.63) and (2.65). Near the fixed point $\widetilde{z}_c = 1/4\pi$ (6.21) can be rewritten as:

$$\left|\frac{I'_\Lambda(|\widetilde{z}|\Lambda)}{I_\Lambda(|\widetilde{z}|\Lambda)}\right| \approx \frac{1}{\widetilde{u}|\widetilde{z}|}\left[1 - \frac{\widetilde{u}}{\Lambda} - \frac{\widetilde{u}^2}{2\Lambda^2} - 2\exp(-|\widetilde{\zeta}|\Lambda)\left(1 - \frac{\widetilde{u}}{2\Lambda} - \frac{3\widetilde{u}^2}{8\Lambda^2}\right)\right] + O(\exp(-2|\zeta|\Lambda)),$$
$$\text{as } T \to 0, \quad (8.3)$$

where we have used the relation $2l = \Lambda$.

In a first step it is necessary to know the respective values of series $U_\pm = 1 + \widetilde{U}_\pm$ and $V_\pm = 1 + \widetilde{V}_\pm$ at $\widetilde{z}_c = 1/4\pi$. We begin by noting that the multiplication theorem allows one to write $I_\Lambda(\widetilde{z}_c \Lambda) \sim \Lambda^\Lambda I_\Lambda(\widetilde{z}_c)$ as $\widetilde{z}_c \ll 1$ [21]. Then, owing to the series definition of $I_\Lambda(\widetilde{z}_c)$ reduced to its first term due to the fact that $\tilde{z}_c \ll 1$, we derive $I_\Lambda(\tilde{z}_c \Lambda) \sim \Lambda^\Lambda (\tilde{z}_c/2)^\Lambda / \Lambda!$. Finally, using the well-known Stirling formula giving $\Lambda!$ when $\Lambda \gg 1$, we have

$$I_\Lambda(\tilde{z}_c \Lambda) \approx \frac{\exp(\Lambda)(\tilde{z}_c/2)^\Lambda}{\sqrt{2\pi\Lambda}}. \tag{8.4}$$

When $\Lambda \gg 1$, the second part of (6.14) giving the expansion of $I_\Lambda(\tilde{z}_c \Lambda)$ vanishes. Then, if introducing in the remaining part of (6.14) the dimensionless quantity $|\tilde{\zeta}|$ defined by (6.10), expressed near $\tilde{z}_c$, i.e., $|\tilde{\zeta}| \approx 1 + \ln(\tilde{z}_c/2)$, and comparing (8.4) and (6.14), we derive $U_+ \approx (1+\tilde{z}_c^2)^{1/4}$ i.e., $1+(\tilde{z}_c/2)^2$. Finally (6.16) allows one to obtain a relationship between series $U_+$ and $V_+$. Thus, near the fixed point $\tilde{z}_c = 1/4\pi$ and when $\Lambda \gg 1$ it is easy to show that $V_+ \approx (1+\tilde{z}_c^2)^{-1/4}$. As a result

$$\binom{U_+(\tilde{z}_c)}{V_+(\tilde{z}_c)} \approx 1 \pm \left(\frac{\tilde{z}_c}{2}\right)^2. \tag{8.5}$$

The sign + of the right-hand side (respectively, the sign −) refers to $U_+$ (respectively, $V_+$). A similar reasoning allows to derive the series $U_-$ and $V_-$ but the function $I_\Lambda(\tilde{z}\Lambda)$ must be now replaced by the other Bessel function $K_\Lambda(\tilde{z}\Lambda)$. Near the fixed point $\tilde{z}_c$ and when $\Lambda \gg 1$, we have symbolically:

$$\binom{U_-(\tilde{z}_c)}{V_-(\tilde{z}_c)} \approx \frac{1}{e}\binom{U_+(\tilde{z}_c)}{V_+(\tilde{z}_c)} \tag{8.6}$$

with the same convention of sign.

In a second step we need to expand the variable

$$|\tilde{\zeta}|\Lambda = \left|\sqrt{1+\tilde{z}^2} - \left|\ln\left(\tilde{z}/\left[1+\sqrt{1+\tilde{z}^2}\right]\right)\right|\right|\Lambda \tag{8.7}$$

in order to derive $\exp(\pm|\tilde{\zeta}|\Lambda)$ near the fixed point $\tilde{z}_c$. At $\tilde{z}_c$ we have exactly $|\tilde{\zeta}| = 0$. Near this critical point (see Fig. 6), $\sqrt{1+\tilde{z}^2} \approx \left|\ln\left(\tilde{z}/\left[1+\sqrt{1+\tilde{z}^2}\right]\right)\right|$. As $\tilde{z}$ remains small $|\tilde{\zeta}|\Lambda$ reduces to

$$|\tilde{\zeta}|\Lambda \approx |1 - |\ln(\tilde{z}/2)||\Lambda. \tag{8.8}$$

Introducing $\tilde{z}_c$ and using the property $\ln(u) \approx u - 1$, with $u \ll 1$, we derive for $\tilde{z} \approx \tilde{z}_c$:

$$|\tilde{\zeta}|\Lambda \approx \left|\ln|\tilde{z}/\tilde{z}_c|^\Lambda\right|. \tag{8.9}$$

In Zones 1 ($x_1\Lambda \ll 1$, $x_1 \ll 1$) and 3 ($x_1\Lambda \gg 1$, $x_1 \gg 1$) we have from the definition of $\rho_s$ (*cf* (2.57), (2.63) and (2.68))

$$\Lambda(|\tilde{z}| - \tilde{z}_c) = \frac{\rho_s}{k_B T} \qquad (8.10)$$

or equivalently by introducing $x_1$ (*cf* (2.75)):

$$\Lambda(|\tilde{z}| - \tilde{z}_c) = \frac{2\tilde{z}_c}{x_1}, \quad \tilde{z}_c = \frac{1}{4\pi}. \qquad (8.11)$$

Using the well-known relationship $(1 \pm u/\Lambda)^\Lambda = \exp(\pm u)$, as $\Lambda \to +\infty$, we derive that, near $\tilde{z}_c$, $|\tilde{z}|/\tilde{z}_c = \exp(2/x_1\Lambda)$. In Zones 2 ($x_2\Lambda \sim 1$, $x_2 \ll 1$) and 4 ($x_2\Lambda \gg 1$, $x_2 \gg 1$) we have from the definition of $\Delta$ (*cf* (2.57) and (2.68))

$$\Lambda(\tilde{z}_c - |\tilde{z}|) = \frac{\Delta}{4\pi k_B T} \qquad (8.12)$$

or equivalently by introducing $x_2$ (*cf* (2.75)):

$$\Lambda(\tilde{z}_c - |\tilde{z}|) = \frac{\tilde{z}_c}{x_2} \qquad (8.13)$$

so that, near $\tilde{z}_c$, $|\tilde{z}_c|/\tilde{z} = \exp(-1/(x_2\Lambda))$ and $|\tilde{z}/\tilde{z}_c|^\Lambda = \exp(-1/x_2)$. The behavior of $|\tilde{\zeta}|\Lambda$ as depicted by Fig. 7 imposes the following properties: i) if $|\tilde{z}| < \tilde{z}_c$ ($T > T_c$) $|\tilde{\zeta}|\Lambda$ decreases when $|\tilde{z}|$ increases and must vary as a function of argument $1/x_2$ because of (8.13); ii) if $|\tilde{z}| > \tilde{z}_c$ ($T < T_c$) $|\tilde{\zeta}|\Lambda$ decreases with $|\tilde{z}|$ and must vary as a function of argument $-1/x_1$ because of (8.11). Thus, if expressing for instance $|\tilde{\zeta}|\Lambda$ from (8.9) when $|\tilde{z}| < \tilde{z}_c$, we have $|\tilde{\zeta}|\Lambda \approx \exp(1/2x_2)$ as $x_2 \ll 1$. As $\exp(1/2x_2) \gg 1$ we can write:

$$|\tilde{\zeta}|\Lambda \approx 2\ln\left(\frac{\exp(1/2x_2)}{2} + \sqrt{1 + \left\{\frac{\exp(1/2x_2)}{2}\right\}^2}\right), |\tilde{z}| < \tilde{z}_c \ (T_c < T), \qquad (8.14)$$

i.e.,

$$|\tilde{\zeta}|\Lambda \approx 2\,\text{argsh}\left(\frac{\exp(1/2x_2)}{2}\right) \text{ (Zones 2 and 4).} \qquad (8.15)$$

If comparing (8.11) (valid in Zones 2 and 4) and (8.13) (valid in Zones 1 and 3), the conversion of this expression near $\tilde{z}_c$ is assumed by exchanging $1/2x_2$ against $-1/x_1$. Under these conditions we have:

$$|\tilde{\zeta}|\Lambda \approx 2\,\text{argsh}\left(\frac{\exp(-1/x_1)}{2}\right) \text{ (Zones 1 and 3).} \qquad (8.16)$$

As a result behaviors can be derived for the four zones of the magnetic diagram given by Fig. 7. In Zone 1 ($x_1\Lambda \ll 1$, $x_1 \ll 1$) we have

$$\left|\widetilde{\zeta}\right|\Lambda \approx \exp(-1/x_1), \; x_1 \ll 1 \; \text{(Zone 1)}. \tag{8.17}$$

In Zone 2 ($x_2\Lambda \sim 1$, $x_2 \ll 1$)

$$\left|\widetilde{\zeta}\right|\Lambda \approx \frac{1}{x_2} + 2\exp(-1/x_2), \; x_2 \ll 1 \; \text{(Zone 2)}. \tag{8.18}$$

In Zones 3 ($x_1\Lambda \gg 1$, $x_1 \gg 1$) and 4 ($x_2\Lambda \gg 1$, $x_2 \gg 1$) we have

$$\left|\widetilde{\zeta}\right|\Lambda \approx 2\ln\left(\frac{1+\sqrt{5}}{2}\right) - \frac{2}{\sqrt{5}x_1}, \; x_1 \gg 1 \; \text{(Zone 3)}, \tag{8.19}$$

$$\left|\widetilde{\zeta}\right|\Lambda \approx 2\ln\left(\frac{1+\sqrt{5}}{2}\right) + \frac{1}{\sqrt{5}x_2}, \; x_2 \gg 1 \; \text{(Zone 4)}. \tag{8.20}$$

At the frontier between Zones 3 and 4 i.e., along the vertical line reaching the Néel line at $T_c$, $x_1$ and $x_2$ become infinite so that:

$$\left|\widetilde{\zeta}\right|\Lambda \approx 2\ln\left(\frac{1+\sqrt{5}}{2}\right), \; T = T_c. \tag{8.21}$$

Now, if expanding (8.2) near $\widetilde{z}_c$ and comparing with (8.1), we can derive $\widetilde{\xi}/L_\tau$ or equivalently $\xi/2a$ near $T_c$. When $\left|\widetilde{\zeta}\right|\Lambda \ll 1$ (Zone 1, $x_1 \ll 1$) or when $\left|\widetilde{\zeta}\right|\Lambda < 1$ (Zone 3, $x_1 \gg 1$ and Zone 4, $x_2 \gg 1$) the factor $\exp(-\left|\widetilde{\zeta}\right|\Lambda)$ appearing in $S_2$ can be approximated by $1 - \left|\widetilde{\zeta}\right|\Lambda + \dots$. As a result (8.1) can be rewritten

$$\left|\frac{I'_\Lambda(|\widetilde{z}|\Lambda)}{I_\Lambda(|\widetilde{z}|\Lambda)}\right| \approx \frac{1}{\widetilde{u}|\widetilde{z}|}\left[1 - 2\left|\widetilde{\zeta}\right|\Lambda\left(1 - \frac{\widetilde{u}}{2\Lambda}\right)\right] + O\left(\frac{\widetilde{u}^2}{\Lambda^2}\right), \; \text{as } T \to 0. \tag{8.22}$$

We immediately derive that, near $T_c$ and due to (2.72), $\widetilde{\xi}/L_\tau$ or equivalently $\xi/2a$ scales as

$$\frac{\xi}{2a} \approx \left(\left|\widetilde{\zeta}\right|\Lambda\right)^{-1}, \frac{\widetilde{\xi}}{L_\tau} \approx \left(\left|\widetilde{\zeta}\right|\Lambda\right)^{-1}, \; \left(\left|\widetilde{\zeta}\right|\Lambda\right)^{-1} \approx (X_i(x_i))^{-1}, \; \text{as } T \to 0, \; i = 1, 2. \tag{8.23}$$

i.e., we directly obtained for $\left|\widetilde{\zeta}\right|\Lambda$ the result of Chubukov *et al.* derived from a renormalization technique and labeled $X_i(x_i)$ [5].

In Zone 1 ($x_1\Lambda \ll 1$, $x_1 \ll 1$), when $|\widetilde{z}|$ deviates from $\widetilde{z}_c$, $S_1$ is reduced to the factor $(\widetilde{u}|\widetilde{z}|)^{-1}$ which tends towards unity because $|\widetilde{z}|$ behaves as $\exp(1/(2x_1\Lambda))$ and becomes infinite. For the three remaining zones $(\widetilde{u}|\widetilde{z}|)^{-1}$ behaves as $(\widetilde{z}_c)^{-1}$. For $S_2$ the situation is more complicated because we already have expressed $|\tilde{\zeta}|\Lambda$ vs $x_1$ and $x_2$. From (8.2) and (8.22) the transformed part of $S_2$ (cf (8.2)) can be decomposed as

$$S_2 = 2f_1 f_2, \quad f_1 = \frac{|\tilde{\zeta}|\Lambda}{\tilde{u}|\tilde{z}|}, \quad f_2 = \frac{1+\tilde{V}_+}{1+\tilde{U}_+}\left(\frac{1+\tilde{V}_-}{1+\tilde{V}_+} + \frac{1+\tilde{U}_-}{1+\tilde{U}_+}\right) + \ldots \quad (8.24)$$

where the series $\tilde{U}_\pm$ and $\tilde{V}_\pm$ are defined by (6.18) and expressed at $\tilde{z}_c$ (cf (8.5) and (8.6)). If considering $\ln(f_1 f_2)$ we have $\ln f_1 = \ln(|\tilde{\zeta}|\Lambda) - \ln(\tilde{u}|\tilde{z}|)$ where $\ln(\tilde{u}|\tilde{z}|) \approx \ln(\tilde{z}_c)$ so that $\ln f_1 \approx \ln(|\tilde{\zeta}|\Lambda/\tilde{z}_c)$ near $\tilde{z}_c$. It remains to evaluate $\ln f_2$. Calculations are tedious but not difficult. We skip the intermediate steps and give the final result:

$$S_2 = \frac{2}{e}|\tilde{\zeta}|\Lambda\left[\frac{1}{|\tilde{z}|} - \frac{1}{2|\tilde{z}|\Lambda} - \frac{3|\tilde{z}|}{8(|\tilde{z}|\Lambda)^2} + \ldots\right]_{|\tilde{z}|\to\tilde{z}_c}. \quad (8.25)$$

As a result, in Zone 1 ($x_1\Lambda \ll 1$, $x_1 \ll 1$), from (8.1), (8.2), (8.11) and (8.17) we derive

$$\left|\frac{I'_{\Lambda/2}(|z|\Lambda)}{I_{\Lambda/2}(|z|\Lambda)}\right| \approx 1 - \frac{8\pi}{e}\exp(-1/x_1)\left(1 - \frac{x_1}{2}\right) + O(x_1^2), \text{ as } T \to 0. \quad (8.26)$$

Comparing (8.1) and (8.26) allows one to write $\xi$. Recalling that $x_1 = k_B T/2\pi\rho_s$, $a = \hbar c/2|J|$ and $\rho_s \approx |J|$ as $g^* = T/T_c$ vanishes with $T$ near $T_c = 0$ K (cf (2.66)), we finally derive:

$$\xi = \frac{e}{8}\frac{\hbar c}{2\pi\rho_s}\exp\left(\frac{2\pi\rho_s}{k_B T}\right)\left(1 + \frac{k_B T}{4\pi\rho_s}\right), \quad x_1 \ll 1 \text{ (Zone 1)}. \quad (8.27)$$

We exactly retrieve the result first obtained by Hasenfratz and Niedermayer [24] and confirmed by Chubukov et al. [5]. In Zones 3 ($x_1 \gg 1$) and 4 ($x_2 \gg 1$) the corrective terms $\tilde{u}/\Lambda$ become negligible near $\tilde{z}_c$ (cf (8.10)-(8.13)). The factor $(\tilde{u}|\tilde{z}|)^{-1} \approx \tilde{z}_c$ can be absorbed out of $r_{\Lambda+1}/r_\Lambda$ i.e., out of $|I'_\Lambda(|\tilde{z}|\Lambda)/I_\Lambda(|\tilde{z}|\Lambda)|$ so that $\xi$ can be derived from (8.19)-(8.21) and (8.23):

$$\tilde{\xi} \approx C^{-1}\frac{\hbar c}{k_B T}\left(1 + \frac{2}{C\sqrt{5}x_1}\right), \quad x_1 \gg 1 \text{ (Zone 3)}, \quad (8.28)$$

$$\tilde{\xi} \approx C^{-1}\frac{\hbar c}{k_B T}\left(1 - \frac{1}{C\sqrt{5}x_2}\right), \quad x_2 \gg 1 \text{ (Zone 4)} \quad (8.29)$$

where we have set:

$$C = 2\ln\left(\frac{1+\sqrt{5}}{2}\right) = 0.962\ 424, \ C^{-1} = 1.039\ 043. \tag{8.30}$$

At the frontier between Zones 3 and 4 i.e., along the vertical line reaching the Néel line at $T_c$, $x_1$ and $x_2$ become infinite so that:

$$\widetilde{\xi} \approx C^{-1}\frac{\hbar c}{k_B T}, \ T = T_c. \tag{8.31}$$

As a result the critical exponent is:

$$\nu = 1. \tag{8.32}$$

In Zone 2 ($\Lambda x_2 \sim 1$, $x_2 \ll 1$), $\exp(-|\widetilde{\zeta}|\Lambda)$ appearing in $S_2$ becomes negligible because $|\widetilde{\zeta}|\Lambda \gg 1$ (cf (8.15)). Recalling that $x_2 = k_B T/\Delta$ (8.3) and (8.18) allows one to write:

$$\widetilde{\xi} \approx \frac{\hbar c}{\Delta}, \ x_2 \ll 1 \text{ (Zone 2)}. \tag{8.33}$$

## Appendix 5: Low-Temperature Expression of the Free Energy Density

Rosenstein *et al.* have shown that, in the vanishing-field limit, the partition function $Z_N(0)$ can be written [27]

$$Z_N(0) = \int D\mathbf{S}\ \delta(\mathbf{S}^2 - 1)\exp\left(-\frac{1}{2\hbar g}\int d^3 x (\partial_\mu \mathbf{S})^2\right) \tag{9.1}$$

where the exchange Hamiltonian given by (2.2) has been expressed in the continuum limit, for a lattice characterized by $J = J_1 = J_2$. $g$ is the coupling constant given by (2.57). The constraint $\mathbf{S}^2 = 1$ is conveniently implemented by introducing a Lagrange multiplier field $\gamma(x)$. We have

$$Z_N(0) = \int D\mathbf{S}D\gamma \exp\left(-\frac{1}{2\hbar}\int d^3 x\left[(\partial_\mu \mathbf{S})^2 + \gamma^2\left(\mathbf{S}^2 - \frac{1}{g}\right)\right]\right) \tag{9.2}$$

where $\mathbf{S}^2$ has been rescaled so that the kinetic energy term is conventionally renormalized. The integrand becomes gaussian in $\mathbf{S}(x)$ and can formally be integrated. This provides an effective action for the $\gamma$-field:

$$Z_N(0) = \int D\gamma \exp\left(-\frac{1}{\hbar}S_{\text{eff}}(\gamma)\right) \tag{9.3}$$

where

$$S_{\text{eff}}(\gamma) = -\frac{\gamma(m)^2}{2g} + \frac{\hbar}{2}\text{tr}\ln\left(\hat{\mathbf{p}}^2 + \gamma(m)^2 \mathbb{1}\right). \tag{9.4}$$

$\hat{p}$ is the momentum operator, tr the trace operator and $\mathbb{1}$ the identity matrix. $\gamma(m)$ and $m$ will be defined later.

In the main text $Z_N(0)$ is expressed with a $l$-summation over eigenvalues $\lambda_l(-\beta J)$ given by (3.5). We have seen (*cf* Fig. 5) that each eigenvalue $\lambda_l(-\beta J)$ is dominant within a range $\delta T_{l_i} = T_{l_i,>} - T_{l_i,<}$ with $T = \sum_{i=0}^{i_{max}}(T_{l_i,>} - T_{l_i,<})$, $T_{l_0,<} = 0$, $T_{l_{i-1},>} = T_{l_i,<}$ ($i \neq 0$) and $T_{l_{i_{max}},>} = T$ or equivalently $\sum_{i=0}^{+\infty}\delta T_{l_i}/T = 1$. Thus $\delta T_{l_i}/T$ appears as the corresponding weight of the eigenvalue $\lambda_{l_i}(\beta|J|)$ at temperature $T$. As a result the contribution per bond to $Z_N(0)$ is $\delta Z_N(0)_{l_i} = \lambda_{l_i}(\beta|J|)\delta T_{l_i}/T$, as $Z_N(0)$ is an even function of $\beta J$. Similarly we can define the elementary contribution to $\ln Z_N(0)$ by $\delta \ln Z_N(0)_{l_i} = \ln(\lambda_{l_i}(\beta|J|))\delta T_{l_i}/T$.

Near $T_c = 0$ K the argument $\beta|J|$ is replaced by $|\tilde{z}|\Lambda$ where $|\tilde{z}|\Lambda$ is defined by (2.63), with $\Lambda = 2l \gg 1$. While approaching $T_c = 0$ K the ranges $T_{l_i,>} - T_{l_i,<}$ are tighter and tighter and the corresponding weight is very small. We tend towards a continuum: there is an increasing number of eigenvalues characterized by a higher and higher index $l$ (replaced by $\Lambda$ near $T_c = 0$ K). In other words we deal with an infinity of eigenvalues showing an infinite argument so that their total contribution dominates the ones characterized by a lower value of $l$ in spite of the fact that they have a larger weight (see Fig. 5). As a result, if $\lambda_\Lambda(|\tilde{z}|\Lambda)$ is the dominant eigenvalue over the range $\delta T_\Lambda = T_{\Lambda,>} - T_{\Lambda,<}$ and $\delta T_\Lambda/T$ the corresponding weight at temperature $T$, $\delta Z_N(0)_\Lambda = \lambda_\Lambda(|\tilde{z}|\Lambda)\delta T_\Lambda/T$ is the $\Lambda$-contribution per bond to $Z_N(0)$. Similarly the $\Lambda$-contribution per bond to $\ln Z_N(0)$ is $\delta \ln(Z_N(0)_\Lambda) = \ln(\lambda_\Lambda(|\tilde{z}|\Lambda))\delta T_\Lambda/T$. As the dominant eigenvalue changes between two consecutive ranges $\delta T_\Lambda$ and $\delta T_{\Lambda+1}$, with $\delta T_\Lambda \approx \delta T_{\Lambda+1}$ as $\Lambda \to +\infty$, we can formally write for a given temperature $T$ close to $T_c = 0$ K

$$\delta \ln(Z_N(0)) \approx \lim_{M \to +\infty}\sum_{\Lambda=0}^{M}\ln(\lambda_\Lambda(|\tilde{z}|\Lambda))\frac{\delta T_\Lambda}{T}, \quad \lim_{M \to +\infty}\sum_{\Lambda=0}^{M}\frac{\delta T_\Lambda}{T} = 1. \qquad (9.5)$$

In fact, only the contributions characterized by $\Lambda \gg 1$ bring a significant value to the $\Lambda$-sum. $\delta \ln Z_N(0)$ appears as a continuous suite of different continuous convergent functions $\ln(\lambda_\Lambda(|\tilde{z}|\Lambda))$, each of them being exclusively valid over ranges $\delta T_\Lambda = T_{>,\Lambda} - T_{<,\Lambda}$ of unequal but close widths. As a result we are dealing with a *Riemann sum*.

From (9.2)-(9.4) involving the effective action $\mathcal{S}_{eff}$ the effective $\Lambda$-contribution to $Z_N(0)_{eff,\Lambda}$ is $\exp(-\mathcal{S}_{eff,\Lambda}(\gamma)/\hbar)$ and that one contributing to $\ln(Z_N(0)_{eff,\Lambda})$ can be written $\ln(Z_N(0)_{eff,\Lambda}) = -\mathcal{S}_{eff,\Lambda}(\gamma)/\hbar$ inside the range $\delta T_\Lambda$. As a result we have:

$$-\frac{1}{\hbar}\delta \mathcal{S}_{eff}(\gamma) = -\frac{1}{\hbar}\lim_{M \to +\infty}\sum_{\Lambda=0}^{M}\mathcal{S}_{eff,\Lambda}(\gamma)\frac{\delta T_\Lambda}{T}. \qquad (9.6)$$

Comparing (9.4)-(9.6) we must obtain the total effective contribution

$$\lim_{M \to +\infty} \sum_{\Lambda=0}^{M} \ln(\lambda_\Lambda(|\tilde{z}|\Lambda)) \frac{\delta T_\Lambda}{T}\bigg|_{\text{eff}} = \frac{\gamma(m)^2}{2\hbar g} - \frac{1}{2} \lim_{M \to +\infty} \sum_{\Lambda=-M}^{M} \ln(\hat{\mathbf{p}}^2 + \gamma(m)^2 \mathbb{1})_\Lambda \frac{\delta T_\Lambda}{T} \quad (9.7)$$

if using the second part of (9.5). $\mathbb{1}$ is the identity matrix (with dim $\mathbb{1} = 2M + 1$, $M \to +\infty$). In (9.7) we have taken into account the fact that $\lambda_\Lambda(|\tilde{z}|\Lambda)$ is the dominant eigenvalue over the range $\delta T_\Lambda$: it means that the trace of the eigenmatrix associated with the operator $\ln(\hat{\mathbf{p}}^2 + \gamma(m)^2 \mathbb{1})$ reduces to the single dominant $\Lambda$-element $\ln(\hat{\mathbf{p}}^2 + \gamma(m)^2 \mathbb{1})_\Lambda$ i.e., $\ln(\mathbf{p}_\Lambda^2 + \gamma(m)^2)$, as $\Lambda \gg 1$.

In the thermodynamic limit, the elementary free energy density per lattice bond can be expressed by (3.9) i.e., if considering the effective action

$$\frac{\delta \mathcal{F}(T)}{8N^2} = -\hbar c C^2 \left(\frac{k_B T}{\hbar c}\right)^3 \lim_{M \to +\infty} \sum_{\Lambda=0}^{M} \ln(\lambda_\Lambda(|\tilde{z}|\Lambda)) \frac{\delta T_\Lambda}{T}\bigg|_{\text{eff}}, \text{ as } N \to +\infty \quad (9.8)$$

where $C$ is given by (2.90). We have explained after (3.8) that the previous sum is a *Riemann sum*. In addition, as we deal with a surface free energy density $\delta \mathcal{F}$, the dimensionless weight $\delta T_\Lambda / T$ must be linked to a surface element through a factor that we shall identify below (*cf* (9.27)). Owing to (9.7) and (9.8) $\delta \mathcal{F}(T)$ must appear as

$$\frac{\delta \mathcal{F}(T)}{8N^2} = -\hbar c C^2 \left(\frac{k_B T}{\hbar c}\right)^3 \left[\frac{\gamma(m)^2}{2\hbar g} - \frac{1}{2} \lim_{M \to +\infty} \sum_{\Lambda=-M}^{M} \ln(\mathbf{p}_\Lambda^2 + \gamma(m)^2) \frac{\delta T_\Lambda}{T}\right], \text{ as } N \to +\infty. \quad (9.9)$$

As a result, in a first step, we have to show that the $\Lambda$-sum over $\ln(\lambda_\Lambda(|\tilde{z}|\Lambda))$ in (9.8) can be transformed into a $\Lambda$-sum over $\ln(\mathbf{p}_\Lambda^2 + \gamma(m)^2)$ in (9.9) when $\delta \mathcal{F}(T)$ is expressed near its extremum: this is the *effective contribution*. In a second step the discrete sum of (9.9) must be written as a *Kurzweil-Henstock integral*, in the infinite $M$-limit. Its value essentially comes from the larger and larger values of $\Lambda$ near $T_c = 0$ K so that this integral must be evaluated owing to the steepest descent method. Under these conditions, in a final step, we can then derive the difference $\mathcal{F}(T) - \mathcal{F}(0)$.

In this purpose we wish to show that the $\Lambda$-sum in (9.9) can be written as an identity recalled by Sachdev [26] (i.e., a Poisson summation):

$$\lim_{M \to +\infty} \left[\frac{1}{L} \sum_{\Lambda=1}^{M} \ln\left(\frac{4\pi^2 \Lambda^2}{L^2} + a^2\right)\right] = \frac{1}{2} \lim_{M \to +\infty} \int_{-2\pi M/L}^{+2\pi M/L} \frac{d\omega}{2\pi} \ln(\omega^2 + a^2)$$

$$+ \frac{1}{L} \ln(1 - \exp(-L|a|)) - \frac{\ln a^2}{2L}, \quad \omega = \frac{2\pi}{L}. \quad (9.10)$$

$a$ is specified below.

The bound of the integral appearing in (9.10) i.e., $2\pi M/L$, represents the $M$-th current value of the momentum $p/\hbar$ (in fact $pc/\hbar$) along the finite $i\tau$-length of the slab. Along the $i\tau$-direction we have a periodic structure due to the propagation of spin waves. $pc/\hbar$ is quantized in integer multiples of $1/L$, $M$ being the corresponding quantum number. For each mode $M$, $p/M = h/\lambda_{DB}$ where $\lambda_{DB} = 2\pi L_\tau$ is the thermal de Broglie wavelength characterizing the spin waves and $L_\tau = \hbar c/k_B T$. As a result (9.10) appears as a summation over Matsubara frequencies $(p/\hbar)_M = \omega_M/c$. As $M$ is integer, we then deal with *bosonic excitations*. Finally we have $L = L_\tau$ where $L$ is the slab thickness.

As a result we wish to connect the quantity $\omega^2 \Lambda^2 + a^2$ in the left-hand side of (9.10) to $\lambda_\Lambda(|\tilde{z}|\Lambda)$ expanded in the infinite $\Lambda$-limit (*cf* (9.5)-(9.8)). Near $T_c = 0$ K, we have (*cf* (3.8)):

$$\lambda_\Lambda(|\tilde{z}|\Lambda) \approx K_1 \left\{ \exp(|\tilde{\zeta}|\Lambda) \sum_{s=0}^{+\infty} \frac{U_s}{(\Lambda)^s} + \exp(-|\tilde{\zeta}|\Lambda) \sum_{s=0}^{+\infty} \frac{U_s}{(-\Lambda)^s} \right\}, \text{ as } T \to 0 \; (\Lambda \to +\infty, |\tilde{z}|\Lambda \gg 1). \quad (9.11)$$

$K_1 = \left( |\tilde{z}|/\sqrt{1+\tilde{z}^2} \right)^{1/2} / (|\tilde{z}|\Lambda)^{1/2}$ expressed near $|\tilde{z}_c| \ll 1$ reduces to $\Lambda^{-1/2}$. The argument $|\tilde{z}|\Lambda$ must be high but not necessarily infinite. The coefficients $U_s$ are defined by (6.12) with $U_s < 1$, and $|\tilde{\zeta}|\Lambda$ is given by (8.7). The corresponding $U_s$-series can be restricted to their first term (i.e., unity) due to the fact that: i) the coefficients $U_s$ are small; ii) the general $s$-term $U_s/(\Lambda)^s$ (with $s > 1$) then becomes very small when $\Lambda \gg 1$.

We define the renormalized quantities

$$\tilde{\lambda}_{DB} = 2\pi \tilde{L}_\tau, \; \tilde{L}_\tau = \frac{L_\tau}{\Lambda}, \; \tilde{p}_\Lambda = \frac{\hbar}{\tilde{L}_\tau}, \frac{\tilde{p}_\Lambda}{\hbar} = \frac{\tilde{\omega}_\Lambda}{c},$$

$$\tilde{p}_\Lambda = p\Lambda, \; \tilde{\omega}_\Lambda = \omega\Lambda, \; \tilde{a} = a\Lambda. \quad (9.12)$$

If expanding $\exp(\pm|\tilde{\zeta}|\Lambda)$ *vs* $|\tilde{\zeta}|\Lambda$ in (9.11) where the $U_s$-series is reduced to $U_0 = 1$ because $\Lambda \gg 1$ and dividing by $\tilde{L}_\tau^2$, (9.11) becomes

$$\frac{\lambda_\Lambda(|\tilde{z}|\Lambda)}{K_1 \tilde{L}_\tau^2} \approx \left( \frac{2}{\tilde{L}_\tau^2} + \left( \frac{|\tilde{\zeta}|\Lambda}{\tilde{L}_\tau} \right)^2 \right) \frac{\cosh(|\tilde{\zeta}|\Lambda)}{1 + (|\tilde{\zeta}|\Lambda)^2/2}, \text{ as } T \to 0 \; (\Lambda \to +\infty). \quad (9.13)$$

Owing to (9.12) we can set for the mode $M = \pm\Lambda$

$$\frac{1}{\tilde{L}_\tau^2} = \left( \frac{\tilde{p}_{\pm\Lambda}}{\hbar} \right)^2, \; \frac{\tilde{p}_{\pm\Lambda}}{\hbar} = \pm\frac{\omega}{c}\Lambda, \; \tilde{a}_{\pm\Lambda}^2 = \left( \frac{\tilde{p}_{\pm\Lambda}}{\hbar} \right)^2 + \left( \frac{|\tilde{\zeta}|\Lambda}{\tilde{L}_\tau} \right)^2. \quad (9.14)$$

If $E = \left( (pc)^2 + (m_L c^2)^2 \right)^{1/2}$ is the relativistic energy and $m_L c^2$ the rest energy we can identify $a_{\pm\Lambda}$ owing to (9.12)

$$|a_{\pm\Lambda}| = \frac{E}{\hbar c}, \; \frac{m_L c^2}{\hbar c} = \frac{|\tilde{\zeta}|\Lambda}{L_\tau}. \quad (9.15)$$

As a result we can write if $T \to 0 \; (\Lambda \to +\infty)$

$$\ln(\lambda_{\pm\Lambda}(|\tilde{z}|\Lambda)) = \ln\left(\left(\frac{\tilde{p}_{\pm\Lambda}}{\hbar}\right)^2 + \tilde{a}_{\pm\Lambda}^2\right) + K_{2,\Lambda}, \quad K_{2,\Lambda} = \ln\left\{K_1\tilde{L}_\tau^2 \frac{\cosh(|\tilde{\zeta}|\Lambda)}{1+(|\tilde{\zeta}|\Lambda)^2/2}\right\} \quad (9.16)$$

where $\tilde{p}_{\pm\Lambda}/\hbar$, $\tilde{a}_{\pm\Lambda}$ and $\tilde{L}_\tau$ are given by (9.14). For the other eigenvalues $\lambda_{\Lambda'}(|\tilde{z}|\Lambda')$, with $\Lambda' \gg 1$ and $\Lambda' < \Lambda \to +\infty$, we have similar expansions for which $\Lambda$ is replaced by $\Lambda'$, near $T_c = 0$ K. Finally (9.13)-(9.16) allow to write:

$$\lim_{M\to+\infty}\sum_{\Lambda=-M}^{M}\ln(\lambda_\Lambda(|\tilde{z}|\Lambda))\frac{\delta T_\Lambda}{T} \approx \lim_{M\to+\infty}\sum_{\Lambda=-M}^{M}\ln\left(\left[\frac{\omega}{c}\right]^2\Lambda^2 + \tilde{a}_\Lambda^2\right)\frac{\delta T_\Lambda}{T} + K_2,$$

$$K_2 = \lim_{M\to+\infty}\sum_{\Lambda=-M}^{M}K_{2,\Lambda}\frac{\delta T_\Lambda}{T}, \quad (9.17a)$$

i.e., with the parity property $\lambda_{-\Lambda}(-|\tilde{z}|\Lambda) = \lambda_\Lambda(|\tilde{z}|\Lambda)$

$$\lim_{M\to+\infty}\sum_{\Lambda=-M}^{M}\ln(\lambda_\Lambda(|\tilde{z}|\Lambda))\frac{\delta T_\Lambda}{T} = 2\lim_{M\to+\infty}\sum_{\Lambda=1}^{M}\ln(\lambda_\Lambda(|\tilde{z}|\Lambda))\frac{\delta T_\Lambda}{T} + \ln(\lambda_0(|\tilde{z}|\Lambda))\frac{\delta T_0}{T},$$

$$\lim_{M\to+\infty}\sum_{\Lambda=-M}^{M}\ln\left(\left[\frac{\omega}{c}\right]^2\Lambda^2 + \tilde{a}_\Lambda^2\right)\frac{\delta T_\Lambda}{T} = 2\lim_{M\to+\infty}\sum_{\Lambda=1}^{M}\ln\left(\left[\frac{\omega}{c}\right]^2\Lambda^2 + \tilde{a}_\Lambda^2\right)\frac{\delta T_\Lambda}{T}, \quad (9.17b)$$

because $\tilde{a}_\Lambda = a_\Lambda\Lambda \to 0$ if $\Lambda \to 0$ so that by combining both equations

$$\lim_{M\to+\infty}\sum_{\Lambda=0}^{M}\ln(\lambda_\Lambda(|\tilde{z}|\Lambda))\frac{\delta T_\Lambda}{T} = \lim_{M\to+\infty}\sum_{\Lambda=1}^{M}\ln\left(\left[\frac{\omega}{c}\right]^2\Lambda^2 + \tilde{a}_\Lambda^2\right)\frac{\delta T_\Lambda}{T} + \frac{1}{2}\ln(\lambda_0(|\tilde{z}|\Lambda))\frac{\delta T_0}{T}. \quad (9.17c)$$

At this step we must note that the expansion given by (9.11) is applied to $\lambda_\Lambda(|\tilde{z}|\Lambda)$'s characterized by high $\Lambda$-values with $|\tilde{z}|\Lambda \gg 1$. It can be extended to finite $\Lambda$-values on condition that the argument $|\tilde{z}|\Lambda$ tends to infinity because this expansion is constructed from the asymptotic expansion of the Bessel function precisely characterized by an infinite argument (see the remark before (6.5)) for any finite or infinite $\Lambda$-value. Near $T_c = 0$ K, the argument $(\omega/c)^2\Lambda^2 + \tilde{a}_{\pm\Lambda}^2$ i.e., $((\omega/c)^2 + a^2)\Lambda^2$ is characterized by $pc/\hbar = \omega/c = 2\pi/L_\tau \ll 1$ and $a \ll 1$ so that $((\omega/c)^2 + a^2)\Lambda^2$ remains finite but can be high. Under these conditions the right-hand side of (9.17b) derived from (9.11) and (9.13)-(9.16) can be extended to finite values of $\Lambda$ (including the case $\Lambda = 0$) because their corresponding contributions are negligible. It simply means that the value of the $\Lambda$-sum is close to the exact one.

For $\Lambda$-values such as $\Lambda \gg 1$ i.e., near $T_c = 0$ K, all the weights $\delta T_\Lambda/T$ tend towards a common finite asymptotic value labeled $\delta\tilde{T}_\infty/\tilde{T} \ll 1$ whereas all the renormalized lengths $\tilde{L}_\tau$ tend towards the common limit $\tilde{L}_{\tau\infty} = L_\tau/\Lambda_\infty$ where $\Lambda_\infty$ is the asymptotic (infinite) limit of Λ. As a result, (9.17a) can be rewritten by dividing the current left-hand side $\ln(\lambda_\Lambda(|\tilde{z}|\Lambda)) = \ln((\omega/c)^2\Lambda^2 + \tilde{a}^2_{\pm\Lambda})$ by $\tilde{L}_\tau$. As just previously explained, only the values $\Lambda \sim \Lambda_\infty \gg 1$ close to the asymptotic limit $\Lambda_\infty$ significantly contribute to the value of the Λ-sum so that the common limit $\tilde{L}_{\tau\infty}$ can be factorized out of the right-hand sum in (9.17a)

$$\lim_{M\to+\infty} \sum_{\Lambda=-M}^{M} \frac{\ln(\lambda_\Lambda(|\tilde{z}|\Lambda))}{\tilde{L}_\tau} \frac{\delta\tilde{T}_\Lambda}{\tilde{T}} \approx \lim_{M\to+\infty} \frac{1}{\tilde{L}_{\tau\infty}} \left[ \sum_{\Lambda=-M}^{M} \ln\left(\left[\frac{\omega}{c}\right]^2 \Lambda^2 + \tilde{a}^2_\Lambda\right) \right] \frac{\delta\tilde{T}_\infty}{\tilde{T}}, \text{ as } T \to 0. \quad (9.18)$$

The term $K_2$ appearing in (9.17a) has disappeared in (9.18) because $K_2/\tilde{L}_\tau$ vanishes as $\tilde{L}_\tau \to +\infty$, even if $K_2$ tends to infinity as $T \to 0$. In (9.18) the term $\Lambda = 0$ can be added in the left-hand side Λ-sum as well as on the right-hand side. Due to the remark made above just before (9.18) each hand-side of (9.18) can then be divided by $\Lambda_\infty$. Recalling the temperature scaling $\tilde{T} = T\Lambda$ so that $\delta\tilde{T}_\Lambda/\tilde{T} \to \delta\tilde{T}_\infty/\tilde{T} = \delta T/T$ as $\Lambda \to +\infty$ and using (9.17c) we simply have owing to (9.12)

$$\lim_{M\to+\infty} \frac{1}{L_\tau} \sum_{\Lambda=0}^{M} \ln(\lambda_\Lambda(|\tilde{z}|\Lambda)) \frac{\delta T}{T} \approx \lim_{M\to+\infty} \frac{1}{L_\tau} \left[ \sum_{\Lambda=1}^{M} \ln\left(\frac{(p_\Lambda/\hbar)^2 + a^2}{\Lambda^{-2}}\right) \right] \frac{\delta T}{T} + \frac{\ln(\lambda_0(|\tilde{z}|\Lambda))}{2L_\tau} \frac{\delta T}{T}, \text{ as } T \to 0. \quad (9.19)$$

As $\lambda_0(|\tilde{z}|\Lambda) = \sinh(|\tilde{z}|\Lambda)/|\tilde{z}|\Lambda$, with $|\tilde{z}|\Lambda = \beta|J|$, we have $\ln(\lambda_0(|\tilde{z}|\Lambda)) \approx \beta|J| - \ln(2\beta|J|)$ i.e., $1/g - \ln(2/g)$. Thus, near $T_c = 0$ K, $\ln(\lambda_0(|\tilde{z}|\Lambda))/L_\tau \approx |J|/\hbar c - \ln(2/g)/L_\tau$. The first term $|J|/\hbar c$ is constant and jointly appears in $F(T)$ and $F(0)$. As a result it can be omitted because it is cancelled when calculating $F(T) - F(0)$. In addition, near $T_c = 0$ K, we have $\ln(2/g)/L_\tau = \ln(1/g)/L_\tau + \ln 2 \cdot k_B T/\hbar c \approx \ln(1/g)/L_\tau$ due to (2.60) because $\ln(2)/L_\tau$ vanishes. As a result $\ln(\lambda_0(|\tilde{z}|\Lambda))/L_\tau \approx -\ln(1/g)/L_\tau$.

Finally it becomes possible to write the right-hand side of (9.19) by means of the Poisson summation given by (9.10) so that the effective contribution to $F(T)$ is

$$\lim_{M\to+\infty} \left[ \frac{1}{L_\tau} \sum_{\Lambda=0}^{M} \ln(\lambda_\Lambda(|\tilde{z}|\Lambda)) \frac{\delta T}{T} \right] = \left[ \frac{1}{2} \lim_{M\to+\infty} \int_{-M/L_\tau}^{+M/L_\tau} \frac{d(p/\hbar)}{2\pi} \ln\left(\frac{(p/\hbar)^2 + (a)^2}{\Lambda^{-2}}\right) \right.$$
$$\left. + \frac{1}{L_\tau} \left\{ \ln\left(1 - \exp\left(-L_\tau\sqrt{k^2 + a^2}\right)\right) - \frac{\ln a^2}{2} - \frac{\ln(1/g)}{2} \right\} \right] \frac{\delta T}{T} \quad (9.20)$$

where we have use the fact that: i) $\tilde{L}_\tau\sqrt{\tilde{k}^2 + \tilde{a}^2} = L_\tau\sqrt{k^2 + a^2}$ owing to (9.12); ii) the term $\ln \tilde{a}^2 = \ln a^2 + \ln \Lambda_\infty^2$ has been reduced to $\ln a^2$ because the term $\ln \Lambda_\infty^2$ jointly appears in $F(T)$ and $F(0)$ and is cancelled when calculating $F(T) - F(0)$.

Thus the effective contribution of the left-hand side of (9.20) can be obtained by minimizing the right-hand side. Due to the fact that the value of the Λ-sum is mainly due to the infinite Λ's, it means that the integral of (9.20) must be evaluated owing to a steepest descent method. This imposes the search of the extremum of the integral argument $(p/\hbar)^2 + a^2$ i.e., $(p/\hbar)^2 + (E/\hbar c)^2$ (due to (9.15)) with respect to $p/\hbar$, where $E$ is the relativistic energy. We must have $\partial E/\partial(p/\hbar) = 0$ so that the argument is minimized when $p/\hbar = 0$ i.e., $L_\tau \to +\infty$ (or $T \to 0$ K) which is a reasonable result. We then derive that $E_{min}/\hbar c = |a|_{min}$ with $|a|_{min} = \|\gamma\|$ where $\|\gamma\|$ is the saddle-point value of the auxiliary $\gamma$-field (cf (9.7)).

Owing to (9.14) and (9.15) we can identify:

$$|a|_{min} = \|\gamma\|, \; \|\gamma\| = \min\left(\frac{|\tilde{\zeta}|\Lambda}{L_\tau}\right), \; \min\left(\frac{|\tilde{\zeta}|\Lambda}{L_\tau}\right) = \frac{m_0 c^2}{\hbar c}, \text{ as } T \to 0. \quad (9.21)$$

Near the fixed point $T = 0$ K, $|\tilde{\zeta}|\Lambda$ tends towards its minimum value given by (8.21) so that

$$\|\gamma\| = \frac{2\ln\alpha}{L_\tau}, \; \|\gamma\| = \frac{m_0 c^2}{\hbar c}. \quad (9.22)$$

Under these conditions we have

$$\frac{m_0 c^2}{\hbar c} = \frac{2\ln\alpha}{L_\tau} \quad (9.23)$$

where $\alpha = (1+\sqrt{5})/2$ is the golden mean and $C = 2\ln\alpha$ is given by (2.90). In the infinite volume system ($L_\tau \to +\infty$ i.e., at $T = 0$ K rigorously) $\|\gamma\| = 0$ i.e., $m_0 = 0$.

As a result, if considering (9.9), (9.19)-(9.23), the elementary free energy density per lattice bond $\delta\mathcal{F}(T)$ can be written near its extremum

$$\frac{\delta\mathcal{F}(T)}{8N^2} = -\hbar c C^2 \left(\frac{k_B T}{\hbar c}\right)^3 \left[\left\{\frac{L_\tau}{2} \lim_{M \to +\infty} \int_{-M/L_\tau}^{+M/L_\tau} \frac{d(p/\hbar)}{2\pi} \ln\left(\frac{(p/\hbar)^2 + \gamma^2}{\Lambda^{-2}}\right)\right.\right.$$
$$\left.\left. + \ln\left(1 - \exp\left(-L_\tau\sqrt{k^2+\gamma^2}\right)\right) - \frac{\ln(\gamma^2/g)}{2}\right\}\frac{\delta T}{T}\right], \text{ as } N \to +\infty \quad (9.24)$$

with $\ln(\gamma^2/g) \approx \gamma^2/g - 1$ as $\gamma^2/g = \left(2\ln\alpha\sqrt{|J|}/(\hbar c)\right)^2 k_B T$ is small due to (9.23) when $T \to 0$. In fact we can write $\ln(\gamma^2/g) \approx \gamma^2/g$ in (9.24) because the term $-1$ disappears when calculating $\mathcal{F}(T) - \mathcal{F}(0)$.

Finally we have to find a relationship between the infinitesimal dimensionless element $dT/T$ and the surface element $d^2k$ (here $kdk$). We have the following definitions

$$u_k = |\tilde{\zeta}|\Lambda, \; u_k = L_\tau\sqrt{k^2+\gamma^2}, \; u_k = \frac{\varepsilon_k}{k_B T} \quad (9.25)$$

where $\varepsilon_k$ is the corresponding relastivistic energy. From (6.7) we have $\delta|\zeta| = \sqrt{1+z^2}\delta|z|/|z|$ and $\delta|\tilde{\zeta}| = \sqrt{1+\tilde{z}^2}\delta|\tilde{z}|/|\tilde{z}|$. Near the critical point, $\sqrt{1+\tilde{z}_c^2} \approx 1$ as $\tilde{z}_c \ll 1$ so that $\delta|\zeta| \approx -\delta T/T$

because $\delta|\tilde{z}|/|\tilde{z}| = \delta|z|/|z| = -\delta T/T$. As a result, if respectively differentiating the first and second parts of (9.25) and assuming $\lim(\delta T_\infty/T) = dT/T$ as $T \to 0$ ($\Lambda \to +\infty$), we derive

$$du_k \approx -\frac{dT}{T}, \quad u_k du_k = L_\tau^2 k dk, \text{ as } T \to 0. \tag{9.26}$$

Near $T = 0$ K we have $u_k \approx \min(|\tilde{\zeta}|\Lambda)$ i.e., $u_k \approx C$ where $C = 2\ln\alpha$ is given by (2.90) so that (9.22) leads to the dimensionless relationship

$$L_\tau^2 C^{-1} k dk \approx -\frac{dT}{T}, \text{ as } T \to 0. \tag{9.27}$$

As a result, if $p$ is the relativistic momentum, we can introduce the following integration measure which is necessary for integrating (9.24)

$$\frac{d^3 p}{(2\pi)^3 \hbar} = \frac{d^2 k}{(2\pi)^2} \frac{d(p/\hbar)}{2\pi} \tag{9.28}$$

where $d^2 k$ reduces to $k dk$ in our case. At this step we introduce the critical coupling $g_c$:

$$\frac{1}{g_c} = \frac{1}{\hbar} \int \frac{d^3 p}{8\pi^3} \frac{1}{p^2}. \tag{9.29}$$

If using (9.26) and taking into account the previous comments, the free energy density per lattice bond can be written as a combination of *Kurzweil-Henstock integrals* in the thermodynamic limit

$$\frac{\mathcal{F}(T)}{8N^2} = \hbar c C \left[ \frac{1}{L_\tau} \int \frac{d^2 k}{4\pi^2} \ln(1 - \exp(-u_k)) + \frac{1}{2\hbar} \int \frac{d^3 p}{8\pi^3} \ln\left(\frac{p^2 + (m_0 c/\hbar)^2}{\hbar^2 \Lambda^{-2}}\right) - \frac{(m_0 c/\hbar)^2}{2g} \right],$$
$$\text{as } N \to +\infty. \tag{9.30}$$

Thus (9.30) is (9.9) after integration. The integral in the right-hand side of (9.30) is badly divergent when $p$ becomes infinite i.e., in the ultraviolet domain. However all divergences disappear if introducing the infinite volume contribution $\mathcal{F}_\infty(T) = \mathcal{F}(0)$ for which $m_0 = 0$ as seen after (9.23). As a result we have:

$$\frac{\mathcal{F}(T) - \mathcal{F}(0)}{8N^2} = \hbar c C \left[ \frac{1}{L_\tau} \int \frac{d^2 k}{4\pi^2} \ln(1 - \exp(-u_k)) + \frac{1}{2\hbar} \int \frac{d^3 p}{8\pi^3} \ln\left(\frac{p^2 + (m_0 c/\hbar)^2}{p^2}\right) \right.$$
$$\left. - \frac{(m_0 c/\hbar)^2}{2g} \right], \text{ as } N \to +\infty. \tag{9.31}$$

If $g < g_c$, owing to the definition of the spin stiffness $\rho_s$ (*cf* (2.68)) we have $1/g = 1/g_c + \rho_s/k_B T$ so that:

$$\frac{\mathcal{F}(T) - \mathcal{F}(0)}{8N^2} = \hbar c C \left[ \frac{1}{L_\tau} \int \frac{d^2k}{4\pi^2} \ln(1 - \exp(-u_k)) + \frac{1}{2\hbar} \int \frac{d^3p}{8\pi^3} \left\{ \ln\left( \frac{p^2 + (m_0 c/\hbar)^2}{p^2} \right) \right. \right.$$
$$\left. \left. - \frac{(m_0 c/\hbar)^2}{p^2} \right\} - \frac{(m_0 c/\hbar)^2}{2} \frac{\rho_s}{k_B T} \right]. \quad (9.32)$$

These integrals can be expressed in terms of polylogarithms or Jonquière functions (*cf* (3.20)). From the corresponding general expression we can derive $\mathcal{F}(T) - \mathcal{F}(0)$ for Zone 1 ($x_1 \ll 1$, *cf* (3.23)) and Zone 3 ($x_1 \gg 1$, *cf* (3.24)-(3.25)).

If $g > g_c$, the following relation between the $T = 0$ gap, $\Delta$, and the coupling $g$ is useful:

$$\frac{1}{g} = \frac{1}{\hbar} \int \frac{d^3p}{8\pi^3} \frac{1}{p^2 + (\Delta/\hbar c)^2}. \quad (9.33)$$

Reporting this relationship in (9.31) we have:

$$\frac{\mathcal{F}(T) - \mathcal{F}(0)}{8N^2} = \hbar c C \left[ \frac{1}{L_\tau} \int \frac{d^2k}{4\pi^2} \ln(1 - \exp(-u_k)) \right.$$
$$\left. + \frac{1}{2\hbar} \int \frac{d^3p}{8\pi^3} \left\{ \ln\left( \frac{p^2 + (m_0 c/\hbar)^2}{p^2 + (\Delta/\hbar c)^2} \right) - \frac{(m_0 c/\hbar)^2 - (\Delta/\hbar c)^2}{p^2 + (\Delta/\hbar c)^2} \right\} \right]. \quad (9.34)$$

As in the previous case $g < g_c$ these integrals can be expressed in terms of polylogarithms (*cf* (3.28)). Thus, from the corresponding general expression, we can derive $\mathcal{F}(T) - \mathcal{F}(0)$ for Zone 2 ($x_2 \ll 1$, *cf* (3.30)) and Zone 4 ($x_2 \gg 1$, *cf* (3.31)-(3.32)).